\documentclass[11pt]{article}

\usepackage[letterpaper,margin=1in,headheight=14pt]{geometry}
\usepackage[T1]{fontenc}
\usepackage[utf8]{inputenc}
\usepackage[sc]{mathpazo}
\usepackage[scaled=0.88]{helvet}
\usepackage[final]{microtype}

\usepackage{amsmath,amssymb,amsthm}
\usepackage{mathtools}
\newtheorem{theorem}{Theorem}
\newtheorem{lemma}[theorem]{Lemma}
\newtheorem{proposition}[theorem]{Proposition}
\newtheorem{corollary}[theorem]{Corollary}
\theoremstyle{definition}
\newtheorem{definition}{Definition}
\newtheorem{assumption}{Assumption}

\theoremstyle{remark}

\usepackage{graphicx}
\usepackage{booktabs,tabularx,multirow,makecell,array,colortbl,threeparttable}
\usepackage[table,dvipsnames]{xcolor}
\usepackage{caption,subcaption}
\usepackage{enumitem}
\setlist{itemsep=2pt,topsep=4pt,parsep=0pt}
\newcolumntype{Y}{>{\raggedright\arraybackslash}X}
\newcolumntype{L}[1]{>{\raggedright\arraybackslash}p{#1}}
\usepackage{algorithm}
\usepackage[noend]{algpseudocode}
\usepackage{listings}
\usepackage{pifont}
\usepackage{multicol}
\usepackage{fancyhdr}
\usepackage{titlesec}
\usepackage{tcolorbox}
\tcbuselibrary{skins,breakable}

\usepackage{tikz}
\usetikzlibrary{arrows.meta,positioning,fit,calc,shapes.geometric,shapes.misc,backgrounds,decorations.pathreplacing,shadows}

\usepackage[numbers,sort&compress]{natbib}
\usepackage[colorlinks=true,linkcolor=pcmblue,citecolor=pcmblue,urlcolor=pcmblue,pdfusetitle]{hyperref}
\usepackage[capitalise,noabbrev,nameinlink]{cleveref}
\crefname{assumption}{Assumption}{Assumptions}
\crefname{definition}{Definition}{Definitions}
\crefname{algorithm}{Algorithm}{Algorithms}

\definecolor{pcmblue}{HTML}{2A78D6}
\definecolor{pcmdeep}{HTML}{184F95}
\definecolor{pcmorange}{HTML}{EB6834}
\definecolor{pcmaqua}{HTML}{1BAF7A}
\definecolor{pcmyellow}{HTML}{EDA100}
\definecolor{pcmred}{HTML}{E34948}
\definecolor{pcmviolet}{HTML}{6250D6}
\definecolor{ink}{HTML}{0B0B0B}
\definecolor{ink2}{HTML}{52514E}
\definecolor{muted}{HTML}{898781}
\definecolor{rule}{HTML}{C3C2B7}
\definecolor{wash}{HTML}{F4F3F0}
\definecolor{bluewash}{HTML}{EAF2FC}
\definecolor{orangewash}{HTML}{FDEEE7}
\definecolor{aquawash}{HTML}{E6F6F0}
\definecolor{yellowwash}{HTML}{FDF4DF}
\definecolor{redwash}{HTML}{FCE9E9}

\titleformat{\section}{\normalfont\sffamily\Large\bfseries\color{ink}}{\thesection}{0.8em}{}
\titleformat{\subsection}{\normalfont\sffamily\large\bfseries\color{ink}}{\thesubsection}{0.7em}{}
\titleformat{\subsubsection}{\normalfont\sffamily\normalsize\bfseries\color{ink}}{\thesubsubsection}{0.6em}{}
\titleformat{\paragraph}[runin]{\normalfont\normalsize\bfseries\color{ink}}{}{}{}[.]
\titlespacing*{\section}{0pt}{2.0ex plus .5ex minus .2ex}{1.0ex}
\titlespacing*{\subsection}{0pt}{1.6ex plus .4ex minus .2ex}{0.7ex}
\titlespacing*{\paragraph}{0pt}{1.0ex plus .3ex}{0.6em}

\renewcommand{\headrulewidth}{0.4pt}
\renewcommand{\headrule}{\hbox to\headwidth{\color{rule}\leaders\hrule height \headrulewidth\hfill}}
\fancypagestyle{plain}{\fancyhf{}\renewcommand{\headrulewidth}{0pt}\fancyfoot[C]{\sffamily\footnotesize\color{ink2}\thepage}}

\lstdefinelanguage{json}{
  morestring=[b]", stringstyle=\color{pcmdeep},
  literate=
   {:}{{{\color{ink2}:}}}1 {,}{{{\color{ink2},}}}1
   {\{}{{{\color{ink}\{}}}1 {\}}{{{\color{ink}\}}}}1
   {[}{{{\color{ink}[}}}1 {]}{{{\color{ink}]}}}1
   {—}{{---}}1 {ü}{{\"u}}1,
}
\newcommand{\tk}{\textsf{Tracekit}}
\newcommand{\code}[1]{\texttt{\small #1}}
\DeclareUrlCommand\vectraw{\urlstyle{tt}}
\DeclareRobustCommand\vect[1]{\vectraw{#1}}

\newcommand{\prov}{\textsf{\color{pcmaqua!80!black}proven}}
\newcommand{\unprov}{\textsf{\color{pcmyellow!80!black}unproven}}
\newcommand{\fail}{\textsf{\color{pcmred}failed}}
\newcommand{\vacc}{\textsf{accept}}
\newcommand{\vunp}{\textsf{accept-unproven}}
\newcommand{\vrej}{\textsf{reject}}
\newcommand{\vbad}{\textsf{bad-input}}
\newcommand{\T}[1]{\textsf{T#1}}
\newcommand{\circled}[1]{\textcolor{pcmblue}{\ding{\numexpr181+#1\relax}}}
\newcommand{\JCS}{\mathsf{JCS}}
\newcommand{\Hh}{\mathsf{H}}
\newcommand{\Sig}{\mathsf{Sig}}
\newcommand{\Vf}{\mathsf{Vf}}
\newcommand{\Sat}{\mathsf{Sat}}
\newcommand{\Ran}{\textsc{Ran}}

\newcommand{\cmark}{\textcolor{pcmaqua!85!black}{\ding{51}}}
\newcommand{\xmark}{\textcolor{pcmred}{\ding{55}}}
\newcommand{\pmark}{\textcolor{pcmyellow!85!black}{$\circ$}}

\newtcolorbox{keybox}[1][]{enhanced,colback=bluewash,colframe=pcmblue,boxrule=0pt,leftrule=2.5pt,
  arc=0pt,outer arc=0pt,left=8pt,right=8pt,top=5pt,bottom=5pt,fontupper=\small,#1}
\newtcolorbox{limitbox}[1][]{enhanced,colback=orangewash,colframe=pcmorange,boxrule=0pt,leftrule=2.5pt,
  arc=0pt,outer arc=0pt,left=8pt,right=8pt,top=5pt,bottom=5pt,fontupper=\small,#1}

\tikzset{
  box/.style={draw=#1, fill=#1!8, rounded corners=2.5pt, line width=0.7pt, align=center,
              font=\sffamily\footnotesize, inner sep=4pt},
  box/.default=pcmblue,
  lbl/.style={font=\sffamily\scriptsize, text=ink2, align=center},
  arr/.style={-{Stealth[length=5pt,width=4pt]}, line width=0.7pt, draw=ink2},
  darr/.style={-{Stealth[length=5pt,width=4pt]}, line width=0.7pt, draw=ink2, dashed},
  zone/.style={draw=rule, rounded corners=4pt, dashed, line width=0.6pt, inner sep=7pt},
}

\hypersetup{pdftitle={PCM: A Proof-Carrying Message Protocol for Verifiable Multi-Agent Communication},
  pdfauthor={Bravish Ghosh}}

\begin{document}
\emergencystretch=2.5em

\thispagestyle{plain}
\begin{center}
{\sffamily\bfseries\LARGE\color{ink} PCM: A Proof-Carrying Message Protocol\\[2pt]
for Verifiable Multi-Agent Communication\par}
\vspace{8pt}
{\large Bravish Ghosh\par}
\vspace{2pt}
{\normalsize Cygnux Labs\par}
{\small\ttfamily \href{mailto:research@cygnuxlabs.com}{research@cygnuxlabs.com}\par}
\end{center}
\vspace{2pt}

\begin{abstract}
\noindent
When one AI agent tells another that \emph{``the tests passed''} or \emph{``a person approved this refund''}, the
receiver has only the sender's word. A2A and MCP authenticate the channel, but the sender is the party in doubt. A
prompt-injected agent can launder fabricated facts and authority through a multi-agent system. We present
\textbf{Proof-Carrying Messages (PCM)}, a protocol in which each claim carries \emph{evidence}: records from an
append-only Merkle log that the agent cannot write, with inclusion proofs, signed by an \emph{attesting signer} the
agent cannot control. Receivers check messages offline and give each claim a three-valued result: \prov, \unprov{} or
\fail. A claim that rests only on the agent's own report is never \prov. Authority travels as grant chains rooted in a
recorded human approval, and the chains can only narrow. We specify four claim types, a typed caveat algebra, replay
rules that survive restarts, and ``strict readings'' that make independent verifiers agree, and we prove the resulting
security properties. On a pre-specified benchmark of 926 messages from real runs, 712 of them lies in 16 classes, PCM
accepts 10 lies, all of them stale evidence (1.4\%, 95\% CI 0.8--2.6\%). Plain A2A accepts all 712, and models of six
deployed protocols accept 504--712. Two independent verifiers agree on all 926 verdicts, and verification takes
$\approx$1\,ms.
\end{abstract}

\vspace{-2pt}
\begin{center}
\includegraphics[width=0.9\linewidth]{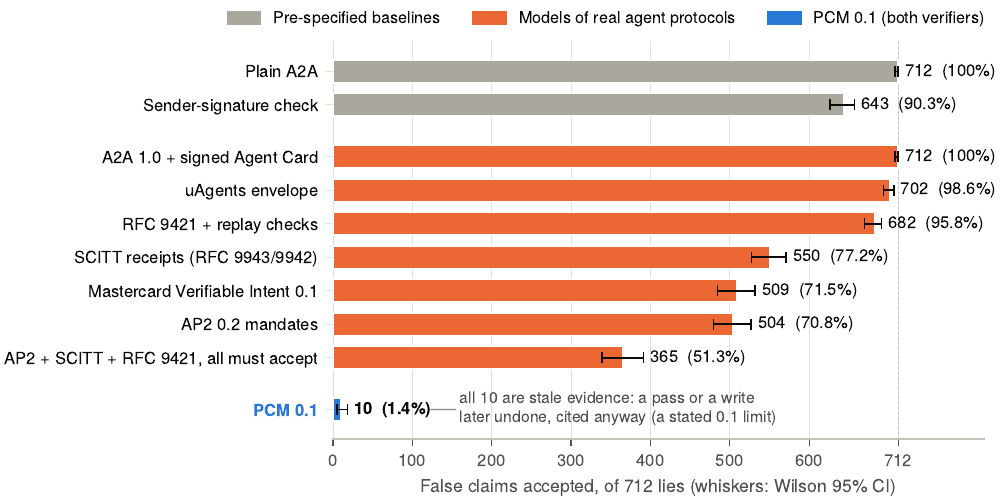}
\vspace{-4pt}
\captionof{figure}{\textbf{How many false claims does a receiving agent accept?} The 712 lies come from real signer
runs and span 16 classes of content, evidence, delivery, authority and encoding lies (\cref{sec:eval}). Grey:
pre-specified baselines. Orange: specification-faithful models of six deployed agent protocols, and three of them
stacked. Blue: PCM~0.1, whose 10 false accepts are all \emph{stale evidence}, a record of an event that really happened
and was later undone (\cref{sec:limits}).}
\label{fig:teaser}
\end{center}

\section{Introduction}
\label{sec:intro}

Large language model (LLM) agents increasingly work in teams. A planning agent delegates to a coding agent, a
customer-service agent asks a payments agent to issue a refund, and a research agent hands results to a writing agent.
Protocols such as Google's Agent2Agent (A2A)~\cite{a2a} and Anthropic's Model Context Protocol (MCP)~\cite{mcp} have
standardised how such messages are framed, addressed and transported. What they leave open is the question every
receiving agent faces: \emph{should I believe what this message says happened?}

Consider agent $A$ asking agent $B$ to refund \$120 on order 81. Alongside the request, $A$ asserts three facts:
\begin{enumerate}[label=(\roman*),leftmargin=2.2em]
\item ``alice approved refunds up to \$200 for this order'';
\item ``the refund tests passed''; and
\item ``I fetched the order from \code{api.example.com}''.
\end{enumerate}
$B$ has only $A$'s word for each. If $A$ was prompt-injected~\cite{greshake2023indirect,perez2022ignore}, if a
hallucination turned a failed tool call into a reported success, or if another process sends the message in $A$'s
name, $B$ pays out anyway. Authenticating the channel does not help: TLS, mTLS, a signed A2A Agent Card or an RFC~9421
HTTP signature~\cite{rfc9421} all prove \emph{who} sent the message, and the sender is the very party whose claims are
in doubt. A lying agent that signs with its own key signs its own lie.

The problem is sharper for LLM agents than for conventional services, for three reasons. First, an agent's behaviour is
driven by untrusted text, so any agent that reads a web page, an email or a tool result may be steered by an
adversary~\cite{greshake2023indirect,debenedetti2024agentdojo,zhan2024injecagent}. Second, LLMs are known to report
tool outcomes that did not occur~\cite{basu2026toolreceipts}, so even an uncompromised agent's self-report is weak
evidence. Third, multi-agent pipelines compose: a single false claim accepted at one hop becomes a premise for every
agent downstream, and authority granted to one agent is easily re-presented by another~\cite{lotfi2026a2abreak,
south2025authenticated}. What is missing is not a better transport but a way for the \emph{content} of a message to
carry its own justification.

\paragraph{Our approach}
We take our cue from proof-carrying code~\cite{necula1996safe,necula1997pcc}. There the producer of untrusted code
ships a proof of a safety property, and the consumer checks it with a small checker instead of trusting the producer.
\emph{Proof-Carrying Messages} (PCM) apply the same shape to agent communication. The ``proof'' is not a formal
derivation but cryptographic evidence that an event was recorded by a party \emph{other than the agent}:
\begin{enumerate}[leftmargin=2.2em]
\item The agent's tool calls go through an \textbf{attesting signer}, a service the agent cannot control. It holds
  every key, decides policy, and records each call in an append-only Merkle log that the agent cannot write. Calls that
  matter run through an attesting gateway or executor, which records the result itself.
\item When the agent sends a message, the signer signs it \emph{as the calling run}, and the message carries the log
  records that back each claim, with RFC~9162 inclusion proofs~\cite{rfc9162} under a signed checkpoint.
\item The receiver checks the signature, the proofs and the records \emph{offline}, against keys it pins, and gives
  each claim one of three results: \prov, \unprov{} or \fail.
\end{enumerate}
Authority is handled in the same spirit. A request travels with a \emph{grant chain} whose root exists only because a
named person approved its exact terms, and each later link can only narrow its parent under a fixed, typed implication
table. No agent ever holds a signing key. The signer checks each delegation before signing it, and the receiver checks
the whole chain again.

\paragraph{Why this is not straightforward}
Signing log entries is old technology. The difficulty lies in the \emph{semantics} that sit between a record and a
claim, and in making those semantics identical across implementations. Building PCM, and attacking it with a
benchmark, surfaced several classes of subtle failure (\cref{sec:security,sec:determinism}):
\begin{itemize}
\item \textbf{Self-reported success.} A tool result the agent's own process reported is the agent's word. Unless the
  verifier reads the record's \emph{tier} from fields the signer sets, a T1 report is indistinguishable from an
  attested T2 execution.
\item \textbf{Names the agent chose.} An agent names its own calls. A person's approval of a call called
  \code{refund-81-approved} says nothing about what the call was; it may have been a \code{read\_file}. Binding an
  approval to what was actually decided needs a three-record linkage and an explicit ``covers'' check.
\item \textbf{Two readers, two answers.} \code{0127.0.0.1} is \code{87.0.0.1} to \code{inet\_aton} and
  \code{127.0.0.1} to the macOS resolver; \code{http://a.example\textbackslash@b.example/} has a different host for a
  browser and for a generic URL parser; \code{pytest \$'\textbackslash'';true \#'} is one command to Python's
  \code{shlex} and two to \code{sh}; \code{1e400} is infinity to one JSON parser and an error to another. Each such gap
  let one verifier prove what another refused, or let a sender prove a claim that was false.
\end{itemize}

\paragraph{Contributions}
\begin{enumerate}[leftmargin=2.2em]
\item \textbf{A protocol} (\cref{sec:overview,sec:formal,sec:claims,sec:grants,sec:verify}). PCM~0.1 specifies a
  closed message envelope signed by an attesting signer; content-addressed intent schemas; evidence as Merkle-proven
  log records with HMAC-committed, selectively opened arguments; four claim types with exact three-valued checkers; a
  typed caveat algebra for person-rooted, attenuating grant chains; and replay rules that survive restarts in bounded
  memory.
\item \textbf{A three-valued semantics that separates doubt from falsity} (\cref{sec:formal}). A claim is \prov{}
  only on records written by a party other than the agent. It is \unprov{} when the evidence is absent, is only the
  agent's word, shows less than the claim (an attempt, not a connection), or cannot be matched with certainty. It is
  \fail{} when it is malformed or contradicted. The verdict lattice lets receivers act on what is proven and treat the
  rest as testimony.
\item \textbf{Security properties with proofs} (\cref{sec:security}): origin authenticity, evidence soundness, T1
  non-elevation, authority attenuation, and at-most-once acceptance across restarts, under stated assumptions. We also
  give a catalogue of 29 attacks, mistakes and trust assumptions, each with the rule that closes or bounds it.
\item \textbf{Determinism by construction} (\cref{sec:determinism}). This is a methodology for security protocols whose
  verdicts depend on parsing. Wherever implementations disagree, PCM adopts the strict reading every verifier can apply
  alike, and a claim outside it becomes \unprov{} with an actionable explanation. 187 conformance vectors pin the result.
\item \textbf{PCM-Lies, a benchmark of false inter-agent claims} (\cref{sec:eval}). It contains 926 messages from real
  signer runs (real \code{pytest} suites, HTTP fetches, file writes and human approvals), with 712 lies in 16 classes,
  labelled from ground truth rather than by any verifier. On it we compare PCM with plain A2A, a sender-signature check
  and specification-faithful models of six deployed protocols: A2A~1.0 with signed Agent Cards, Fetch.ai uAgents,
  RFC~9421 with replay checks, SCITT receipts, AP2 mandates and Mastercard Verifiable Intent.
\item \textbf{Open artefacts}: the specification, JSON Schemas, the frozen tool registry, 187 vectors, an independent
  standalone verifier, the reference implementation (\tk{}~1.5.0~\cite{ghosh2026tracekit}) and the benchmark, all under the MIT licence.
\end{enumerate}

\paragraph{Headline results}
PCM accepts 10 of 712 lies (1.4\%; 95\% CI 0.8--2.6\%), and all 10 are of one kind, stale evidence: a suite passed, the
code changed and the suite failed, and the sender cites the pass. Every other lie, 702 of 712, is either rejected (626)
or explicitly flagged \vunp{} (76), across the claim, evidence, authority, delivery, encoding and message scopes. All
72 claims backed only by the agent's own report are \unprov. Two independent verifiers, written separately, agree on
the verdict and reason class of all 926 messages. The deployed protocols we model accept 70.8--100\% of the same lies;
stacking AP2, SCITT and RFC~9421 still accepts 51.3\%. A PCM message is 1.9--11.7\,KB at the median (4.2--6.1\,KB with one claim) and verifies in 0.8--1.2\,ms at the
median.

\paragraph{What PCM does not do}
PCM does not make a compromised agent honest: a prompt-injected agent can still ask for anything its grants allow, and
the grant, not the agent, sets the limit. PCM authenticates the \emph{signer's records}, so the signer is the trusted
computing base. A claim says that something \emph{happened}, not that it still holds. We state these limits precisely
in \cref{sec:model,sec:limits} and show that they explain every remaining error in the evaluation.

\paragraph{Roadmap}
\Cref{sec:background} gives background. \Cref{sec:model} fixes the system and threat model, and \cref{sec:overview}
gives a protocol overview. \Cref{sec:formal} formalises claims and verdicts, \cref{sec:claims} details the four claim
types, \cref{sec:grants} covers grants, and \cref{sec:verify} gives the verification and replay algorithms.
\Cref{sec:security} states and proves security properties, \cref{sec:determinism} presents determinism by
construction, \cref{sec:impl} describes the implementation, and \cref{sec:eval} evaluates PCM. \Cref{sec:limits}
discusses limitations, \cref{sec:related} relates PCM to prior work, and \cref{sec:conclusion} concludes.

\section{Background}
\label{sec:background}

\paragraph{Agent protocols}
A2A~\cite{a2a} defines how agents discover each other (through an \emph{Agent Card}, which A2A~1.0 lets the card's
owner sign as a JWS), exchange messages made of typed \emph{parts}, and manage long-running tasks. MCP~\cite{mcp}
connects a model to tools and resources. Both carry content and can run over authenticated transports. Neither defines
a way for the receiver of a message to check, independently of the sender, a statement about what the sender did. ANP
signs requests with RFC~9421 HTTP Message Signatures~\cite{rfc9421} under the sender's DID key, adds a time window and a
nonce cache, and states explicitly that authentication alone does not imply business authority. In the payments
domain, AP2~\cite{ap2} and Mastercard's Verifiable Intent~\cite{mastercardvi} carry signed \emph{mandates}: SD-JWT
credentials that record who authorised a purchase, under which constraints and until when.

\paragraph{Transparency logs}
Certificate Transparency (CT)~\cite{rfc6962,rfc9162} introduced append-only Merkle logs whose operators sign a
\emph{checkpoint} (a tree size and root hash). Any party can verify that a leaf is in the tree with an \emph{inclusion
proof} of $\lceil\log_2 n\rceil$ hashes. Leaves are hashed as $\mathsf{SHA256}(\texttt{0x00}\,\|\,\ell)$ and interior
nodes as $\mathsf{SHA256}(\texttt{0x01}\,\|\,a\,\|\,b)$, which separates the two domains. The C2SP specifications fix
a text format for signed notes~\cite{c2sp-note}, for checkpoints~\cite{c2sp-checkpoint} and for witness
cosignatures~\cite{c2sp-cosig}. A \emph{witness} cosigns a checkpoint only if it is consistent with every checkpoint
the witness saw before, which makes a log that shows different histories to different parties detectable. SCITT
(RFC~9943, with COSE Receipts, RFC~9942)~\cite{rfc9943,rfc9942} generalises this to signed statements about
supply-chain artefacts.

\paragraph{Canonical JSON}
The JSON Canonicalization Scheme (JCS, RFC~8785)~\cite{rfc8785} maps a JSON value to a unique byte string: object
members sorted by UTF-16 code units, numbers in the ECMAScript shortest round-trip form, and minimal string escaping.
PCM hashes and signs only JCS bytes, so two implementations that parse the same value sign and verify the same bytes.

\paragraph{Capability tokens}
Macaroons~\cite{birgisson2014macaroons}, Biscuit~\cite{biscuit} and UCAN~\cite{ucan} are bearer or holder-of-key
credentials that a holder can \emph{attenuate} offline by adding caveats. Biscuit expresses caveats and policies in
Datalog. In all three, the delegating party holds the key that attenuates.

\paragraph{Proof-carrying code and provenance}
Proof-carrying code~\cite{necula1996safe,necula1997pcc} ships a machine-checkable proof with untrusted code, so that a
small trusted checker replaces trust in the producer. Supply-chain provenance frameworks such as
in-toto~\cite{torresarias2019intoto} and SLSA~\cite{slsa} sign statements about build steps. SLSA in particular holds
that provenance should be generated by the build \emph{platform}, not by the build's own (possibly compromised) steps.
PCM's T1/T2 distinction (\cref{sec:formal}) is the per-claim analogue of that principle.

\paragraph{Running example}
Throughout, agent $A$ (identity \code{uid:990101}) sends agent $B$ (\code{uid:990102}) the intent
\code{refund.request} with arguments \code{\{"amount": 12000, "order": 81\}} (amounts are integer cents). The
message carries a root grant that alice (\code{uid:990001}) approved, \code{amount <= 20000} and \code{order == 81} for
one hour and not delegable, together with an \code{approved\_by} claim that cites her approval. Every example in this
paper is taken from a conformance vector produced by the reference implementation's real signer
(\cref{app:example}).

\section{System and Threat Model}
\label{sec:model}

\subsection{Parties}

\begin{table}[t]
\centering\small
\caption{Parties in PCM and what each is trusted for. ``Writes'' lists what enters the log or the message on that
party's authority.}
\label{tab:parties}
\begin{tabularx}{\linewidth}{@{}L{3.1cm} Y Y L{2.3cm}@{}}
\toprule
\textbf{Party} & \textbf{Role} & \textbf{Writes} & \textbf{Trusted?}\\
\midrule
Agent (sender) & Calls tools, composes the message body & \code{from}, \code{to}, \code{intent}, \code{claims},
  \code{grants}, \code{payload}; T1 tool results & \xmark\\
Attesting signer & Holds keys, decides policy, appends records, signs messages, grants and checkpoints & \code{id},
  \code{ts}, \code{nonce}, \code{sig}; record \code{seq}, \code{source}, \code{tier}; approvals; effects & \cmark{} (TCB)\\
Attesting gateway / executor & Runs a tool call on the agent's behalf and reports the result & T2 tool results
  (marked by the signer from the authenticated caller) & \cmark{} (per tool class)\\
Observer & Egress proxy or kernel sensor that reports network effects & \code{effect.observed} records & \cmark\\
Person & Human approver named in the signer's approvals configuration & approval answers & \cmark{} (for what they approve)\\
Witness & Cosigns checkpoints consistent with its view & cosignatures & optional\\
Receiver & Verifies, keeps replay state, decides to act & nothing & ---\\
\bottomrule
\end{tabularx}
\end{table}

\Cref{tab:parties} lists the parties. An \emph{agent} is any program that calls tools and sends messages; PCM never
trusts it. A \emph{run} is one session of an agent within a log. It has an identifier and a \emph{tenant} (an isolation
unit within one log), and the signer binds it to the transport identity that registered it, written
\code{scheme:subject}, for example \code{uid:990101} or an mTLS SPIFFE ID. The \emph{attesting signer} is a service
the agent reaches over an authenticated channel. It keeps the run's events in an append-only Merkle log and signs, with
its \emph{log key}, the log's checkpoints, PCM messages and PCM grants. The \emph{receiver} pins the log keys it trusts
in a trust document and keeps replay state.

\subsection{Adversary}

We consider an adversary that \textbf{fully controls the sending agent}. It may be a prompt-injected LLM, a
hallucinating LLM, malicious code in the agent's process, or another process that holds the agent's credential. The
adversary
\begin{itemize}
\item composes arbitrary message bodies, claims, evidence selections and grant chains, and may submit any body to the
  signer for signing;
\item reports arbitrary T1 tool results, chooses the identifiers (\code{tool\_use\_id}) of its own calls, and names
  tools as it likes;
\item writes any file the agent can write, including test-runner configuration in its working directory;
\item records, replays, reorders, drops and redirects messages on the network, and sends messages to any receiver;
\item operates other agents, other runs, other tenants and even other signers, whose keys a receiver may also pin;
\item reads every message, grant and evidence record it has ever been sent.
\end{itemize}
The adversary does \textbf{not} control the signer, its keys or its log; cannot make a T2 executor or an observer
report something that did not happen; and cannot impersonate a configured person to the signer. These are trust
assumptions about the deployment, and we make them explicit as \cref{as:signer} below. A \emph{signer-bug} variant, in
which the signer signs grants it should have refused, is evaluated separately (\cref{sec:eval}).

\subsection{Goals}

\begin{description}[leftmargin=0pt,style=unboxed,font=\normalfont\bfseries]
\item[G1 Origin.] A receiver accepts a message only if an honest signer signed that exact body, as the run that
  appears in \code{from}.
\item[G2 Evidence soundness.] A claim is \prov{} only if the sender's run contains signer-appended records, written by
  a party other than the agent, that establish exactly what the claim states.
\item[G3 No self-attestation.] No combination of the agent's own reports (T1) makes a claim \prov.
\item[G4 Attenuated authority.] Any authority a message presents is no wider than what a configured person approved:
  never widened at a delegation step, never past its expiry, and never moved to another tenant or log.
\item[G5 Freshness of delivery.] A receiver acts on each message at most once, including across restarts and replicas,
  and only if the message is addressed to it.
\item[G6 Determinism.] Any two conforming verifiers return the same verdict, claim results and conformance fields on
  the same inputs.
\item[G7 Honest progress.] Honest senders who follow the specification get their claims \prov{} and their messages
  accepted.
\end{description}

\subsection{Non-goals}

PCM does not (N1) make the agent's \emph{intentions} trustworthy, since a compromised agent may send any request its
grants allow; (N2) prove that a state is \emph{current}, since a claim is existential, ``a call happened'', and a later
event can undo it; (N3) prove \emph{absence}, since the sender chooses which records to disclose; (N4) provide
confidentiality, which TLS does; or (N5) protect against a compromised signer or a leaked log key
(\cref{sec:keycompromise}).

\subsection{Assumptions}

\begin{assumption}[Cryptography]\label{as:crypto}
Ed25519 with the strict verification rules of \cref{sec:sigrules} is existentially unforgeable under chosen-message
attack (EUF-CMA). SHA-256 is collision-resistant. HMAC-SHA256 is a binding commitment when keyed by a uniformly random
32-byte salt.
\end{assumption}

\begin{assumption}[Honest signer]\label{as:signer}
The signer for each pinned log (i)~signs a message only as the calling run, and sets \code{id}, \code{ts} and
\code{nonce} itself; (ii)~assigns each record's \code{seq} and \code{ts} as it appends it; (iii)~sets a record's
top-level \code{source}/\code{tier} to \code{gateway}/\code{T2} only for results reported by an authenticated attesting
gateway or executor; (iv)~writes \code{approval} and \code{approval.request} records only from authenticated answers
and its own requests, and \code{effect.observed} only from observer reports; (v)~issues a root grant only under
\cref{sec:rootgrants}'s conditions and a child only under \cref{sec:delegation}'s; and (vi)~never lets an agent read
its keys or write its log.
\end{assumption}

\begin{assumption}[Honest attesting parties]\label{as:t2}
An attesting gateway or executor records a call under a tool name that the frozen placement table maps to class $k$ only
if the tool it ran is of class $k$, and records \code{ok} only if the call succeeded.
\end{assumption}

\begin{assumption}[Clocks and receivers]\label{as:clock}
The clocks of signer and receiver agree within the receiver's replay window. A receiver keeps one replay state per
identity, shared by all replicas that answer for that identity.
\end{assumption}

Assumption~\ref{as:signer} makes the signer the trusted computing base, much as a CT log operator, a SLSA build
platform or a hardware root of trust is trusted in its own setting. \Cref{sec:keycompromise} analyses what follows if it
fails. We note that the assumption is \emph{operational}: a deployment satisfies it by running the signer as a
different user, on a different host or in a different trust domain from the agents it records.

\section{Protocol Overview}
\label{sec:overview}

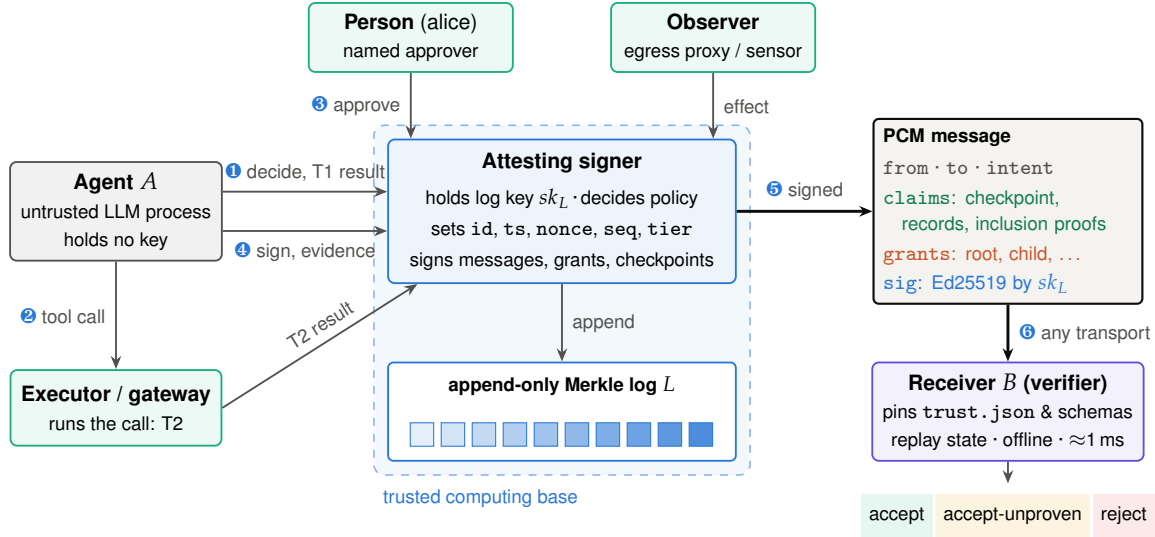
\begin{figure}[t]
\centering
\begin{tikzpicture}[x=1cm,y=1cm]
  \node[box=pcmaqua, minimum width=27mm, minimum height=9mm] (person) at (3.0,2.65)
     {\textbf{Person} (alice)\\[-1pt]{\scriptsize named approver}};
  \node[box=pcmaqua, minimum width=27mm, minimum height=9mm] (obs) at (7.0,2.65)
     {\textbf{Observer}\\[-1pt]{\scriptsize egress proxy / sensor}};
  \node[box=ink2, minimum width=27mm, minimum height=12mm] (agent) at (-0.9,0.35)
     {\textbf{Agent $A$}\\[-1pt]{\scriptsize untrusted LLM process}\\[-1pt]{\scriptsize holds no key}};
  \node[box=pcmaqua, minimum width=27mm, minimum height=10mm] (exec) at (-0.9,-2.25)
     {\textbf{Executor / gateway}\\[-1pt]{\scriptsize runs the call: \T2}};
  \node[box=pcmblue, minimum width=46mm, minimum height=19mm] (signer) at (5,0.35)
     {\textbf{Attesting signer}\\[2pt]
      {\scriptsize holds log key $sk_L$\,\textperiodcentered\,decides policy}\\
      {\scriptsize sets \texttt{id}, \texttt{ts}, \texttt{nonce}, \texttt{seq}, \texttt{tier}}\\
      {\scriptsize signs messages, grants, checkpoints}};
  \node[draw=pcmblue, fill=white, minimum width=46mm, minimum height=13mm, rounded corners=2pt, line width=0.7pt] (log) at (5,-2.3) {};
  \node[font=\sffamily\scriptsize, anchor=north] at ($(log.north)+(0,-0.6mm)$) {\textbf{append-only Merkle log $L$}};
  \foreach \i in {0,...,9} {
    \pgfmathtruncatemacro{\shade}{12+\i*8}
    \node[draw=pcmblue, fill=pcmblue!\shade, minimum width=3.1mm, minimum height=3.1mm, inner sep=0pt]
      at ($(log.south west)+(4.6mm+\i*4.1mm,3.6mm)$) {};
  }
  \node[font=\sffamily\tiny, text=ink2, anchor=south east] at ($(log.south east)+(-1mm,0.5mm)$) {};
  \node[draw=ink, fill=wash, rounded corners=2pt, text width=33mm, minimum height=21mm, line width=0.8pt,
        font=\sffamily\scriptsize, align=left] (msg) at (10.9,0.35)
     {\textbf{PCM message}\\[2pt]
      \textcolor{ink2}{\texttt{from} \textperiodcentered{} \texttt{to} \textperiodcentered{} \texttt{intent}}\\[1pt]
      \textcolor{pcmaqua!70!black}{\texttt{claims}: checkpoint,}\\
      \textcolor{pcmaqua!70!black}{\quad records, inclusion proofs}\\[1pt]
      \textcolor{pcmorange!85!black}{\texttt{grants}: root, child, \ldots}\\[1pt]
      \textcolor{pcmblue}{\texttt{sig}: Ed25519 by $sk_L$}};
  \node[box=pcmviolet, minimum width=35mm, minimum height=13mm] (recv) at (10.9,-2.3)
     {\textbf{Receiver $B$ (verifier)}\\[-1pt]{\scriptsize pins \texttt{trust.json} \& schemas}\\[-1pt]{\scriptsize replay state \textperiodcentered{} offline \textperiodcentered{} $\approx$1\,ms}};
  \node[font=\sffamily\scriptsize, anchor=north] (out) at (10.9,-3.25)
     {\colorbox{aquawash}{\strut accept}\,\colorbox{yellowwash}{\strut accept-unproven}\,\colorbox{redwash}{\strut reject}};
  \begin{scope}[on background layer]
    \node[zone, draw=pcmblue!55, fill=bluewash!55, fit=(signer)(log), inner sep=5pt] (tcb) {};
  \end{scope}
  \node[lbl, anchor=north west, text=pcmblue] at ($(tcb.south west)+(0,-0.5mm)$) {trusted computing base};
  \draw[arr] ($(agent.east)+(0,2.6mm)$) -- node[lbl, above] {\circled{1} decide, T1 result} ($(signer.west)+(0,2.6mm)$);
  \draw[arr] ($(agent.east)+(0,-2.6mm)$) -- node[lbl, below] {\circled{4} sign, evidence} ($(signer.west)+(0,-2.6mm)$);
  \draw[arr] (agent) -- node[lbl, left] {\circled{2} tool call} (exec);
  \draw[arr] (exec.east) -- node[lbl, above, sloped, pos=0.55] {T2 result} ($(signer.south west)+(4mm,0)$);
  \draw[arr] (person.south) -- node[lbl, left] {\circled{3} approve} (person.south |- signer.north);
  \draw[arr] (obs.south) -- node[lbl, right] {effect} (obs.south |- signer.north);
  \draw[arr] (signer.south) -- node[lbl, right] {append} (log.north);
  \draw[arr, line width=1.1pt, draw=ink] (signer.east) -- node[lbl, above] {\circled{5} signed} (msg.west);
  \draw[arr, line width=1.1pt, draw=ink] (msg.south) -- node[lbl, right] {\circled{6} any transport} (recv.north);
  \draw[arr] (recv.south) -- (out.north);
\end{tikzpicture}
\caption{\textbf{PCM architecture.} \circled{1}--\circled{3}: every tool call of agent $A$ is decided and recorded by
the attesting signer. Results come from the agent's own process (\T1) or from an attesting executor or gateway (\T2);
approvals come from named people; network effects come from observers. \circled{4}: when $A$ sends a message, it asks
the signer to sign the body as its run, and for records with inclusion proofs. \circled{5}--\circled{6}: the receiver
verifies everything offline against keys it pins. The agent never holds a key.}
\label{fig:arch}
\end{figure}

\Cref{fig:arch} shows a PCM deployment. Agent $A$ runs inside its own process with no access to signing keys. Each tool
call first goes to the signer, which records a \code{policy.decision} and answers \code{allow}, \code{flag}, \code{ask}
or \code{deny}. A call that needs a person's answer (\code{ask}) produces an \code{approval.request} record and then
an \code{approval} record. Calls that the deployment routes through an attesting executor or gateway are run there, and
the executor reports the result to the signer over its own authenticated channel, so the signer marks the
\code{tool.result} record as \T2. Calls the agent runs itself are reported by the agent, and the signer marks them \T1.
Observers report network effects. All of these become records in the log. Each record is a JSON event whose hash is
$\Hh(e) = \texttt{"sha256:"}\,\|\,\mathrm{hex}(\mathsf{SHA256}(\JCS(e)))$, and whose \code{seq} is its leaf index.

To send a message, $A$ composes the body (\code{from}, \code{to}, \code{intent}, optional \code{claims},
\code{grants} and \code{payload}) and asks the signer to sign it. The signer refuses unless \code{from} is exactly
the calling run's identity, run and tenant; it sets \code{pcm}, \code{id}, \code{ts} and \code{nonce} from its own
clock and randomness, signs, and (in \tk) records that it signed (\cref{sec:sigrules}). For each claim, $A$ fetches from the
signer the records it wants to disclose, each with an inclusion proof under the latest checkpoint, and places them in
the claim's \code{evidence} \emph{before} signing, so that the message signature covers them too.

\begin{figure}[t]
\centering
\resizebox{\linewidth}{!}{\begin{tikzpicture}[x=1cm,y=1cm,
  f/.style={draw=#1, fill=#1!9, rounded corners=1.5pt, line width=0.5pt, inner sep=2.4pt, anchor=north west,
            font=\ttfamily\fontsize{6.9}{8.2}\selectfont, text=ink},
  f/.default=ink2,
  hdr/.style={font=\sffamily\scriptsize\bfseries, anchor=north west, inner sep=0pt, text=#1},
]
  \def\gap{0.13}
  \node[hdr=ink] (h) at (0.2,0) {application/vnd.pcm+json\quad\textmd{\textcolor{ink2}{closed object \textperiodcentered{} signed bytes $=\JCS(\text{message}\setminus\texttt{sig}\,\cup\,\{\texttt{"t"}:\texttt{"pcm.message.v0"}\})$}}};
  \node[f=pcmblue] (r1a) at (0.2,-0.42) {"pcm": 0};
  \node[f=pcmblue, anchor=west] (r1b) at ($(r1a.east)+(\gap,0)$) {"id": "0ac65318\ldots"};
  \node[f=pcmblue, anchor=west] (r1c) at ($(r1b.east)+(\gap,0)$) {"ts": "2026-10-28T02:24:35.496285Z"};
  \node[f=pcmblue, anchor=west] (r1d) at ($(r1c.east)+(\gap,0)$) {"nonce": "705e9bb5\ldots"};
  \node[f] (r2a) at (0.2,-0.95) {"from": \{"agent": "uid:990101", "run": "ec130a62\ldots", "tenant": "default"\}};
  \node[f, anchor=west] (r2b) at ($(r2a.east)+(\gap,0)$) {"to": \{"agent": "uid:990102"\}};
  \node[f] (r3) at (0.2,-1.48) {"intent": \{"type": "refund.request@sha256:d8636b6f\ldots", "args": \{"amount": 12000, "order": 81\}\}};
  \node[hdr=pcmaqua!65!black] (ch) at (0.45,-2.12) {"claims": [ \ldots ] \quad\textmd{\textcolor{ink2}{at most 8; each \{type, args, evidence?\}}}};
  \node[f=pcmaqua] (c1) at (0.45,-2.5) {"type": "approved\_by", "args": \{"person": "alice", "call": "grant-1", "covers": true\}};
  \node[f=pcmaqua] (c2) at (0.7,-3.03) {"checkpoint": "example.com/pcm-test-log\textbackslash n9\textbackslash nQhInJS9+\ldots\textbackslash n\textbackslash n--- example.com/pcm-test-log C7keZoQn\ldots"};
  \node[f=pcmaqua] (c3) at (0.7,-3.56) {"records": [ \{"event": \{\ldots policy.decision, seq 3\ldots\}, "hash": "sha256:444e\ldots", "proof": [4 hashes]\},};
  \node[f=pcmaqua] (c4) at (0.7,-4.04) {\phantom{"records": [ }\{\ldots approval.request, seq 4\ldots\}, \{\ldots approval, seq 5\ldots\} ]\quad\textrm{\textcolor{ink2}{(+ "salt", "args" to open a decision)}}};
  \begin{scope}[on background layer]
    \node[draw=pcmaqua, rounded corners=3pt, line width=0.7pt, fit=(ch)(c1)(c2)(c3)(c4), inner sep=4pt] {};
  \end{scope}
  \node[hdr=pcmorange!85!black] (gh) at (0.45,-4.85) {"grants": [ \ldots ] \quad\textmd{\textcolor{ink2}{root first, at most 8; tag \texttt{pcm.grant.v0}}}};
  \node[f=pcmorange] (g1) at (0.45,-5.23) {\{"issuer": \{"kind": "person", "person": "alice", "approval\_id": "apr-afa8\ldots"\}, "subject": "uid:990101",};
  \node[f=pcmorange] (g2) at (0.7,-5.71) {"caveats": [amount <= 20000, order == 81], "expires\_at": +1\,h, "delegable": false, "parent": null, "sig": \{\ldots\}\}};
  \begin{scope}[on background layer]
    \node[draw=pcmorange, rounded corners=3pt, line width=0.7pt, fit=(gh)(g1)(g2), inner sep=4pt] {};
  \end{scope}
  \node[f] (p) at (0.2,-6.55) {"payload": \{"kind": "text" | "data", \ldots\}};
  \node[f=pcmblue, anchor=west] (s) at ($(p.east)+(\gap,0)$) {"sig": \{"key": "example.com/pcm-test-log+0bb91e66", "sig": base64(64 bytes)\}};
  \begin{scope}[on background layer]
    \node[draw=ink, rounded corners=4pt, line width=0.8pt, fit=(h)(r1a)(r2a)(r3)(c1)(c4)(g2)(p)(s), inner sep=6pt] (env) {};
  \end{scope}
  \node[f=pcmblue, font=\sffamily\scriptsize, anchor=north west] (l1) at ($(env.south west)+(0,-0.15)$) {set by the signer};
  \node[f=ink2, font=\sffamily\scriptsize, anchor=west] (l2) at ($(l1.east)+(0.2,0)$) {set by the agent};
  \node[f=pcmaqua, font=\sffamily\scriptsize, anchor=west] (l3) at ($(l2.east)+(0.2,0)$) {evidence};
  \node[f=pcmorange, font=\sffamily\scriptsize, anchor=west] (l4) at ($(l3.east)+(0.2,0)$) {authority};
  \node[font=\sffamily\scriptsize, text=ink2, anchor=west] at ($(l4.east)+(0.2,0)$) {every member except \texttt{sig} is signed};
\end{tikzpicture}}
\caption{\textbf{Anatomy of a PCM message}, from the conformance vector \vect{accept-approved-by}. Every object is
closed (an unknown member makes the envelope invalid). The signature covers the JCS encoding of every member except
\code{sig}, together with the domain-separation tag. Evidence and grants are therefore bound to the message, and
nothing can be added after signing.}
\label{fig:anatomy}
\end{figure}
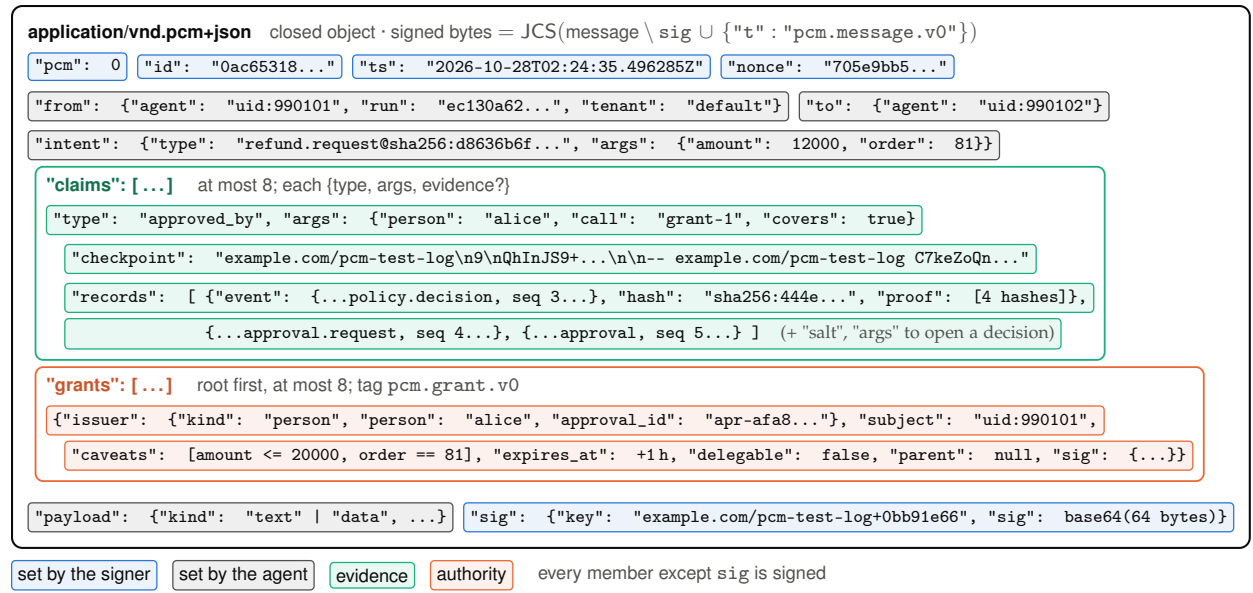

\subsection{The envelope}

\Cref{fig:anatomy} shows the message structure. The envelope is a closed JSON object of media type
\code{application/vnd.pcm+json}, parsed strictly: valid UTF-8, no duplicate names, integers within $\pm(2^{53}-1)$, no
lone surrogates, nesting at most 128 deep, and no number that rounds to no finite IEEE~754 double. A text that fails
any of these is \vbad{} and is refused before any check. Identifiers, timestamps and digests follow fixed ASCII
patterns. A timestamp has exactly six fractional digits, so timestamps compare correctly as strings, and the replay
check refuses a timestamp that names no real time (for example, month 13). Integer fields must be JSON integers:
\code{true}, \code{0.0} and \code{"0"} are not the integer~0.

\subsection{Intents and content-addressed schemas}

An \code{intent} is a schema reference and typed arguments. Both sides hold a copy of the \emph{schema document}, a
small closed JSON object that types each argument field (\code{string}, \code{int}, \code{money} in minor units of a
currency, \code{id} with a pattern, \code{enum}, \code{bool}, \code{time}). It may also list the claim types the intent
requires, demand a grant, and name the tool the intent stands for. On the wire a schema is named
$\texttt{name@}\Hh(\textit{document})$:
\begin{lstlisting}
refund.request@sha256:d8636b6fbbd07430d057ea9ead9c27f0df1b918b10d34ab9257a4971d487846c
\end{lstlisting}
so an edited document is a different schema. Argument matching is strict and performs no conversions: \code{120.0},
\code{true} and \code{"120"} are not integers, every member must be a declared field, and every non-optional field
must be present.

\subsection{Claims, evidence and grants at a glance}

A \textbf{claim} is a typed statement about the sender's own run: \code{tests\_passed\{suite\}},
\code{approved\_by\{person, call, covers?\}}, \code{reached\_host\{host\}} or
\code{file\_written\{path, digest\}}. Its optional \textbf{evidence} is a C2SP checkpoint note signed by the log key
together with 1--16 records of the sender's run, each with its hash and RFC~9162 inclusion proof. A record of a
\code{policy.decision} may also carry an \emph{opening}, a salt and the arguments, which opens the HMAC commitment to
the call's arguments that the signer stored instead of the arguments themselves (\cref{sec:openings}). A
\textbf{grant} states that a subject may send an intent within typed caveats until a time. The chain is root first, and
the root's issuer is a person (\cref{sec:grants}).

\subsection{Outcomes}

The verifier returns a structured result (\cref{alg:verify}): signature, replay, schema and grant sub-results, one
result per claim, and a verdict. A \vacc{} means that everything required is proven. A \vunp{} means that nothing
failed but something rests on the sender's word, and a receiver \textsc{should} treat it as testimony. A \vrej{} must
never be acted on. The result also names any checks the caller \emph{skipped} (no replay state, no receiver
identity); a non-empty \code{skipped} marks an audit result that must not drive action.

\section{Formal Model of Claims and Verdicts}
\label{sec:formal}

This section fixes the objects a verifier reasons about and the three-valued semantics of claims. Later sections
instantiate it for each claim type (\cref{sec:claims}) and for grants (\cref{sec:grants}).

\subsection{Notation and primitives}

$\JCS(v)$ is the RFC~8785 encoding of a JSON value $v$, and $\Hh(v) = \texttt{"sha256:"}\|\mathrm{hex}(\mathsf{SHA256}
(\JCS(v)))$. A signature scheme $(\mathsf{KG},\Sig,\Vf)$ is Ed25519 with the strict verification rules of
\cref{sec:sigrules}. For a closed object $o$ with a \code{sig} member, the signed bytes under tag $t$ are
\[
  \mathsf{bytes}_t(o) \;=\; \JCS\big((o\setminus\{\texttt{sig}\}) \cup \{\texttt{"t"}: t\}\big),
  \qquad t\in\{\texttt{pcm.message.v0},\;\texttt{pcm.grant.v0}\}.
\]
Because messages and grants are closed objects without a \code{"t"} member on the wire, a \code{"t"} member makes the
envelope invalid. The tag therefore cannot be supplied by the sender, and a signature under one tag never verifies
under the other (vector \vect{reject-grant-signature-as-message}).

\subsection{Logs, records and evidence}

\begin{definition}[Log, record, run]
A \emph{log} $L$ is a sequence of records $r_0, r_1, \ldots$ appended by $L$'s signer, where $r_i.\code{seq} = i$.
Each record carries a \code{tenant}, a \code{run\_id}, a \code{type} (such as \code{policy.decision},
\code{tool.result}, \code{approval}, \code{approval.request} or \code{effect.observed}), a \code{data} object, and (for \code{tool.result}) top-level \code{source} and \code{tier} fields. The \emph{run}
$\rho = (\code{run}, \code{tenant})$ of $L$ is the subsequence of records with that run and tenant. The leaf hash of
$r_i$ is $\ell_i = \mathsf{SHA256}(\texttt{0x00}\,\|\,\mathsf{bytes}(\Hh(r_i)))$, and the Merkle tree head at size $n$ is
$\mathsf{MTH}(\ell_0..\ell_{n-1})$ as defined in RFC~9162.
\end{definition}

\begin{definition}[Checkpoint]
A \emph{checkpoint} is a C2SP signed note whose text is exactly three lines, an origin, a decimal tree size $n$ and a
base64 root $R$, followed by 1--64 signature lines. It is \emph{valid for} a trust document $\mathcal{T}$ and a
message $m$ when: (i)~every line is well formed, including lines of unpinned keys; (ii)~the Ed25519 log key that
$\mathcal{T}$ pins under the origin's name has a valid line; (iii)~the origin equals the log name in $m.\code{sig.key}$;
and (iv)~at least $\mathcal{T}.\code{witnesses\_required}$ pinned witnesses have valid cosignatures over
$\texttt{cosignature/v1\textbackslash ntime }t\texttt{\textbackslash n}\,\|\,\textit{text}$.
\end{definition}

\begin{definition}[Evidence item]\label{def:evidence}
An evidence item is $E = (\kappa, \langle (e_j, h_j, \pi_j, o_j) \rangle_{j=1}^{k})$ with checkpoint $\kappa$,
$1\le k\le 16$, event $e_j$, claimed hash $h_j$, inclusion proof $\pi_j$ (at most 64 hashes) and optional opening $o_j$.
$E$ \emph{verifies} for message $m$, written $\mathsf{EvOK}(E,m,\mathcal{T})$, iff $\kappa$ is valid for $\mathcal{T}$
and $m$, and for every $j$:
\begin{enumerate}[label=(E\arabic*),leftmargin=3em]
\item $\Hh(e_j) = h_j$;
\item every entry of $\pi_j$ is canonical base64, and $\pi_j$ proves leaf $\mathsf{SHA256}(\texttt{0x00}\,\|\,h_j)$ at
  index $e_j.\code{seq}$ in the tree $(n,R)$ of $\kappa$;
\item no earlier $j'<j$ has $e_{j'}.\code{seq} = e_j.\code{seq}$ (a record is listed once);
\item $e_j.\code{tenant} = m.\code{from.tenant}$ and $e_j.\code{run\_id} = m.\code{from.run}$;
\item $e_j$ validates against the event schema of the evidence profile (\code{tracekit.event.v2});
\item if $o_j$ is present, it has both \code{salt} and \code{args}, $e_j.\code{type} = \code{policy.decision}$, and
  the opening verifies (\cref{def:opening}).
\end{enumerate}
\end{definition}

Condition (E3) matters more than it appears. Without it, a sender could list one decision twice, once opened and
once not, and make the checker's answer depend on the order of the list. Condition (E4) binds evidence to the sender's
own run \emph{and} tenant, because run identifiers are chosen by clients and are unique only within a log and tenant.
The origin check in the checkpoint definition binds evidence to the log that signed the message
(\cref{sec:attacks}, items 1 and 18).

\subsection{Committed arguments and openings}
\label{sec:openings}

A signer that never stores tool arguments in clear records, in each \code{policy.decision}, a commitment
\[
  \code{args\_commitment} = \texttt{"hmac-sha256:"}\,\|\,\mathrm{hex}\big(\mathsf{HMAC}_{s}\big(\mathrm{ASCII}(\Hh(\{\code{"tool"}:\,\textit{tool},\;\code{"args"}:\,\textit{args}\}))\big)\big)
\]
with a fresh 32-byte salt $s$. The signer gives $s$ only to the run's own agent.

\begin{definition}[Opening]\label{def:opening}
An opening $o = (s, a)$ of a decision record $e$ \emph{verifies} iff the recomputed commitment from $e.\code{data.tool}$,
$a$ and $s$ equals $e.\code{data.args\_commitment}$, compared in constant time. Only the arguments of opened decisions
are visible to a claim checker.
\end{definition}

Openings give PCM \emph{selective disclosure}: a sender reveals the arguments of exactly the decisions that a claim
needs, and every other decision in the disclosed evidence stays hidden behind its commitment (\cref{sec:privacy}).

\subsection{Who wrote a record: tiers}

The central distinction in PCM is between records whose content \emph{the agent} determined and records whose content
\emph{someone else} determined.

\begin{definition}[Attested record]\label{def:attested}
A record $e$ is \emph{attested} iff either (a)~its \code{type} is \code{approval}, \code{approval.request} or
\code{effect.observed}: records that only the signer writes, from authenticated answers, its own requests and
observer reports; or (b)~$e.\code{type} = \code{tool.result}$ with \emph{top-level} $e.\code{source} =
\code{"gateway"}$ and $e.\code{tier} = \code{"T2"}$. Any other \code{tool.result} is \T1: the agent's own report.
\end{definition}

The tier is read from top-level fields that the signer sets from the authenticated caller, and never from
\code{data}, which the reporting party fills. An agent that writes \code{"tier": "T2"} into its own result's data
changes nothing.

\subsection{Three-valued claim semantics}

Let $\mathbb{R} = \{\fail \prec \unprov \prec \prov\}$, totally ordered, and let $\bigsqcup$ denote the maximum (``best
of'') over a set of candidates, with $\bigsqcup\emptyset = \fail$.

\begin{definition}[Claim checker]\label{def:checker}
For each claim type $\tau$, PCM fixes (i)~an argument table, (ii)~a predicate $\mathsf{Malformed}_\tau(\textit{args})$,
and (iii)~a function $\mathsf{Check}_\tau(\textit{args}, E, m) \in \mathbb{R}$ over the records of a verifying
evidence item. The result of a claim $c = (\tau, \textit{args}, E^?)$ in message $m$ is
\[
\mathsf{Res}(c) =
\begin{cases}
\fail & \text{if \textit{args} fails its table, or } \mathsf{Malformed}_\tau(\textit{args});\\
\unprov & \text{else if $c$ carries no evidence (or the verifier is Level~1);}\\
\fail & \text{else if } \neg\,\mathsf{EvOK}(E,m,\mathcal{T});\\
\mathsf{Check}_\tau(\textit{args},E,m) & \text{otherwise.}
\end{cases}
\]
\end{definition}

The ordering of the cases is deliberate. A malformed claim, such as \code{pytest || true} or a relative path, is
\fail{} \emph{with or without evidence}, at every conformance level, so a sender cannot downgrade a provably bad claim
to \unprov{} by omitting its evidence.

The informal reading of the three values is:
\begin{itemize}
\item \prov: the evidence verifies, and an attested record in it says what the claim says.
\item \unprov: there is no evidence; or the only matching record is \T1; or the matching call is of a tool the verifier
  cannot place in the needed class; or the only matching record shows an \emph{attempt} (a \code{connect}), not a
  connection; or the verifier cannot tell whether a record matches (a host that will not parse under
  \cref{sec:hosts}, or a \code{Write} of a relative path that may be the claimed file).
\item \fail: the claim is malformed; the evidence does not verify; or no record in it says what the claim says.
\end{itemize}

\paragraph{The run predicate}
Three of the four checkers ask whether a call \emph{ran and succeeded}. \Cref{alg:ran} defines the shared predicate
$\Ran(\mathit{match},k)$ over one evidence item, parameterised by a matcher on (tool, opened args) and an optional tool
class $k$. It formalises three rules. Only opened, non-denied decisions with an identifier can match. A tool is placed
in a class only by the frozen placement table (\cref{sec:placement}), never by the shape of its arguments. Among
matching results, only an attested (\T2) result with \code{ok} = \code{true} proves the claim.

\begin{algorithm}[t]
\caption{$\Ran(\mathit{match}, k)$ over the records $\mathcal{E}$ of one verified evidence item}
\label{alg:ran}
\small
\begin{algorithmic}[1]
\State $\mathit{cands} \gets \{\fail\}$ \Comment{``no call in the evidence \ldots''}
\ForAll{$d\in\mathcal{E}$ with an opening, $d.\code{type}=\code{policy.decision}$, opened args an object}
  \If{$d.\code{data.tool}$ or $d.\code{data.decision\_id}$ absent \textbf{or} $d.\code{data.decision}=\code{deny}$} \textbf{continue} \EndIf
  \State $a \gets \mathit{match}(d.\code{data.tool}, d.\textit{args})$ \Comment{$a\in\{\textsc{yes},\textsc{no},\textsc{cannot-tell}\}$}
  \If{$a=\textsc{no}$} \textbf{continue}
  \ElsIf{$a=\textsc{cannot-tell}$} $\mathit{cands}\gets\mathit{cands}\cup\{\unprov\}$; \textbf{continue}
  \EndIf
  \If{$k\neq\bot$ \textbf{and} $\mathsf{Place}(d.\code{data.tool})\neq k$} $\mathit{cands}\gets\mathit{cands}\cup\{\unprov\}$; \textbf{continue} \EndIf
  \ForAll{$r\in\mathcal{E}$, $r.\code{type}=\code{tool.result}$, $(r.\code{decision\_id}, r.\code{tool\_use\_id}) = (d.\code{decision\_id}, d.\code{tool\_use\_id})$}
     \If{$r.\code{data.ok}\neq\code{true}$} $\mathit{cands}\gets\mathit{cands}\cup\{\fail\}$ \Comment{``\ldots and failed''}
     \ElsIf{$r.\code{source}=\code{gateway} \wedge r.\code{tier}=\code{T2}$} \Return \prov
     \Else{} $\mathit{cands}\gets\mathit{cands}\cup\{\unprov\}$ \Comment{``the agent's own report (T1)''}
     \EndIf
  \EndFor
\EndFor
\State \Return $\bigsqcup \mathit{cands}$
\end{algorithmic}
\end{algorithm}

\subsection{Verdicts}

\begin{definition}[Verdict]\label{def:verdict}
Let $\mathcal{S}$ be the intent's schema document with required claim types $Q\subseteq\{\text{claim types}\}$, and let
$\mathsf{Met}(q)$ hold iff some claim of type $q$ is \prov{}, where for $q=\code{approved\_by}$ the claim must
additionally carry $\code{covers}=\code{true}$ (\cref{sec:approvedby}). Given that the envelope parsed and the
signature verified, the verdict is
\[
\mathsf{V}(m) =
\begin{cases}
\vrej & \text{if } \mathsf{Bad}(m) \text{ or } \big(\textit{strict} \wedge \exists q\in Q.\,\neg\mathsf{Met}(q)\big),\\
\vunp & \text{else if } \exists c.\ \mathsf{Res}(c)=\unprov, \text{ or } \exists q\in Q.\,\neg\mathsf{Met}(q),\\
\vacc & \text{otherwise,}
\end{cases}
\]
where $\mathsf{Bad}(m)$ holds iff the replay check fails, the schema check has errors, the grants result is \fail, or
some claim has $\mathsf{Res}(c)=\fail$.
An invalid envelope or a failed message signature yields \vrej{} immediately, with every later part \code{null}: no
claim work is done for a forged message.
\end{definition}

Two simple facts follow directly from the definitions, and we use them later.

\begin{proposition}[Level~1 never proves]\label{prop:level1}
A Level~1 verifier, which does not check evidence, never assigns \prov{} to any claim, and therefore returns \vacc{} only
for messages that carry no claims and whose schema requires none.
\end{proposition}
\begin{proof}
By \cref{def:checker}, a Level~1 verifier maps every well-formed claim to \unprov{} (no evidence or ``evidence not
checked''). A claim then forces at least \vunp{} by \cref{def:verdict}, and so does any required type, since
$\mathsf{Met}$ needs a \prov{} claim.
\end{proof}

\begin{proposition}[Order independence]\label{prop:order}
$\mathsf{Res}(c)$ does not depend on the order in which $E$ lists its records.
\end{proposition}
\begin{proof}[Proof sketch]
(E1), (E2) and (E4)--(E6) are per-record conditions, and (E3) rejects any duplicate \code{seq} regardless of which copy
appears first. $\Ran$ takes the maximum over candidates, which is order-free. The one checker that selects among
several records, \code{approved\_by}, is specified to fail whenever the qualifying records name more than one
(tool, args) pair (\cref{sec:approvedby}), so the choice of record cannot change the result. The detail string names
the lowest-\code{seq} witness, which is also order-free.
\end{proof}

\section{The Four Claim Types}
\label{sec:claims}

PCM~0.1 has a deliberately closed vocabulary of four claim types (\cref{tab:claims}). A new type can be added only
together with its arguments in the schema language, an exact checker over a named evidence profile with \prov,
\unprov{} and \fail{} rules, and conformance vectors for each result. We present each checker and the attack that
shaped it.

\begin{table}[t]
\centering\small
\caption{The four claim types of PCM~0.1 and when each is \prov, \unprov{} or \fail. ``Attested'' is
\cref{def:attested}. Each row is pinned by the vectors listed in the specification.}
\label{tab:claims}
\renewcommand{\arraystretch}{1.15}
\begin{tabularx}{\linewidth}{@{}L{2.4cm} Y Y Y@{}}
\toprule
\textbf{Claim} & \cellcolor{aquawash}\textbf{\prov{} when} & \cellcolor{yellowwash}\textbf{\unprov{} when} & \cellcolor{redwash}\textbf{\fail{} when}\\
\midrule
\code{tests\_passed}\newline\code{\{suite\}} &
an opened, non-denied decision of a \code{shell}-class tool whose command equals \code{suite} exactly, with a \T2
result \code{ok} &
no evidence; result is \T1; tool not placed as \code{shell} &
\code{suite} is no single test command (\cref{sec:testcmd}); the \T2 result is not \code{ok}; no matching call\\
\addlinespace
\code{approved\_by}\newline\code{\{person, call,}\newline\code{covers?\}} &
an \code{approval} by the configured \code{person} (not an agent, not self) of \code{call}, \emph{plus} the
\code{approval.request} it answers \emph{plus} the \code{policy.decision} that request names; with \code{covers},
the approval also covers this request &
no evidence; with \code{covers}: the schema names no tool, or the decision is not opened &
no such triple; records disagree on what was approved; with \code{covers}: another tool, other args, or a
\code{pcm.grant} that did not issue this root grant\\
\addlinespace
\code{reached\_host}\newline\code{\{host\}} &
an observer's \code{net} effect with \code{connections} $>0$ to the host; or an opened \code{http}-class call whose URL
host equals it, with a \T2 result \code{ok} &
no evidence; only a \code{connect} attempt or a zero-connection minute; result is \T1; a host that will not parse
(\cref{sec:hosts}) &
no record of the host; a URL whose scheme has no host (\code{file:}, \code{gopher:}); the \T2 result is not \code{ok}\\
\addlinespace
\code{file\_written}\newline\code{\{path, digest\}} &
an opened \code{Write} whose full \code{content} hashes to \code{digest} and whose absolute path equals \code{path}
under the claim's path rules, with a \T2 result \code{ok} &
no evidence; result is \T1; the call's path is relative and may be the file; a drive path spelled otherwise &
\code{path} is relative or has a \code{..} segment; the digest differs; a tool other than \code{Write}\\
\bottomrule
\end{tabularx}
\end{table}

\subsection{\texttt{tests\_passed}: exactly one test command}
\label{sec:testcmd}

The checker is $\Ran(\mathit{match}, \code{shell})$, where $\mathit{match}$ holds when the call's shell command (the
first non-null of the argument members \code{command}, \code{cmd}, \code{commands}) equals \code{suite}
\emph{exactly}. The difficulty is not the match but the claim: \code{pytest || true}, \code{pytest; true} and
\code{pytest \&} all exit 0 whether the tests pass or not, and each of them can really run through an honest executor.
A claim about such a command is a lie even when it is backed by a genuine \T2 record. PCM therefore makes
\code{suite} \emph{malformed}, and hence \fail{} with or without evidence, unless it is a \textbf{test command}:
\begin{enumerate}[label=(\alph*),leftmargin=2em]
\item \textbf{Plain words.} It contains none of \code{; \& | ( ) < > \$ ` \textbackslash{} \# \{ \} * ? [ ]}, no
  control character other than tab, and no U+007F; and its quotes balance. Splitting it into words at unquoted spaces
  and tabs then gives the same words under \code{sh} and under any word splitter.
\item \textbf{A listed runner, by its bare name.} The words begin with one of 42 fixed runner sequences
  (\code{pytest}, \code{python -m pytest}, \code{go test}, \code{cargo test}, \code{npm test}, \ldots;
  \cref{app:runners}), compared exactly. \code{./pytest}, \code{/tmp/x/pytest}, \code{uv run pytest} and
  \code{CI=1 pytest} are not test commands.
\item \textbf{Runs tests.} No word is \code{-h} or \code{-V}, and no word of three or more characters is a prefix of
  \code{--help}, \code{--version} or \code{--collect-only}, because argparse-based runners accept unambiguous
  prefixes.
\end{enumerate}
Rule~(a) is stricter than a shell parser on purpose. An earlier draft cut command lines at operators using Python's
\code{shlex}. That splitter has no \code{\$'\ldots'} quoting, so it reads \code{pytest \$'\textbackslash'';true \#'}
as one command, while \code{sh} runs \code{pytest "'"} and then \code{true}. Likewise a \code{\#} inside a word is a
comment to \code{shlex} but literal to \code{sh} (\code{pytest -q tests/x.py\#|| true}). Rather than chase one
shell's grammar, PCM~0.1 forbids every character on which a shell and a word splitter can differ.

\begin{limitbox}
\textbf{What \prov{} means for \texttt{tests\_passed}.} The signer recorded that \emph{this exact command line} ran
through an attesting executor and exited~0. Runners accept flags that select no tests and still exit 0
(\code{go test -run NONE}, \code{mvn test -DskipTests}, \code{cargo test --no-run}), and they read configuration from
a working directory the agent may have written. A receiver \textsc{must} compare \code{suite} with the suites it
accepts, and must not rely on the runner list to tell a real test run from another.
\end{limitbox}

\subsection{\texttt{approved\_by}: binding approval to what was decided}
\label{sec:approvedby}

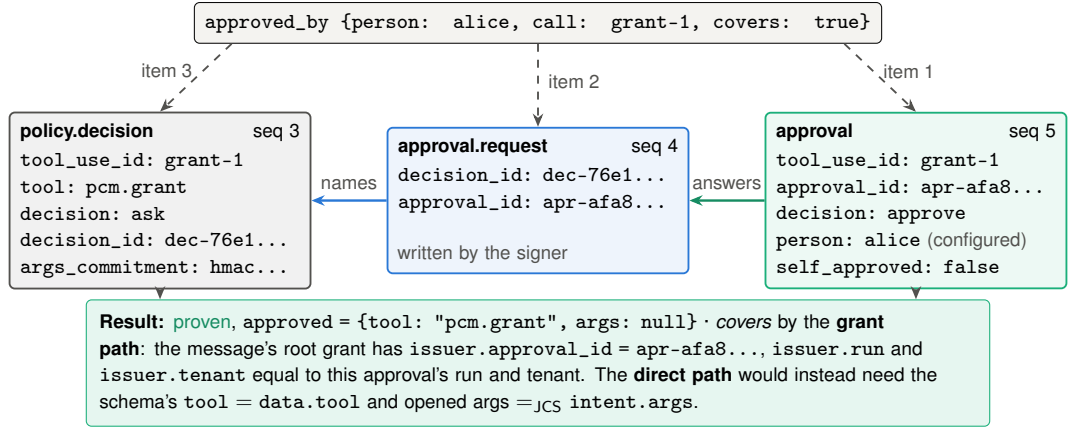
\begin{figure}[t]
\centering
\begin{tikzpicture}[
  rec/.style={draw=#1, fill=#1!8, rounded corners=2pt, line width=0.7pt, text width=37mm, align=left,
              font=\sffamily\scriptsize, inner sep=4pt},
  key/.style={font=\ttfamily\scriptsize, text=ink},
]
  \node[rec=ink2] (dec) at (0,0) {\textbf{policy.decision} \hfill seq 3\\[1pt]
     \texttt{tool\_use\_id: grant-1}\\ \texttt{tool: pcm.grant}\\ \texttt{decision: ask}\\
     \texttt{decision\_id: dec-76e1\ldots}\\ \texttt{args\_commitment: hmac\ldots}};
  \node[rec=pcmblue] (req) at (5.0,0) {\textbf{approval.request} \hfill seq 4\\[1pt]
     \texttt{decision\_id: dec-76e1\ldots}\\ \texttt{approval\_id: apr-afa8\ldots}\\[9pt]
     {\color{ink2}\scriptsize written by the signer}};
  \node[rec=pcmaqua] (apr) at (10.0,0) {\textbf{approval} \hfill seq 5\\[1pt]
     \texttt{tool\_use\_id: grant-1}\\ \texttt{approval\_id: apr-afa8\ldots}\\ \texttt{decision: approve}\\
     \texttt{person: alice} \textcolor{ink2}{(configured)}\\ \texttt{self\_approved: false}};
  \draw[arr, draw=pcmblue, line width=0.9pt] (req.west) -- node[lbl, above] {names} (dec.east);
  \draw[arr, draw=pcmaqua!80!black, line width=0.9pt] (apr.west) -- node[lbl, above] {answers} (req.east);
  \node[draw=ink, fill=wash, rounded corners=2pt, font=\ttfamily\scriptsize, inner sep=4pt] (claim) at (5.0,2.35)
     {approved\_by \{person: alice, call: grant-1, covers: true\}};
  \draw[darr] (claim.south west) ++(5mm,0) -- node[lbl, left, pos=0.4] {item 3} (dec.north);
  \draw[darr] (claim.south) -- node[lbl, right, pos=0.45] {item 2} (req.north);
  \draw[darr] (claim.south east) ++(-5mm,0) -- node[lbl, right, pos=0.4] {item 1} (apr.north);
  \node[draw=pcmaqua, fill=aquawash, rounded corners=2pt, font=\sffamily\scriptsize, text width=11.6cm, align=left, inner sep=5pt]
     (res) at (5.0,-2.15) {\textbf{Result:} \prov, \texttt{approved = \{tool: "pcm.grant", args: null\}} \textperiodcentered{}
     \emph{covers} by the \textbf{grant path}: the message's root grant has \texttt{issuer.approval\_id = apr-afa8\ldots},
     \texttt{issuer.run} and \texttt{issuer.tenant} equal to this approval's run and tenant.
     The \textbf{direct path} would instead need the schema's \texttt{tool} $=$ \texttt{data.tool} and opened args $=_{\JCS}$ \texttt{intent.args}.};
  \draw[arr] (dec.south) -- (dec.south |- res.north);
  \draw[arr] (apr.south) -- (apr.south |- res.north);
\end{tikzpicture}
\caption{\textbf{How \code{approved\_by} is proven} (vector \vect{accept-approved-by}). The \code{call} id is the
agent's own name, so an approval record alone says nothing about \emph{what} was approved. The checker requires the
three-record linkage. The result's conformance field \code{approved} reports the decision's tool and opened args, and
\code{covers: true} asks the verifier to check that this approval stands for \emph{this} request.}
\label{fig:approvedby}
\end{figure}

An agent names its own calls. It can decide a \code{read\_file} call under the id \code{refund-81-approved}, have a
person approve it, and claim ``alice approved \code{refund-81-approved}''. An early draft proved \code{approved\_by}
from the approval record alone, so this claim was \prov{} (\cref{sec:attacks}, item 27). PCM~0.1 instead requires, in
one evidence item, the three linked records of \cref{fig:approvedby}:
\begin{enumerate}[leftmargin=2em]
\item an \code{approval} with \code{decision} = \code{approve}, \code{tool\_use\_id} = \code{call}, \code{person}
  equal to the claim's \code{person} (compared only with the signer-configured name, never with the raw approver
  identity), \code{approver\_kind} absent or \code{person}, and \code{self\_approved} not \code{true};
\item the \code{approval.request} with the same \code{approval\_id};
\item the \code{policy.decision} that request names, with \code{tool\_use\_id} = \code{call} and a non-empty
  \code{tool}.
\end{enumerate}
An approval without the other two is \fail, not \unprov: the sender chooses its evidence and can always include them.

\paragraph{Records that disagree}
To make the result independent of record order (\cref{prop:order}), the checker collects every qualifying approval.
If the requests of one approval name two or more decisions, the claim is \fail. Otherwise the checker collects the
qualifying decisions as pairs $(\textit{tool}, \textit{args or null})$. If two pairs differ in $\JCS$ (so
\code{\{"amount": true\}} and \code{\{"amount": 1\}} differ, though some languages compare them equal), the claim is
\fail; if exactly one pair exists, the claim is \prov{} and the result carries
$\code{approved} = \{\code{tool}, \code{args}\}$ as a \emph{conformance field} that every verifier must return
identically.

\paragraph{What the approval covers}
Even a correctly linked approval may be of a \emph{different} request: alice approved a \$5 refund, and the message
asks for \$50. Leaving this comparison to receivers proved unsafe in our benchmark (finding F3,
\cref{sec:eval-findings}), so PCM~0.1 adds \code{covers}. With $\code{covers}=\code{true}$, a linked claim is \prov{}
only by one of two paths:
\begin{itemize}
\item \textbf{Grant path.} The decision's tool is \code{pcm.grant}, and the approval issued the message's root grant:
  the root's \code{issuer.approval\_id}, \code{run} and \code{tenant} are those of the approval.
\item \textbf{Direct path.} The intent's schema names a \code{tool} equal to the decision's tool, the evidence opens
  the decision, and its args equal \code{intent.args} in $\JCS$.
\end{itemize}
Otherwise the claim is \fail{} (a \code{pcm.grant} decision whose approval did not issue the root, another tool, or
other args) or \unprov{} (the schema names no tool, or the decision is not opened, so the verifier cannot tell).
A schema that lists \code{approved\_by} is met \emph{only} by a \prov{} claim with \code{covers}. The direct path covers
every message that asks for the same tool with the same args, so one-time authority needs a field unique to the
request (such as a \code{request\_id}) or a grant, whose caveats and expiry bound it.

\subsection{\texttt{reached\_host}: a connection, not an attempt}
\label{sec:reachedhost}

The claim is \prov{} by either of two kinds of attested record. The first is an egress observer's \code{net} effect for
the host with \code{connections} $> 0$. The second is $\Ran(\mathit{match},\code{http})$, where $\mathit{match}$
compares the host of the call's \code{url} with the claim's host in canonical form. Two subtleties shaped the rule. A
kernel sensor reports a process's \emph{call} to \code{connect}, which may fail, and an egress observer records a
minute with zero connections when its upstream connect failed; both show an \emph{attempt}, so both give \unprov{}
(and a \emph{denied} \code{connect} counts for nothing). And a call that received an HTTP error status reached the
host, but the evidence profile records only \code{ok}, so PCM~0.1 cannot prove it from the call record (finding F4;
\cref{sec:limits}).

\subsubsection{Canonical hosts}
\label{sec:hosts}

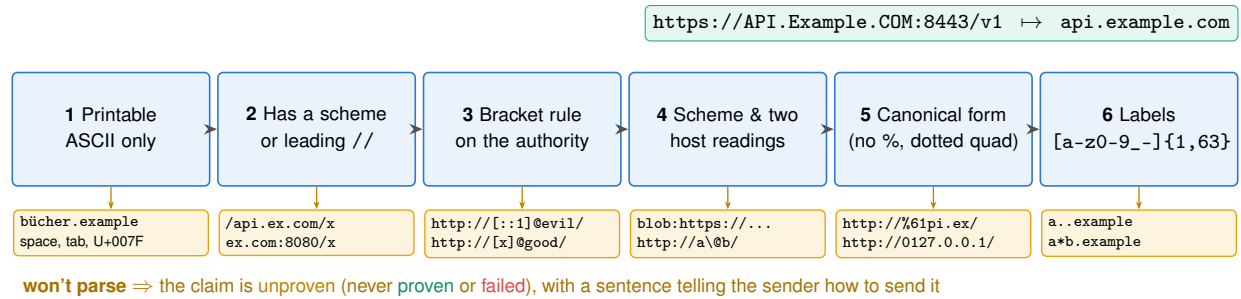
\begin{figure}[t]
\centering
\begin{tikzpicture}[
  st/.style={draw=pcmblue, fill=bluewash, rounded corners=2pt, line width=0.6pt, text width=24mm, minimum height=15mm,
             align=center, font=\sffamily\scriptsize, inner sep=3pt},
  no/.style={draw=pcmyellow, fill=yellowwash, rounded corners=2pt, line width=0.5pt, text width=24mm, align=left,
             font=\sffamily\tiny, inner sep=2.5pt},
]
  \foreach \i/\t/\d in {1/{\textbf{1} Printable\\ASCII only}/{\texttt{bücher.example}\\ space, tab, U+007F},
                        2/{\textbf{2} Has a scheme\\or leading \texttt{//}}/{\texttt{/api.ex.com/x}\\ \texttt{ex.com:8080/x}},
                        3/{\textbf{3} Bracket rule\\on the authority}/{\texttt{http://[::1]@evil/}\\ \texttt{http://[x]@good/}},
                        4/{\textbf{4} Scheme \& two\\host readings}/{\texttt{blob:https://\ldots}\\ \texttt{http://a\textbackslash @b/}},
                        5/{\textbf{5} Canonical form\\(no \%, dotted quad)}/{\texttt{http://\%61pi.ex/}\\ \texttt{http://0127.0.0.1/}},
                        6/{\textbf{6} Labels\\\texttt{[a-z0-9\_-]\{1,63\}}}/{\texttt{a..example}\\ \texttt{a*b.example}}} {
    \node[st] (s\i) at ({(\i-1)*2.72},0) {\t};
    \node[no, below=3mm of s\i] (n\i) {\d};
    \draw[-{Stealth[length=3pt]}, draw=pcmyellow!80!black, line width=0.5pt] (s\i.south) -- (n\i.north);
  }
  \foreach \i [evaluate=\i as \j using int(\i+1)] in {1,...,5} { \draw[arr] (s\i.east) -- (s\j.west); }
  \node[draw=pcmaqua, fill=aquawash, rounded corners=2pt, font=\ttfamily\scriptsize, inner sep=3pt, above=4mm of s6.north east, anchor=south east]
    (ok) {https://API.Example.COM:8443/v1 $\;\mapsto\;$ api.example.com};
  \node[font=\sffamily\scriptsize, text=pcmyellow!70!black, anchor=north west] at ($(n1.south west)+(0,-1.5mm)$)
    {\textbf{won't parse} $\Rightarrow$ the claim is \unprov{} (never \prov{} or \fail), with a sentence telling the sender how to send it};
\end{tikzpicture}
\caption{\textbf{Host canonicalisation} (\cref{sec:hosts}). Every step is specified character by character, so that
verifiers in any language agree. Where two common readers disagree on a URL's host, the host \emph{will not parse}
rather than match a host the call may not have reached. Examples below each step are refused there; each is a
conformance vector.}
\label{fig:hosts}
\end{figure}

A host has many spellings: \code{API.Example.COM:8443}, \code{api.example.com.}, \code{[::ffff:7f00:1]},
\code{xn--bcher-kva.example}. Worse, two widely used readers can assign different hosts to one URL. PCM compares hosts
in a single canonical form produced by a six-step reading (\cref{fig:hosts}). Its guiding rule is that \emph{where two
readers named in the specification disagree, the host will not parse}. A host that will not parse yields \unprov, never
\prov{} or \fail, together with a fixed sentence that tells the sender how to resend. Three steps encode lessons
that were costly to learn:
\begin{itemize}
\item \textbf{ASCII only.} IDNA~2003, UTS~46 and browsers map some Unicode names differently (\code{faß.example} is
  \code{fass.example} to IDNA~2003 and \code{xn--fa-hia.example} to a browser). IDNA~2003 maps U+1806 to nothing, so
  \code{exa}\,U+1806\,\code{mple.com} proved \code{example.com} in an earlier draft. PCM~0.1 compares hosts in printable
  ASCII (A-label form) only, so no verifier depends on its IDNA or Unicode version.
\item \textbf{One bracket rule, checked by the verifier itself.} CPython's \code{urlsplit} has, across releases, made
  no bracket check, checked the first bracketed part of the authority, and checked the part after the last \code{@}.
  Verifiers on two releases gave one message different verdicts (\vacc{} versus \vunp{} on
  \code{http://[::1]@api.example.com/}). PCM fixes one rule that every reader agrees with and requires verifiers to
  apply it \emph{before} any library reads the URL.
\item \textbf{IPv4 only as a dotted quad.} \code{inet\_aton} reads \code{0127.0.0.1} as octal (\code{87.0.0.1}),
  while the macOS resolver reads it as decimal and connects to \code{127.0.0.1}. Our benchmark found that a sender on a
  macOS executor could thereby make PCM prove an address it never contacted (finding F1). PCM~0.1 accepts an IPv4
  address only as four decimal octets without leading zeros. \code{127.1}, \code{0x7f.0.0.1}, \code{2130706433} and
  \code{1.2.3.256} will not parse. IPv4-mapped and NAT64 IPv6 addresses compare as the IPv4 address inside them.
\end{itemize}

\subsection{\texttt{file\_written}: whose path rules?}

A claim \code{file\_written\{path, digest\}} is \prov{} by an opened \code{Write} call (the tool name decides; an
\code{op} argument is the agent's word) whose full \code{content} hashes to \code{digest} and whose path matches. The
path rules are chosen \emph{by the claim}, since the record does not say which operating system ran the call:

\begin{center}\small
\begin{tabularx}{\linewidth}{@{}L{4.7cm} L{2.6cm} Y@{}}
\toprule
\textbf{Claim path} & \textbf{Rules} & \textbf{Consequences}\\
\midrule
\code{C:\textbackslash work\textbackslash a.txt}, \code{C:/work/a.txt} & Windows (drive path) & \code{\textbackslash} and
  \code{/} separate; proven only by a call of the identical string\\
\code{/work/a.txt} & POSIX & only \code{/} separates; \code{/work\textbackslash a.txt} is one file in \code{/}\\
\code{work/a.txt}, \code{\textbackslash work\textbackslash a.txt}, \code{C:a.txt} & relative & \fail{}, with or
  without evidence\\
\code{/work/../a.txt} & has a \code{..} segment & \fail{}: ``name the file's own path''\\
\bottomrule
\end{tabularx}
\end{center}

A call whose own path is relative and is a suffix of the claimed path gives \unprov: the record does not hold the
working directory, so it may or may not be the same file. An earlier draft read every \code{\textbackslash} as
\code{/}, so a \code{Write} of the POSIX file \code{/work\textbackslash report.txt} proved \code{/work/report.txt}. A
proven \code{file\_written} says that a call wrote these exact bytes. A receiver that holds the file can check the
digest itself, and this check is the one freshness test PCM~0.1 makes fully sound (\cref{sec:limits}).

\subsection{Tool placement as frozen protocol data}
\label{sec:placement}

Whether a call is a \code{shell} or an \code{http} call is decided by its tool name alone, through a frozen table of 54
names in six classes (\code{http} 14, \code{payment} 12, \code{sql} 10, \code{shell} 8, \code{fs} 7, \code{email} 3),
published with its SHA-256 (\code{15f8410f\ldots c8fe07}). A \code{command} or \code{url} member in the arguments does
not place a tool. A tool not in the table is unplaced and gives \unprov. The table is \emph{protocol data}: an earlier
draft derived it from an implementation's policy packs, so installing a pack changed the verdicts of messages already
sent, and two verifiers of different versions disagreed (\cref{sec:attacks}, item 29).

\section{Grants: Person-Rooted, Attenuating Authority}
\label{sec:grants}

Claims say what happened. \emph{Grants} say what the sender may ask for. A grant is a closed object signed under tag
\code{pcm.grant.v0}: \emph{subject} $S$ may send intent $I$ while the intent's args satisfy the caveats $C$, until
\code{expires\_at}. It names its \code{issuer} (a person, for a root; the parent's subject, for a child), the issuing
run and tenant, whether it is \code{delegable}, and its \code{parent}, which is $\Hh$ of the full signed parent
including its signature (\code{null} for a root).

\begin{figure}[t]
\centering
\begin{tikzpicture}[
  g/.style={draw=#1, fill=#1!7, rounded corners=2.5pt, line width=0.7pt, text width=47mm, align=left,
            font=\sffamily\scriptsize, inner sep=4pt},
]
  \node[box=pcmaqua, text width=24mm] (alice) at (-0.5,0) {\textbf{alice} (person)\\[-1pt]{\scriptsize approves a \texttt{pcm.grant} call with these exact terms}};
  \node[g=pcmorange] (root) at (4.9,0) {\textbf{root grant} \hfill \texttt{parent: null}\\[1pt]
    issuer: person alice (approval \texttt{apr-\ldots})\\
    subject: agent $A$\\ intent: \texttt{refund.request@sha256:\ldots}\\
    \texttt{amount <= 20000}, \texttt{order == 81}\\
    expires: $t_0{+}1$h \textperiodcentered{} delegable: \textbf{true}};
  \node[g=pcmaqua] (child) at (12.0,1.15) {\textbf{child} \hfill \texttt{parent: H(root)}\\[1pt]
    issuer: agent $A$ \textperiodcentered{} subject: agent $C$\\
    \texttt{amount <= 15000}, \texttt{order == 81}\\
    expires: $t_0{+}30$m \hfill \cmark{} \textbf{narrows}};
  \node[g=pcmred] (bad) at (12.0,-1.25) {\textbf{child} \hfill \texttt{parent: H(root)}\\[1pt]
    \texttt{amount <= 30000}, \texttt{amount >= 100},\\ \texttt{order == 81}\\
    \xmark{} \texttt{amount <= 20000: dropped or widened}};
  \draw[arr, line width=0.9pt] (alice) -- node[lbl, above] {consumed} node[lbl, below] {approval} (root);
  \draw[arr, draw=pcmaqua!80!black, line width=0.9pt] (root.east) -- node[lbl, above, sloped, pos=0.5] {signer signs} (child.west);
  \draw[arr, draw=pcmred, line width=0.9pt] (root.east) -- node[lbl, below, sloped, pos=0.5] {signer refuses} (bad.west);
  \node[lbl, text width=15.5cm, align=center] at (6.0,-2.75) {The verifier re-checks every link: signed by the
    \emph{message's} log key, same intent and tenant, \texttt{parent} hash, delegable parent, issuer $=$ previous subject,
    expiry shrinks, caveats narrow, and \texttt{intent.args} satisfies \emph{every} grant.};
\end{tikzpicture}
\caption{\textbf{A grant chain only narrows} (illustrative values; the vector \vect{accept-approved-by} uses a
non-delegable root). No agent holds a key. The signer signs a child only after checking it against its parent, and
the verifier checks the whole chain again. A child that widens one caveat while tightening another is refused.}
\label{fig:grants}
\end{figure}

\subsection{Typed caveats}

A caveat $(f, \mathit{op}, v)$ limits one field $f$ of the intent's schema. It is valid only when $f$ exists, the
operator suits the field's type, and every value passes the field's rule:

\begin{center}\small
\begin{tabular}{@{}ll@{}}
\toprule
\textbf{Field type} & \textbf{Operators}\\
\midrule
\code{int}, \code{money} & \code{==}, \code{<=}, \code{>=}, \code{<}, \code{>}\\
\code{time} & \code{==}, \code{<=}, \code{>=} (compared as fixed-format strings)\\
\code{string}, \code{id}, \code{enum} & \code{==}, \code{in} (1--64 distinct values)\\
\code{bool} & \code{==}\\
\bottomrule
\end{tabular}
\end{center}

\noindent For comparison every caveat is normalised to one of $\{{=}v,\ \in\!S,\ {\le}b,\ {\ge}b\}$, with
$<b \mapsto\ \le b{-}1$ and $>b \mapsto\ \ge b{+}1$ on integers. Arguments \emph{satisfy} a caveat when the field is
present and the single-value caveat $= \textit{args}[f]$ implies it under the table below. Types must match exactly,
so \code{true} does not satisfy \code{<= 5}, and \code{"81"} does not satisfy \code{== 81}. A signer need not know the
schema, so it can sign a caveat whose value has the wrong JSON type; such a grant fails at the verifier twice over, as
an invalid caveat and as one that no args satisfy.

\subsection{Narrowing: a sound implication table}
\label{sec:narrowing}

\begin{table}[t]
\centering\small
\caption{When a child caveat $c$ implies a parent caveat $p$ on the same field (\cref{lem:implication}). A child
\emph{narrows} its parent iff every parent caveat is implied by at least one child caveat. Missing caveats count as
widening; the child may add caveats on any field.}
\label{tab:implication}
\renewcommand{\arraystretch}{1.2}
\begin{tabular}{@{}l cccc@{}}
\toprule
& \multicolumn{4}{c}{\textbf{child caveat} $c$}\\
\cmidrule(l){2-5}
\textbf{parent} $p$ & $=v$ & $\in T$ & $\le b'$ & $\ge b'$\\
\midrule
$=x$     & $v = x$, same JSON type & $T=\{x\}$ (strings) & never & never\\
$\in S$  & $v$ a string in $S$ & $T\subseteq S$ & never & never\\
$\le b$  & same type, $v\le b$ & never & same type, $b'\le b$ & never\\
$\ge b$  & same type, $v\ge b$ & never & never & same type, $b'\ge b$\\
\bottomrule
\end{tabular}
\end{table}

For a caveat $c$ on field $f$ of type $\theta$, let $\Sat(c)\subseteq\mathrm{Dom}(\theta)$ be the set of values of
$f$ that satisfy it, and for a set of caveats $C$, let $\Sat(C)$ be the set of argument objects that satisfy every
caveat in $C$.

\begin{lemma}[Soundness of the implication table]\label{lem:implication}
If \cref{tab:implication} says that $c$ implies $p$, then $\Sat(c)\subseteq\Sat(p)$.
\end{lemma}
\begin{proof}
By cases on the table's non-``never'' cells. ($=v$ vs.\ $=x$): $\Sat(=v)=\{v\}$ and $v=x$ with the same JSON type, so
$\{v\}=\{x\}$. ($\in T$ vs.\ $=x$): $T=\{x\}$ gives $\Sat(\in T)=\{x\}$. ($=v$ vs.\ $\in S$): $v\in S$ gives
$\{v\}\subseteq S$. ($\in T$ vs.\ $\in S$): $T\subseteq S$. ($=v$ vs.\ $\le b$): every satisfying value equals $v\le
b$. ($\le b'$ vs.\ $\le b$): any $w\le b'\le b$ satisfies $w\le b$, by transitivity of the total order on integers and
on fixed-format timestamps compared as strings. The $\ge$ cases are symmetric. Requiring the same JSON type excludes
cross-type coincidences (\code{true} vs.\ \code{1}) on which some languages' equality would otherwise make the
inclusion false.
\end{proof}

The ``never'' cells are conservative. For example, bounds never imply an equality, even though
$\{\le 81, \ge 81\}$ together pin $81$. Each child caveat is matched individually, so a child must restate an equality
as an equality. This keeps the check a table lookup, comparable across languages, rather than a solver.

\begin{corollary}[Narrowing is attenuation]\label{cor:narrow}
If child caveats $C'$ narrow parent caveats $C$, then $\Sat(C')\subseteq\Sat(C)$. Narrowing is transitive.
\end{corollary}
\begin{proof}
Take $a\in\Sat(C')$ and $p\in C$. By narrowing there is $c\in C'$ on the same field that implies $p$; $a$ satisfies $c$,
so by \cref{lem:implication} it satisfies $p$. Transitivity follows from the transitivity of $\subseteq$.
\end{proof}

\subsection{Root grants come from a person}
\label{sec:rootgrants}

A caveat-free grant is the widest grant, so the root of authority must be human. The signer issues a root grant
(\code{parent: null}) only if all of the following hold: an approvals configuration names who may answer; the agent
made a tool call \code{pcm.grant} whose args are \emph{exactly} the grant's subject, intent, caveats, expiry and
delegability, and that call was approved and the approval consumed with exactly those args; the approver is a person
mapped to a name, not an agent approver, not the run's own agent or its person (self-approval), and not the grant's
subject; \code{expires\_at} is in the future and at most 24 hours away; the call has no result yet (one grant per
approval); and the run is not held or quarantined. The issuer then becomes \code{\{kind: person, id, person, run,
tenant, approval\_id\}}. A policy that never asks a person about \code{pcm.grant} leaves nothing to consume, so no root
grant can exist: the design fails closed.

\subsection{Delegation}
\label{sec:delegation}

The subject of a delegable grant may ask the signer for a child. The signer refuses unless the parent verifies under
\emph{its own} log key, belongs to the caller's tenant, names the caller as subject, is delegable and unexpired, and
the child expires no later than the parent and narrows it. The child is issued by \code{\{kind: agent, id: caller\}},
for the parent's intent, with \code{parent} = $\Hh(\text{parent})$. A child needs no further approval, since by
\cref{cor:narrow} it grants nothing new.

\subsection{Chain verification}

\Cref{tab:chain} lists the rules a verifier checks over the chain $g_0,\ldots,g_n$, which is given root first. Two
rules deserve comment. First, \emph{every} grant must be signed by the key that signed the \emph{message}. Identities
and tenants are names inside one signer (\code{default} is everyone's default tenant), so without this rule a grant
signed by log~$X$ could ride in a message from log~$Y$ whose agent happens to share the subject string. Second, the
args must satisfy \emph{every} grant, not only the last. This is redundant given narrowing, and kept deliberately as
defence in depth.

\begin{table}[t]
\centering\small
\caption{Grant-chain rules (S7.6). Any failing rule makes the grants result \fail{} and the verdict \vrej.}
\label{tab:chain}
\begin{tabularx}{\linewidth}{@{}L{2.3cm} Y L{5.4cm}@{}}
\toprule
\textbf{Rule} & \textbf{Check} & \textbf{Example vectors (\vect{reject-} omitted)}\\
\midrule
needed & no grants: \code{none}, or \fail{} if the schema has \code{grant: true} & \vect{grant-needed}\\
signature & each \code{sig.key} = message \code{sig.key}; verifies with tag \code{pcm.grant.v0} & \vect{grant-tampered}, \vect{grant-from-another-log}\\
intent & each grant's intent = \code{intent.type} & \vect{grant-other-intent}\\
caveats & valid for the intent's schema & \vect{grant-bad-caveat}\\
expiry & message \code{ts} $\le$ each \code{expires\_at} & \vect{grant-expired}\\
tenant & each \code{issuer.tenant} = \code{from.tenant} & \vect{grant-cross-tenant}\\
args & \code{intent.args} satisfies every caveat of every grant & \vect{grant-amount-over-caveat}, \vect{grant-caveat-bool-for-int}\\
root & $g_0$: no parent, issuer kind person with a name, issuer $\neq$ subject & \vect{grant-root-by-agent}, \vect{grant-issuer-is-subject}\\
link & $g_{i}.\code{parent} = \Hh(g_{i-1})$ & \vect{grant-broken-parent}\\
delegable & $g_{i-1}$ delegable & \vect{grant-below-non-delegable}\\
who delegates & $g_i.\code{issuer} = $ agent $g_{i-1}.\code{subject}$ & \vect{grant-child-wrong-issuer}\\
expiry shrinks & $g_i.\code{expires\_at}\le g_{i-1}.\code{expires\_at}$ & \vect{grant-child-outlives-parent}\\
narrows & caveats of $g_i$ narrow $g_{i-1}$ (\cref{tab:implication}) & \vect{grant-widen-and-tighten-child}, \vect{grant-in-set-widened-child}\\
subject & $g_n.\code{subject}$ = \code{from.agent} & \vect{grant-subject-not-sender}\\
\bottomrule
\end{tabularx}
\end{table}

The verifier reports the root's issuer (kind, identity, person name) but does not decide whether to trust that
person; a receiver checks \code{grants.root} against the people it accepts. PCM~0.1 has no revocation. A grant holds
until \code{expires\_at}, which is why the signer caps root lifetimes at 24 hours and why expiry is judged against the
signer-set message \code{ts}. A later audit therefore reaches the same answer, and the replay check
(\cref{sec:replay}) prevents a message captured while its grant was valid from being acted on later.

\section{Verification, Replay and Conformance}
\label{sec:verify}

\subsection{Signatures and keys}
\label{sec:sigrules}

Log keys are pinned as C2SP signed-note verifier keys, \code{<name>+<keyid>+<base64(0x01\,||\,pk)>}, where the key id
is the first four bytes of $\mathsf{SHA256}(\textit{name}\,\|\,\texttt{0x0A}\,\|\,\texttt{0x01}\,\|\,\textit{pk})$. A
message names its key by \code{sig.key} $=$ \code{<name>+<keyid>}, and the verifier looks up the one pinned Ed25519
key with that name and id. A trust document may pin \emph{at most one} Ed25519 key per log name. With one key per
name, the key that signed a message names exactly one log, so the evidence-origin check and the replay sender are
unambiguous.

Ed25519 verification is not uniquely specified in practice, and libraries differ on edge cases~\cite{chalkias2020taming}.
PCM therefore fixes it: the signature is 64 bytes, $S < L$ (the group order), the public key is a canonical encoding
($y < 2^{255}-19$) of a point not of small order, and the \emph{cofactorless} equation $[S]B = R + [k]A$ holds. The
small-order check is not pedantry. Many libraries skip it, and a pinned small-order key (the identity point) then
verifies the forged signature $R=\mathcal{O}$, $S=0$ on \emph{any} message (vector \vect{reject-small-order-key}).
Every base64 value in PCM must be \emph{canonical}: standard alphabet, padded, no whitespace, and zero unused bits, so
that decoding and re-encoding gives back the same text. Otherwise one signature has several texts that all verify, and
a grant's text, and hence its hash in a child's \code{parent}, could change without a new signature.

The attesting signer enforces the sender side (\cref{as:signer}). It signs only as the calling run's own identity, run
and tenant, refusing a caller that merely holds the run's credential (a log reader or an observer). It sets \code{pcm},
\code{id}, \code{ts} and \code{nonce} itself and refuses a body that contains them or \code{sig} or \code{t}. It
refuses to sign for a held or quarantined run. It \textsc{should} (and \tk{} does) record each message it signs in the sender's run (its hash, id,
receiver and intent reference, but never its args), so an auditor sees every message the agent had signed, in order
among its tool calls. The signer does \emph{not} check claims or grants; that is the verifier's job.

\subsection{The verification algorithm}

\Cref{alg:verify} is the complete verifier. Its order is part of the specification. The envelope and the signature
come first, so no claim work is done for a forged message. The recipient check runs before replay. Nonces are spent
only by messages whose schema check passed (below).

\begin{algorithm}[t]
\caption{$\textsc{Verify}(m, \mathcal{T}, \mathcal{D}, \textit{state}^?, \textit{me}^?, \textit{strict}, \textit{now})$}
\label{alg:verify}
\small
\begin{algorithmic}[1]
\Require strictly parsed message $m$; trust document $\mathcal{T}$; schema documents $\mathcal{D}$
\State $\textit{skipped}\gets[\,\textsf{replay}$ if $\textit{state}=\bot$; $\textsf{recipient}$ if $\textit{me}=\bot\,]$
\If{$m$ violates the closed envelope schema} \Return \vrej{} with all parts \code{null} \EndIf
\State $k \gets$ the pinned Ed25519 key named $m.\code{sig.key}$
\If{$k=\bot$ \textbf{or} $\neg\Vf_k(\mathsf{bytes}_{\texttt{pcm.message.v0}}(m), m.\code{sig.sig})$} \Return \vrej{} (\textsf{signature}) \EndIf
\State check every $\mathcal{S}\in\mathcal{D}$ \Comment{a malformed schema is a configuration error, not a verdict}
\State $\mathcal{S}\gets$ the $\mathcal{S}\in\mathcal{D}$ with $\texttt{name@}\Hh(\mathcal{S}) = m.\code{intent.type}$;\ \ $\textit{errs}\gets\textsc{StrictArgs}(\mathcal{S}, m.\code{intent.args})$
\If{$\textit{me}\neq\bot$ \textbf{and} $m.\code{to.agent}\neq\textit{me}$} $\textit{replay}\gets$ \textsf{fail} (``addressed to \ldots''); state untouched
\ElsIf{$\textit{state}\neq\bot$} $\textit{replay}\gets\textsc{Replay}(\textit{state}, m, \textit{now}, \textit{record}{=}(\textit{errs}=\emptyset))$
\Else{} $\textit{replay}\gets$ \textsf{not checked}
\EndIf
\State $\textit{grants}\gets\textsc{Chain}(m.\code{grants}, m, \mathcal{S})$ if $\mathcal{S}\neq\bot$ \Comment{\cref{tab:chain}}
\ForAll{claims $c\in m.\code{claims}$} $\textit{res}[c]\gets\mathsf{Res}(c)$ \Comment{\cref{def:checker}}
\EndFor
\State \Return $\langle\mathsf{V}(m), \textit{signature}, \textit{replay}, \textit{schema}, \textit{grants}, \textit{res},
\textit{reasons}, \textit{skipped}\rangle$ \Comment{\cref{def:verdict}}
\end{algorithmic}
\end{algorithm}

A receiver passes its own identity as \textit{me}, keeps one replay state for its whole life, never acts on \vrej,
treats \vunp{} as the sender's word, and treats any result with non-empty \textit{skipped} as an audit result only.
Command-line verifiers keep no replay state, so they always report \textsf{replay} as skipped.

\subsection{Replay in bounded memory, across restarts}
\label{sec:replay}

A \emph{sender} is the pair $(\code{sig.key}, \code{from.agent})$, never the agent identity alone. An identity is a
name inside one signer, so \code{uid:990101} of one log and \code{uid:990101} of another are two senders, and one must
not raise the other's floor or spend its nonces (vector \vect{accept-replay-same-agent-other-log}). The state has
parameters $W$ (window, default 300\,s), $P$ (per-sender capacity, default 4096) and \textit{started} (the creation time
of the state). It holds, for each sender $\sigma$, a FIFO list $\textit{seen}[\sigma]$ of $(\textit{nonce}, t)$ pairs
in arrival order and a floor $\textit{floor}[\sigma]$.

\begin{algorithm}[t]
\caption{$\textsc{Replay}(\textit{state}, m, \textit{now}, \textit{record})$, with $t=m.\code{ts}$ and $\sigma=(m.\code{sig.key}, m.\code{from.agent})$}
\label{alg:replay}
\small
\begin{algorithmic}[1]
\If{$t$ names no real UTC time} \Return \textsf{fail} \EndIf
\If{$W\neq\bot$ \textbf{and} $(|\textit{now}-t| > W$ \textbf{or} $t<\textit{started})$} \Return \textsf{fail} \Comment{``ask for it again''}
\EndIf
\If{$t\le\textit{floor}[\sigma]$} \Return \textsf{fail} \Comment{older than what this receiver remembers}
\EndIf
\If{$m.\code{nonce}\in\textit{seen}[\sigma]$} \Return \textsf{fail} \Comment{replayed}
\EndIf
\If{\textit{record}}
  \State append $(m.\code{nonce}, t)$ to $\textit{seen}[\sigma]$
  \If{$W\neq\bot$} \textbf{while} head of $\textit{seen}[\sigma]$ has $t' < \textit{now}-W$: pop it \EndIf
  \While{$|\textit{seen}[\sigma]| > P$}
     \State $(\_, t')\gets$ pop head;\quad $\textit{floor}[\sigma]\gets\max(\textit{floor}[\sigma], t')$
  \EndWhile
\EndIf
\State \Return \textsf{fresh}
\end{algorithmic}
\end{algorithm}

Three design decisions are visible in \cref{alg:replay}.
\begin{enumerate}[leftmargin=2em]
\item \textbf{Restart without saved state.} A receiver refuses any message signed before its state was created, so a
  restarted receiver with empty state cannot be fed a message it may already have accepted; the sender simply asks again.
\item \textbf{Bounded memory.} Eviction raises a per-sender floor, and messages at or below the floor are refused. Memory
  is $O(P)$ per sender regardless of traffic. Eviction is in \emph{arrival} order, not timestamp order. Both orders
  are safe (\cref{thm:replay}), but they give different verdicts on some sequences, and PCM pins one so that verifiers
  agree (vector \vect{reject-replay-floor-arrival-order}).
\item \textbf{Which messages spend a nonce.} A message addressed to another receiver is not this receiver's to spend,
  and one whose schema the receiver lacks may come again once the receiver has the right document. Both leave the
  state unchanged. A message rejected for its grants or claims is recorded, since it would be rejected again anyway.
\end{enumerate}
Grant expiry is judged at the signer-set \code{ts}, so an audit is reproducible. That is exactly why a receiver that
\emph{acts} must use replay state: without it, a message captured while its grant was valid would remain \vacc{}
forever.

\subsection{Conformance levels and test vectors}

PCM defines two levels. \textbf{Level~1} checks the envelope, signature, schema, grants, replay and recipient, the
verdict, and the malformed-claim rules, and gives every claim with evidence the result \unprov{} (``evidence not
checked''). \textbf{Level~2} adds evidence, the evidence profile and the four claim checkers. A verifier conforms on a
vector when its \code{verdict}, \code{signature.ok}, \code{replay.ok}, \code{schema.ok}, \code{grants.result}, every
\code{claims[i].result}, every \code{claims[i].approved} and \code{skipped} equal the expected values. Human-readable
\code{detail}, \code{reasons} and \code{problems} strings are not compared. \Cref{tab:vectors} summarises the 187
vectors, all produced by the reference implementation's real signer.

\begin{table}[t]
\centering\small
\caption{The PCM~0.1 conformance vectors by level and expected verdict. A vector is Level~2 iff some claim carries
evidence. One \vbad{} vector holds raw text (\code{1e400}) that must be refused as a parse error.}
\label{tab:vectors}
\begin{tabular}{@{}lrrrrr@{}}
\toprule
& \vacc & \vunp & \vrej & \vbad & \textbf{total}\\
\midrule
Level 1 (no evidence) & 11 & 2 & 35 & 1 & 49\\
Level 2 (evidence)    & 19 & 44 & 75 & -- & 138\\
\midrule
\textbf{all}          & \textbf{30} & \textbf{46} & \textbf{110} & \textbf{1} & \textbf{187}\\
\bottomrule
\end{tabular}
\end{table}

\subsection{Transport bindings and versioning}

PCM's authenticity does not depend on the transport, and confidentiality is left to TLS. In A2A, a PCM message is one
part of media type \code{application/vnd.pcm+json}, and parts outside it are not covered by its signature. In MCP, it
is an embedded resource in a tool result or in arguments. Over HTTP, it is the request body. These bindings are
drafts. The wire version \code{"pcm": 0} covers every 0.x draft and is bound to the signature tags, so a change to what
is signed needs a new wire version. Every object is closed, so a verifier rejects what it does not know.

\section{Security Analysis}
\label{sec:security}

We now state the guarantees PCM provides against the adversary of \cref{sec:model} and prove them under
\cref{as:crypto,as:signer,as:t2,as:clock}. ``Negligible'' is with respect to the security parameter of Ed25519 and
SHA-256. Throughout, the verifier is a conforming Level~2 verifier and $\mathcal{T}$ pins log key $pk_L$ for log $L$.

\subsection{Domain separation}

\begin{lemma}[Signed byte strings are disjoint]\label{lem:domains}
The byte strings that $sk_L$ signs for (a)~messages, (b)~grants and (c)~checkpoint notes are pairwise distinct for all
inputs.
\end{lemma}
\begin{proof}
(a) and (b) are $\JCS$ encodings of objects whose \code{"t"} members differ (\code{pcm.message.v0} versus
\code{pcm.grant.v0}). $\JCS$ is injective on the JSON data model under strict parsing (no duplicate names, finite
numbers, integers within $2^{53}$), so the encodings differ. A checkpoint note's text consists of lines terminated by
the byte \texttt{0x0A}, whereas a $\JCS$ encoding contains no raw line feed (it escapes control characters in strings
and emits no whitespace between tokens). A note therefore never equals (a) or (b).
\end{proof}

\subsection{Origin and evidence}

\begin{theorem}[Origin authenticity, G1]\label{thm:origin}
If \textsc{Verify} reports \code{signature.ok} for $m$, then, except with negligible probability, the honest signer of
$L$ produced that signature on $\mathsf{bytes}_{\texttt{pcm.message.v0}}(m)$ in response to a signing request from the
run $(m.\code{from.run}, m.\code{from.tenant})$ registered by $m.\code{from.agent}$, and it chose $m.\code{id}$,
$m.\code{ts}$ and $m.\code{nonce}$ itself.
\end{theorem}
\begin{proof}
Since $\mathcal{T}$ pins exactly one Ed25519 key per log name, \code{sig.key} determines $pk_L$. Suppose the adversary
outputs $m$ that verifies but whose signed bytes the signer never signed as a message. By \cref{lem:domains} those bytes
are not the bytes of any grant or note the signer signed, so the adversary has forged a signature on a fresh message,
which contradicts EUF-CMA (\cref{as:crypto}). The strict verification rules exclude the known malleability and
small-order-key forgeries that EUF-CMA proofs for Ed25519 assume away~\cite{chalkias2020taming,brendel2021provable}.
So the signer signed these exact bytes, and by $\JCS$ injectivity every member except \code{sig}, including
\code{claims}, \code{grants} and \code{payload}, is what the signer saw. By \cref{as:signer}(i) the signer signs only
for the calling run, with \code{from} equal to that run's registered identity, run and tenant, and with \code{id},
\code{ts} and \code{nonce} of its own choosing.
\end{proof}

\begin{theorem}[Evidence soundness, G2]\label{thm:evidence}
If $\mathsf{EvOK}(E,m,\mathcal{T})$ holds, then, except with negligible probability, for every record $(e_j,h_j,\pi_j)$ in
$E$ the honest signer of $L$ appended a record equal to $e_j$ at index $e_j.\code{seq}$ of its log, in the run
$(m.\code{from.run}, m.\code{from.tenant})$. If moreover $e_j$ carries an opening $(s,a)$ that verifies, the arguments the
signer committed to for that decision are $a$.
\end{theorem}
\begin{proof}
The checkpoint carries a valid line of $pk_L$, and by the checkpoint definition its origin is the log that signed $m$. By
\cref{lem:domains} and EUF-CMA, the signer issued the note $(n,R)$, and an honest signer issues notes only for its
log's actual tree head at size $n$. The inclusion proof $\pi_j$ verifies leaf $\mathsf{SHA256}(\texttt{0x00}\|h_j)$ at
index $e_j.\code{seq}$ against $(n,R)$. Since RFC~6962/9162 hashing separates leaves (\texttt{0x00}) from interior nodes
(\texttt{0x01}), any proof for a leaf different from the true leaf at that index yields a SHA-256 collision on the path
to $R$. Hence $h_j$ is the hash of the record the signer appended there. By (E1), $h_j = \Hh(e_j)$, and by
collision resistance and $\JCS$ injectivity, $e_j$ \emph{is} that record. (E4) checks its run and tenant. For an
opening, a second pair $(s',a')\neq(s,a)$ that reproduces the commitment contradicts the binding property in
\cref{as:crypto}.
\end{proof}

\begin{corollary}[No self-attestation, G3]\label{cor:t1}
No claim whose evidence contains, as its only matching records, \T1 tool results is \prov.
\end{corollary}
\begin{proof}
By inspection of the checkers. \code{tests\_passed} and \code{file\_written} return \prov{} only from the \T2 branch of
$\Ran$ (\cref{alg:ran}), which requires top-level \code{source}=\code{gateway} and \code{tier}=\code{T2}. \code{reached\_host} returns
\prov{} only from that line or from an \code{effect.observed} record. \code{approved\_by} needs \code{approval} and
\code{approval.request} records. By \cref{thm:evidence} every record is a genuine record of $L$, and by
\cref{as:signer}(iii)--(iv) the signer sets these types and fields only from authenticated attesting parties,
observers and persons. A \T1 result can only add \unprov{} or \fail{} candidates.
\end{proof}

\cref{cor:t1} is the property that plain signing does not provide: an agent that signs its own results with its own key
can mark them in any way it likes. PCM's guarantee rests on who \emph{set the field}, not on who signed the bytes.

\begin{theorem}[Approval binding]\label{thm:approval}
If an \code{approved\_by\{person: p, call: c, covers: true\}} claim in $m$ is \prov, then, except with negligible
probability, the configured person named $p$, not an agent approver and not self-approving, answered \code{approve} to a
request that the signer created about a decision $d$ with \code{tool\_use\_id} $c$ in the sender's run. In addition,
either (grant path) that approval is the one whose consumption issued $m$'s root grant, so the person approved exactly
the root grant's terms; or (direct path) $d$'s tool is the tool the intent's schema names, and the arguments the signer
committed for $d$ equal $m.\code{intent.args}$ in $\JCS$.
\end{theorem}
\begin{proof}
By \cref{thm:evidence} the three linked records are genuine, and by \cref{as:signer}(iv) the approval reflects an
authenticated answer from the identity the configuration maps to $p$, with \code{approver\_kind} and
\code{self\_approved} set by the signer. The linkage by \code{approval\_id} and \code{decision\_id} identifies $d$
uniquely: if the requests named two decisions, or the qualifying decisions differed in $\JCS$, the claim would be
\fail. On the grant path, the root grant's signature (\cref{thm:origin}'s argument applied with the grant tag) and
\cref{as:signer}(v) imply that the signer minted it after consuming exactly this approval, whose decision had exactly
the grant's terms as arguments. On the direct path, the opening binds $d$'s committed arguments (\cref{thm:evidence}),
and the verifier compared them with $m.\code{intent.args}$ in $\JCS$.
\end{proof}

\subsection{Authority}

For a chain $g_0..g_n$ write $C_i$ for the caveats of $g_i$.

\begin{theorem}[Attenuation, G4]\label{thm:attenuation}
If the grants result of $m$ is \code{ok}, then, except with negligible probability: (i)~$g_0$ was issued by the honest
signer of the log that signed $m$, after a configured person other than $g_0$'s subject approved exactly $g_0$'s terms;
(ii)~$m.\code{intent.args}\in\Sat(C_n)\subseteq\cdots\subseteq\Sat(C_0)$; (iii)~$m.\code{ts}\le
g_n.\code{expires\_at}\le\cdots\le g_0.\code{expires\_at}$; (iv)~every $g_i$ is for $m$'s intent and tenant; and
(v)~$g_n$'s subject is $m$'s sender.
\end{theorem}
\begin{proof}
Every grant verifies under the message's own key with the grant tag, so by EUF-CMA and \cref{lem:domains} the honest
signer of $L$ signed it. For $g_0$, the signer's root-grant conditions (\cref{sec:rootgrants}, \cref{as:signer}(v)) give
(i), and the verifier independently checks that $g_0$'s issuer is a person with a name, distinct from the subject, and
that $g_0$ has no parent. The link rule, $g_i.\code{parent} = \Hh(g_{i-1})$, with collision resistance and canonical
base64 (so a signed grant has a single text) ensures that the chain is the one the signer delegated. The narrowing rule
and \cref{cor:narrow} give the inclusions in (ii), and the args rule checks membership in every $\Sat(C_i)$ directly.
(iii)--(v) are the expiry, expiry-shrinks, intent, tenant and subject rules of \cref{tab:chain}.
\end{proof}

Note that clause~(ii) holds even if a buggy signer signs a widening child: the verifier's own narrowing check rejects
it. In our benchmark, all 11 grants deliberately re-signed with the log key to model a signer bug were rejected
(\cref{sec:eval}).

\subsection{Delivery}

\begin{theorem}[At-most-once acceptance, G5]\label{thm:replay}
Let a receiver identity keep one replay state (shared by its replicas, \cref{as:clock}) with a set window $W\neq\bot$, and suppose the
signer's \code{ts} never exceeds the receiver's clock at first receipt (so clock skew is absorbed by $W$). Then for each
sender $\sigma$ and message $m$, the receiver returns a non-\vrej{} verdict for $m$ at most once, including across
restarts that create a fresh state.
\end{theorem}
\begin{proof}
A non-\vrej{} verdict requires no schema error, so the first such verdict for $m$ \emph{recorded} $(\nu,t)$ in
$\textit{seen}[\sigma]$ at some time $\textit{now}_1\ge t$, where $\nu$ and $t$ are $m$'s nonce and timestamp. Consider
a later presentation of $m$ (same $\nu$ and $t$, since any change breaks the signature, \cref{thm:origin}) at
$\textit{now}_2\ge\textit{now}_1$. (a)~If $(\nu,t)$ is still in $\textit{seen}[\sigma]$, the nonce check rejects. (b)~If the window
evicted it at some $\textit{now}'\in[\textit{now}_1,\textit{now}_2]$, then $t<\textit{now}'-W\le\textit{now}_2-W$, so
$|\textit{now}_2-t|>W$ and the window check rejects. (c)~If capacity evicted it, $\textit{floor}[\sigma]\ge t$ from then on
(floors never decrease), so the floor check rejects. (d)~If the receiver restarted with a fresh state at $\textit{started} >
\textit{now}_1\ge t$, then $t<\textit{started}$ and the start check rejects. The recipient check additionally prevents a
message addressed to $B$ from being accepted by $C$.
\end{proof}

\subsection{Attack catalogue}
\label{sec:attacks}

\Cref{tab:attacks} condenses the 29 attacks, mistakes and trust assumptions that the specification records (S14). Most were found while
building or reviewing the reference implementation, or exposed by the benchmark; items 24 and 25 are trust assumptions
rather than attacks, and item 13 is bounded, not closed. Most items are pinned by conformance vectors. We highlight three patterns. First, \emph{identity is relative to a signer}: run ids, tenants
and agent names are only unique within one log, which produced three separate attacks (items 1, 18 and the replay-sender
rule). Second, \emph{the agent chooses names}, of calls, of tools and of files, so every checker reads a name only
through a mapping the agent does not control. Third, \emph{an attempt is not an outcome}: a \code{connect}, a denied
call or a zero-connection minute never proves a connection.

\begin{table}[p]
\centering\scriptsize
\caption{Attacks and mistakes closed by PCM~0.1 rules (condensed from S14). ``Class'' marks where the check lives:
\textsf{E}nvelope/signature, \textsf{V}idence, \textsf{C}laim checker, \textsf{G}rant, \textsf{R}eplay, \textsf{D}eterminism.}
\label{tab:attacks}
\renewcommand{\arraystretch}{1.12}
\begin{tabularx}{\linewidth}{@{}r c L{4.0cm} Y@{}}
\toprule
\# & \textbf{Class} & \textbf{Attack or mistake} & \textbf{Rule that closes it}\\
\midrule
1 & V & Reuse a victim's run id and tenant on another signer; replay its evidence & checkpoint origin $=$ message's log; one key per log name; records of sender's run and tenant\\
2 & C & Agent reports its own tool result as success (\T1) & only \T2 or signer-written records prove; tier read from signer-set top-level fields\\
3 & C & Harmless tool given a \code{command}/\code{url} member to pose as shell/http & placement by frozen table of tool names only\\
4 & G & Grant from one tenant authorises another & issuer tenant must equal \code{from.tenant}; no cross-tenant delegation\\
5 & E & Log key signs checkpoints, messages and grants & tags \code{pcm.message.v0}/\code{pcm.grant.v0}; JCS bytes never parse as a note\\
6 & G & Root grant minted by an agent, self-approved, or by the subject & root only from a configured person's approval of the exact terms\\
7 & G & Child widens, drops or re-shapes a caveat; outlives parent & implication table; expiry shrinks; intent fixed\\
8 & G & One consumed approval mints several roots & refuse once the \code{pcm.grant} call has a result\\
9 & E & Held or quarantined run keeps signing & signer refuses held runs\\
10 & C & \code{pytest || true}, \code{pytest \&}, \code{bash -c '\ldots'} & suite must be one plain-word test command from a fixed list\\
11 & C & \code{\textbackslash} read as \code{/}; relative or \code{..} paths; edits without content & POSIX vs.\ drive-path rules by claim; absolute only; full content digest\\
12 & D & Hostile inputs crash a verifier or are read leniently & total functions; strict patterns; ASCII classes; key validity checks\\
13 & C & Cite a pass after a later failure (stale evidence) & \textit{bounded only}: age and digest checks by receiver (\cref{sec:limits})\\
14 & R & Replay to another replica, after restart, below the floor & shared state per identity; refuse \code{ts} $<$ \textit{started}; arrival-order floor\\
15 & E & Hostile schema patterns (ReDoS); untrusted schema documents & receiver loads only schemas it chose and reviews them as code\\
16 & V & Salts disclose low-entropy args & salts only to the run's own agent; open only what a claim needs\\
17 & G & No revocation & 24\,h cap on roots; expiry at signer \code{ts}; replay check mandatory to act\\
18 & G & Grant of log $X$ rides in a message of log $Y$ & every grant signed by the message's key; sender $=$ (key, agent)\\
19 & D & Splitter vs.\ shell disagree (\code{\#} in a word, \code{\$'\ldots'}) & forbid every character on which they can differ\\
20 & D & Host spellings, IDNA versions, bracket checks, octal IPv4 & ASCII-only canonical host; own bracket rule; dotted quads only\\
21 & E & Lenient base64 makes signatures malleable & canonical base64 everywhere\\
22 & C & \code{approved\_by} matched against any tenant identity & compare only with the configured person's name\\
23 & C & Runner matched by base name (\code{/tmp/x/pytest}) & first word compared exactly, as a bare name\\
24 & E & Key compromise; rotation & remove leaked key; one key per name; witnesses for checkpoints (\cref{sec:keycompromise})\\
25 & E & The signer is the TCB & run signer in another trust domain; pin only signers you trust\\
26 & C & HTTP action named \code{Bash} proves \code{tests\_passed} & a profile log must not record a tool under another class's name\\
27 & C & Approval of a call the agent named, or of other args & three-record linkage; \code{approved} field; \code{covers} check\\
28 & C & Connect attempt taken as a connection & only \code{net} with \code{connections} $>0$ proves\\
29 & D & Placement table changes with policy packs & table frozen as protocol data with a digest\\
\bottomrule
\end{tabularx}
\end{table}

\subsection{Key compromise and the trusted computing base}
\label{sec:keycompromise}

Whoever holds a pinned log key \emph{is} that log's signer to the receiver. They can sign messages as any identity, run
and tenant of the log, with any \code{ts} and \code{nonce}; they can sign root grants with any issuer and expiry, since
the verifier cannot check how long a grant was meant to live; and they can sign checkpoints over fabricated records. In
PCM~0.1 the one log key signs all three, so a single leak forges all three. Witnesses limit the damage for checkpoints:
with $\code{witnesses\_required}>0$, forged evidence needs a checkpoint that witnesses cosigned, and witnesses cosign only
a history consistent with what they saw, so a forgery either fails or forks the log in a way that the honest signer's
next checkpoint exposes. Messages and grants carry no cosignature. A receiver that learns of a leak removes the key,
which fails everything the key signed (honest senders ask again), and treats what it accepted since the leak as
unverified. A dedicated key for messages and grants and a declared rotation schedule are open items for the next draft.

More generally, a verifier checks that the signer \emph{signed}, not that the signer was \emph{right}. This is the same
trust structure as CT and SLSA provenance. PCM narrows the TCB to one service with a small interface (decide, record,
sign, delegate, hand out evidence) that can run as another user, on another host or in another trust domain, instead of
the whole agent stack. In particular, the LLM, its prompt, its tools' outputs and its planning code are all outside the
TCB.

\section{Determinism by Construction}
\label{sec:determinism}

A verification protocol is only as strong as its \emph{weakest conforming verifier}. If two verifiers can disagree on a
message, an adversary can probe for the more permissive one, and a sender cannot know whether its honest message will
be accepted. Determinism (G6) is therefore a security property, not a convenience. It is also hard to achieve:
PCM's verdicts depend on parsing JSON, base64, URLs, IP addresses, shell command lines, file paths and regular
expressions, and widely used libraries disagree on every one of them.

\subsection{The strict-reading principle}

Let a field have readers $\rho_1,\ldots,\rho_k$: the parsers that real verifiers, clients, resolvers or shells apply
to it. PCM defines its reading $\hat\rho$ as follows.
\[
\hat\rho(x) = \begin{cases} \rho_1(x) & \text{if } x \in \mathcal{G} \text{ and } \rho_1(x)=\cdots=\rho_k(x),\\
\bot \ (\text{``won't parse''}) & \text{otherwise,}\end{cases}
\]
where $\mathcal{G}$ is a syntactically simple set, specified character by character, inside which the readers are known
to agree. Where the field is one the sender cannot simply rewrite (a host in a URL that was really fetched, the path of a call
that really ran), a claim whose decisive field reads as $\bot$ is \unprov, never \prov{} and never \fail. The guarantee
follows directly. \emph{If PCM proves a fact from a field, then every reader in the specification's list would have
read that field the same way}, so a sender cannot exploit the gap between, for example, the resolver that made the
connection and the library the verifier uses. For such fields $\bot$ maps to \unprov, so the sender loses the proof but not the
message. Fields the sender writes in the claim itself are held to a stricter standard: a \code{tests\_passed} suite
outside the plain-word grammar, or a relative \code{file\_written} path, is malformed and \fail{}, and a number outside
the double range makes the whole text \vbad. Every $\bot$ comes with a fixed sentence that tells the sender how to resend in
$\mathcal{G}$ (for instance, ``send it as four decimal numbers 0--255 without leading zeros''), which serves G7.

\Cref{tab:divergence} lists the divergences that shaped PCM~0.1. Most were found by testing the reference
implementation against an adversarial review or by the benchmark. In each case an earlier draft deferred to ``what the
library does'', and in each case that let at least one verifier prove something another refused, or prove something
false.

\begin{table}[t]
\centering\scriptsize
\caption{Reader divergences that PCM~0.1 resolves by a strict reading. Each row is pinned by at least one conformance
vector (name abbreviated). ``Effect before'' is what the divergence allowed in an earlier draft or a naive
implementation.}
\label{tab:divergence}
\renewcommand{\arraystretch}{1.18}
\begin{tabularx}{\linewidth}{@{}L{3.25cm} L{3.4cm} Y L{2.75cm}@{}}
\toprule
\textbf{Input} & \textbf{Readers disagree} & \textbf{Effect before $\rightarrow$ PCM~0.1 rule} & \textbf{Vector}\\
\midrule
\code{http://0127.0.0.1/} & \code{inet\_aton}: \code{87.0.0.1}; macOS resolver: \code{127.0.0.1} & proved a host never
  contacted (F1) $\rightarrow$ IPv4 only as a dotted quad & \vect{...-ipv4-leading-zero}\\
\code{1e400} & one JSON loader: error; another: $+\infty$ & different verdict classes (F5) $\rightarrow$ parse error,
  \vbad & \vect{bad-input-number-past-double}\\
\code{http://[::1]@}\newline\code{api.example.com/} & CPython \code{urlsplit}, across releases, three bracket checks & \vacc{} vs.\
  \vunp{} $\rightarrow$ one bracket rule, applied first & \vect{...-bracket-ipv6-userinfo}\\
\code{[::ffff:01.2.3.4]} & \code{ipaddress} $<$3.9.5 accepts; $\ge$3.9.5 refuses & reached \code{1.2.3.4} vs.\ nothing
  $\rightarrow$ refuse leading zero in an embedded quad & \vect{...-ipv6-leading-zero}\\
\code{faß.example}, U+1806 & IDNA 2003 vs.\ UTS 46 vs.\ browsers & proved \code{example.com} from a look-alike
  $\rightarrow$ printable ASCII hosts only & \vect{...-deviation}, \vect{...-idn}\\
\code{http://a.example\textbackslash}\newline\code{@b.example/} & browser: \code{a.example}; generic parser: \code{b.example} &
  either host provable $\rightarrow$ two readings must agree & \vect{...-ambiguous}\\
\code{/api.example.com/x} & client with a base URL: a path; curl: a host & host provable without a scheme
  $\rightarrow$ require a scheme or \code{//} & \vect{...-relative-path}\\
\code{pytest}\newline\code{\$'\textbackslash'';true \#'} & \code{shlex}: one command; \code{sh}: \code{pytest "'"; true} & exit 0
  regardless of tests $\rightarrow$ plain-word suites only & \vect{...-ansi-c-quote}\\
\code{pytest -q x.py\#}\newline\code{|| true} & \code{shlex}: comment; \code{sh}: literal \code{\#} then \code{||} & same
  $\rightarrow$ no \code{\#} at all & \vect{...-hash-inside-word}\\
\code{\{"amount": true\}}\newline vs.\ \code{1} & Python/JS equality: equal; $\JCS$: different & approval of one args
  proved another $\rightarrow$ compare in $\JCS$ & \vect{...-args-true-and-1}\\
non-canonical base64 & lenient vs.\ strict decoders & malleable signatures and grant hashes $\rightarrow$
  canonical only & \vect{reject-sig-non-canonical-base64}\\
small-order Ed25519 key & libraries skip key checks & forgery on any message $\rightarrow$ strict key and $S$ checks &
  \vect{reject-small-order-key}\\
\code{.} in a pattern & Python: not \code{\textbackslash n}; ECMAScript: not \code{\textbackslash r}, U+2028/9 too &
  envelope valid for one verifier only $\rightarrow$ patterns use \code{[\^{}\textbackslash n]} & \vect{accept-identity-line-break-chars}\\
tool placement & table derived from installed policy packs & verdicts changed with configuration $\rightarrow$ frozen
  table with a digest & (all \code{shell}/\code{http} vectors)\\
\bottomrule
\end{tabularx}
\end{table}

\subsection{Mechanisms}

Four mechanisms implement the principle.
\begin{enumerate}[leftmargin=2em]
\item \textbf{Closed, strictly parsed objects.} Every JSON object in PCM is closed and every pattern matches the whole
  value with ASCII classes. Parsing refuses duplicate names, lone surrogates, numbers outside the double range and
  integers past $2^{53}$, so every verifier works from the same parsed value, and $\JCS$ gives that value the same
  bytes everywhere.
\item \textbf{Specification by steps, not by library.} Where a reading is intricate (hosts, paths, test commands), the
  specification states it step by step. Notes name the standard-library functions that behave as a step does (so a
  Python verifier can call \code{urlsplit} \emph{after} applying the bracket rule itself), but the step, not the library,
  is the rule.
\item \textbf{Protocol data, frozen and hashed.} The tool placement table (54 names) and the runner list (42 sequences)
  are part of the protocol and shipped with every verifier (the table is also identified by its SHA-256), never taken from a message or an
  operator.
\item \textbf{Conformance fields.} The result fields that receivers act on (\code{verdict}, sub-results, each claim's
  result, \code{approved}, \code{skipped}) are compared on every vector, while human-readable strings are not. This
  separates what must be identical from what may be localised or improved.
\end{enumerate}

\paragraph{Evidence that it works}
Two verifiers were written separately: \tk's \code{tracekit.pcm.verify} and the standalone \code{pcm\_verify}, which
shares no code with \tk{} and depends only on an RFC~8785 library and \code{cryptography}. On the first full benchmark
run they disagreed on 5 of 879 messages, all on \code{1e400}-style inputs (F5). After the strict-number rule was
added, they agree on the verdict, reason class and conformance fields of all 926 messages of the final dataset,
including messages sent as raw text through each verifier's own strict loader (\cref{sec:eval}). Both also pass all
187 vectors.

\section{Implementation}
\label{sec:impl}

\paragraph{Reference implementation}
\tk~1.5.0~\cite{ghosh2026tracekit} (signer RPC contract 20) implements every part of PCM~0.1 marked \emph{Implemented} in the specification. Its
attesting signer serves four PCM calls: \code{pcm\_sign} signs a message body as the calling run, \code{pcm\_grant}
issues a root grant from a consumed person approval, \code{pcm\_delegate} issues a narrowing child, and
\code{pcm\_evidence} returns records of the caller's own run with inclusion proofs. The signer assigns every record's
\code{ts} from its own clock when it appends the record. It writes a checkpoint at most once per second on request (and
every 10\,s while the log grows), so an honest sender has evidence about one second after a call. It records an
\code{ext} record for every message and grant it signs. Its executor runs tool calls under the signer's custody
protocol (decide, ask a person, consume the approval, complete as the executor's identity) and reports \T2 results. A
log may also pin a hybrid SLH-DSA-SHA2-128s key (FIPS~205) under the log's name, in which case every checkpoint needs a
valid post-quantum signature line as well.
The command \code{tracekit pcm demo} runs the refund example end to end with a real signer:

\begin{lstlisting}[language={},basicstyle=\ttfamily\scriptsize]
alice approves agent A's grant: refund.request, amount <= $200.00, order == 81, until 18:00Z, can be passed on: no
A -> B  refund.request {amount $120.00, order 81} + approved_by(alice)
    B: ACCEPT           signature ok . replay ok . schema ok . grant ok (alice, 1 step) . approved_by proven (...)
A -> B  same request + tests_passed(pytest tests/test_refund.py), no evidence
    B: ACCEPT-UNPROVEN  tests_passed: unproven (no evidence)
A -> B  approved_by with one record hash changed
    B: REJECT           approved_by: failed (record 1: hash does not match the event)
A -> B  refund.request {amount $500.00, order 81} + approved_by(alice)
    B: REJECT           grant: grant 0: amount 50000 breaks amount <= 20000
A asks the signer to sign as B
    signer: forbidden: the signer signs only as the calling run's own agent (uid:990101, run ..., tenant default)
\end{lstlisting}

\paragraph{Standalone verifier}
\code{pcm-verify}~0.1.0 is an independent Level~2 verifier of about 1{,}200 lines of Python (CPython $\ge$\,3.11.4). It
contains no \tk{} code and depends only on \code{rfc8785} and \code{cryptography}. It ships a copy of the frozen tool
table and checks its digest. Its exit codes map to verdicts (0 \vacc, 3 \vunp, 1 \vrej, 2 unreadable input or
configuration error). After a proven \code{tests\_passed} or \code{file\_written}, it prints how old the oldest evidence
record is, to support the receiver's age check (\cref{sec:limits}). Its test suite (556 tests) checks it on every vector
and on every example that the specification and README tie to a vector.

\paragraph{Generated artefacts}
JSON Schemas (draft 2020-12) for the message, grant, claim, evidence and schema document, and the event schema of the
evidence profile, are generated from \tk's RPC shapes. The 187 vectors are regenerated from a real signer by a script,
and the specification's examples are quoted from them verbatim and checked by the verifier's tests, so the prose cannot
drift from the vectors.

\paragraph{Known gaps of the reference implementation}
The specification records where \tk{} does less than it asks. Replay state is per process, so a deployment with several
replicas per receiver identity must route each sender to one replica. And until a \code{pcm.grant} call has its
result, repeated grant requests on one consumed approval can mint grants that differ only in \code{ts}, which the
specification permits but discourages. Transport bindings are drafts: \tk's A2A gateway forwards a PCM part without
verifying it.

\section{Evaluation: PCM-Lies}
\label{sec:eval}

We ask how often a receiving agent accepts a false claim from a sending agent, and at what cost to honest senders. The
evaluation addresses six research questions:
\begin{description}[leftmargin=0pt,style=unboxed,font=\normalfont\bfseries]
\item[RQ1 (Safety).] How many lies does a PCM receiver accept, and of which kinds?
\item[RQ2 (Honest progress).] Are honest claims backed by attested evidence proven?
\item[RQ3 (No self-attestation).] Are claims backed only by the agent's own reports ever proven?
\item[RQ4 (Determinism).] Do independent verifiers agree?
\item[RQ5 (Comparison).] What do existing agent protocols let a receiver catch, and why?
\item[RQ6 (Cost).] How large are PCM messages, and how long does verification take?
\end{description}

\subsection{Methodology}

\paragraph{Pre-specified design}
Before the first benchmark run, we wrote down the threat model, a taxonomy of 16 lie classes, four hypotheses with pass
bars, the two baselines, and the cost metrics:
\begin{itemize}
\item \textbf{H1}: no lie is accepted (0 false accepts);
\item \textbf{H2}: at least 99\% of honest claims backed by \T2 or signer records are \prov;
\item \textbf{H3}: every claim backed only by \T1 is \unprov;
\item \textbf{H4}: the two verifiers give the same verdict and reason class on every message.
\end{itemize}
The design was a dated planning note written before the first benchmark commit. It has no independent public
timestamp, so we present it as a pre-specified design rather than a registered one, and list every departure from it in
\cref{sec:threats}. The six real-protocol receivers were added after the results and are covered by no hypothesis.

\paragraph{Ground truth from real runs}
Every message comes from runs of a real \tk~1.5.0 signer, and all \T2 calls go through the executor's own custody code.
The tools really run. \code{pytest} runs 12 suites that pass, 12 that fail, 5 that pass and then fail after a code
change, and 10 look-alike test files, all in a temporary project. HTTP fetches go to a local server on \code{127.0.0.1}
and \code{::1}. Files are written in a temporary directory. Refunds and \code{read\_file} calls are answered by person
identities (alice, carol, erin), by an unmapped approver (dave) and by an agent approver. Root grants come from a
person's approval of \code{pcm.grant}, and delegations from \code{pcm\_delegate}. Egress-observer records are
included. Each message is labelled \textsc{true}, \textsc{false} or \textsc{unknowable} (the only evidence is \T1) at
construction time, \emph{from what happened}, never by a verifier. For tests, the label is the exit code of the suite's
latest run. For hosts, it is the connection's destination as the machine's resolver gives it. For files, it is the inode
and the digest at send time. For approvals, it is who answered what. A \emph{lie} is a \textsc{false} message, or an
\textsc{unknowable} one whose claim was in fact false. Most lies are edits a sender can make to its own message body
\emph{before} the signer signs it, which is within the threat model. L6 also edits messages after signing. Eleven grants
are re-signed with the log key to model a buggy signer; they are outside the threat model and reported apart.

\begin{table}[t]
\centering\footnotesize
\setlength{\tabcolsep}{4pt}
\caption{The PCM-Lies dataset (926 messages). Columns are claim types: \textsf{T} \code{tests\_passed}, \textsf{H}
\code{reached\_host}, \textsf{F} \code{file\_written}, \textsf{A} \code{approved\_by}, \textsf{G} grant only,
\textsf{X} three claims. L4 and AB-t1 are \textsc{unknowable} (\T1-only); L16 mixes 96 lies with 35 true encodings.}
\label{tab:dataset}
\renewcommand{\arraystretch}{1.08}
\begin{tabular}{@{}l l rrrrrr r@{}}
\toprule
\textbf{Class} & \textbf{Lie (mutation)} & \textsf{T} & \textsf{H} & \textsf{F} & \textsf{A} & \textsf{G} & \textsf{X} & \textbf{Total}\\
\midrule
\multicolumn{9}{@{}l}{\textit{Content lies: the sender says something false about the world}}\\
L1 & fabricated event (no such call or approval) & 12 & 12 & 12 & 12 & & & 48\\
L2 & wrong outcome (ran and failed; claims success) & 12 & 12 & 12 & 12 & & & 48\\
L3 & wrong object (other host, file, args or tool) & 12 & 12 & 12 & 23 & & & 59\\
L4 & \T1-only evidence (agent's own report) & 12 & 12 & 12 & & & & 36\\
\multicolumn{9}{@{}l}{\textit{Evidence and delivery lies: the sender forges or reuses proof}}\\
L5 & edited record bytes & 14 & 16 & 12 & 12 & 12 & & 66\\
L6 & forged or stripped signature & 12 & 12 & 12 & 12 & 12 & & 60\\
L7 & records of another run of the same log & 12 & 10 & 10 & 10 & & & 42\\
L8 & records of another log or tenant & 13 & 10 & 10 & 10 & 12 & & 55\\
L9 & proof or checkpoint for another tree & 12 & 12 & 12 & 12 & & & 48\\
L10 & stale proof (older checkpoint, later undone) & 10 & 10 & 10 & 10 & & & 40\\
L11 & replay (nonce, receiver, restart, window, floor) & 10 & 10 & 10 & 10 & 10 & & 50\\
\multicolumn{9}{@{}l}{\textit{Authority lies}}\\
L12 & grant widened on delegation & & & & & 18 & & 18\\
L13 & expired grant & & & & & 10 & & 10\\
L14 & grant or approval from a non-person or other tenant & & & & 10 & 13 & & 23\\
L15 & forged or missing delegation link & & & & & 13 & & 13\\
\multicolumn{9}{@{}l}{\textit{Encoding lies}}\\
L16 & look-alike hosts, IP forms, \%-escapes, numbers & 19{+}4 & 30{+}16 & 18{+}7 & 20{+}4 & 9{+}4 & & 131\\
\midrule
honest & genuine claims and grants & 12 & 20 & 12 & 22 & 13 & 4 & 83\\
AB-strip & ablation: honest with evidence removed & 12 & 20 & 12 & 12 & & 4 & 60\\
AB-t1 & ablation: \T1 twin of an honest claim & 12 & 12 & 12 & & & & 36\\
\midrule
\multicolumn{8}{@{}l}{\textbf{Lies} (L1--L16, excluding the 35 true L16 messages)} & \textbf{712}\\
\bottomrule
\end{tabular}
\end{table}

\paragraph{Receivers}
Each message is checked by ten receivers. The first two are the pre-specified PCM verifiers, \tk's \code{pcm.verify}
and the standalone \code{pcm\_verify}. Then come the two pre-specified baselines: \textbf{plain A2A}, which checks
nothing, and \textbf{sender signature only}, which checks the PCM signer's message signature and nothing else. The last
six are post hoc models of what a deployed protocol lets a receiver check, each built from that protocol's specification
rather than by running its SDK:
\begin{itemize}
\item \textbf{A2A~1.0 + signed Agent Card} checks the card's JWS; messages themselves are unsigned.
\item \textbf{uAgents} checks the envelope signed by the sender's key and its target.
\item \textbf{RFC~9421 + replay} (as in ANP, Visa TAP and Web Bot Auth) checks the sender's HTTP signature over method,
  target and body digest, with a 300\,s window and a nonce cache.
\item \textbf{SCITT receipts} (RFC~9943/9942) check that each claim's records are in a pinned log.
\item \textbf{AP2~0.2 mandates} (Trusted Agent Provider model) check the open mandate's signer, constraints, expiry and
  key binding, the checkout mandate, and the KB-JWT audience, nonce and issued-at time.
\item \textbf{Mastercard Verifiable Intent~0.1} applies AP2's checks with a person key certified by an issuer.
\end{itemize}
The crucial modelling decision is \emph{who holds the key}. In every one of these protocols the sender signs with its
own key or its platform's, so a lying sender signs its own lie. Only AP2 and VI mandates and SCITT receipts carry a
signature that the sender cannot make, and there the sender carries what really exists (a mandate that no real grant or
approval stands behind is one the sender signs itself). Signatures are real Ed25519 over each model's own byte layout.
No check depends on literal JWS, SD-JWT or COSE syntax.

\paragraph{Metrics}
A \emph{false accept} is a lie whose verdict is \vacc. We report the false-accept rate with Wilson 95\% intervals, full
confusion matrices, per-class and per-scope breakdowns, and, for PCM, how many lies are flagged \vunp{} rather than
rejected. Every message is checked with the receiver's identity and a fresh replay state, except the L11 replays, which
carry their own receiver state.

\subsection{RQ1: Safety}

\begin{table}[t]
\centering\small
\caption{Pre-specified hypotheses, final dataset (926 messages; \tk~1.5.0). Both PCM verifiers give identical numbers.}
\label{tab:hyp}
\renewcommand{\arraystretch}{1.12}
\begin{tabularx}{\linewidth}{@{}L{4.3cm} Y L{1.4cm}@{}}
\toprule
\textbf{Hypothesis (pass bar)} & \textbf{Result} & \textbf{Outcome}\\
\midrule
H1: no lie accepted (0) & 10 of 712 accepted (1.4\%, 95\% CI 0.76--2.57\%), all stale evidence (L10, freshness scope);
  verdicts on lies: 626 \vrej, 76 \vunp, 10 \vacc & \textcolor{pcmred}{\textbf{fail}}\\
H2: honest claims proven ($\ge$99\%) & 76 of 78 claims (97.4\%); 81 of 83 messages; both misses are HTTP 404 fetches & \textcolor{pcmred}{\textbf{fail}}\\
H3: \T1-only claims unproven (100\%) & 72 of 72 & \textcolor{pcmaqua!80!black}{\textbf{pass}}\\
H4: verifiers agree (100\%) & 926 of 926 (verdict, reason class, conformance fields) & \textcolor{pcmaqua!80!black}{\textbf{pass}}\\
\midrule
Baseline: plain A2A & 712 of 712 lies accepted & ---\\
Baseline: sender signature only & 643 of 712 (90.3\%, CI 87.9--92.3\%) & ---\\
\bottomrule
\end{tabularx}
\end{table}

\begin{figure}[t]
\centering
\includegraphics[width=0.93\linewidth]{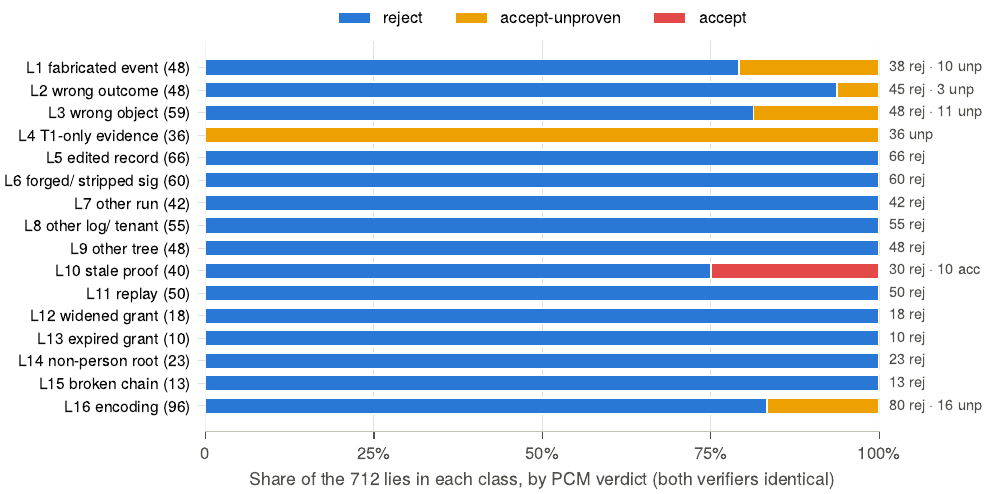}
\caption{\textbf{PCM's verdict on each lie class.} Only L10 (stale proofs) produces \vacc{}, for the 10 messages whose
cited evidence is real but later undone. L4 is entirely \vunp{} by design (\T1-only). In L1--L3 and L16, \vunp{} marks
claims the verifier cannot decide (no evidence, a host that will not parse), which a receiver treats as the sender's
word.}
\label{fig:verdicts}
\end{figure}

\Cref{tab:hyp} reports the pre-specified hypotheses, and \cref{fig:verdicts} shows PCM's verdict on every lie class. H1
fails, and it fails in exactly one way. All 10 false accepts are \code{L10.T.06}--\code{10} and
\code{L10.F.06}--\code{10}. In the first five, a suite passed, the code changed and the suite failed, and the sender cites
the pass under the checkpoint from before the failure. In the other five, a file was written twice and the sender cites
the first write. In both cases the claim is genuinely \prov: the call happened, through an attesting executor, and the
claim type asserts that a call happened, not that nothing happened since. This is the existential semantics of
PCM~0.1 (non-goal N2), stated in the specification \emph{before} the benchmark (S14.13). The pre-specified taxonomy
nonetheless labels these messages as lies, and we keep that label. \Cref{sec:limits} discusses the receiver-side
mitigations and the design for fresh evidence.

\emph{Every other lie, 702 of 712, is not accepted}: 626 are rejected and 76 are flagged \vunp. The flagged ones are
the 36 \T1-only lies of L4 (correct by H3), and 40 lies whose claim the verifier cannot decide (for instance, a
fabricated claim with no evidence, or a host that will not parse). A receiver that follows the specification treats
these as testimony and does not act on them as fact. By scope, PCM accepts 0 of 192 claim lies, 0 of 265 evidence
lies, 0 of 90 authority lies, 0 of 50 delivery lies, 0 of 96 encoding lies and 0 of 9 message lies
(\cref{tab:scope}). All 11 signer-bug grants are rejected by the verifier's own chain checks.

\subsection{RQ2 and RQ3: Honest progress and no self-attestation}

Of the 83 honest messages, 69 get \vacc, 12 get \vunp{} and 2 get \vrej. The 12 \vunp{} messages are by design: they
cite a genuine, proven \code{approved\_by} \emph{without} \code{covers} under a schema that requires a covering
approval, and the verdict tells the sender ``claim it with covers true''. Counting claims as pre-specified, 76 of 78
honest claims are \prov{} (97.4\%), below the 99\% bar. The two misses (\code{honest.H.15}, \code{.16}) are fetches that
the host answered with HTTP 404. The host was reached, but the executor records a status of 400 or more as an error, and
\code{reached\_host} needs an \code{ok} \T2 result, so the claim is \fail{} and an honest message is rejected. This is
again a stated limit of the 0.1 evidence profile (S6.6), with a fix designed for 0.2 (\cref{sec:limits}).

H3 passes: all 72 claims whose only evidence is \T1 (36 in L4 and 36 in the AB-t1 ablation) are \unprov. The second
ablation behaves as designed: stripping the evidence from 60 honest messages yields \vunp{} for all 60, never \vrej, so
removing proof downgrades a message but cannot make an honest sender look like a liar.

PCM is deliberately conservative on the 35 \emph{true} encoding messages of L16: 5 \vacc, 12 \vunp, 18 \vrej. Eleven of
the 12 \vunp{} messages are IPv4 hosts spelled in non-canonical forms (hex, short, 32-bit, leading zeros), which now
come with an instruction to resend as a dotted quad; this is the price of the F1 fix. The twelfth is the wire form of an
honest \code{approved\_by} sent without \code{covers}. The rejects are file claims that are true
on a case-insensitive APFS volume, under an NFD name or behind a symlinked directory (PCM compares paths as written);
integers sent as \code{12000.0} or \code{1.2e4}, which fail strict schema matching; and a UTF-8 byte-order mark, which
both strict loaders refuse.

\subsection{RQ4: Determinism}

H4 passes. On all 926 messages, \tk's verifier and the standalone verifier agree on the verdict, the reason class and
every conformance field, including messages sent as raw text through each verifier's own strict loader. The benchmark
had already earned its keep here: the run before the fixes found five disagreements, all caused by \code{1e400} (F5).
Median verification time is 1.22\,ms for \tk{} and 0.80\,ms for the standalone verifier.

\subsection{RQ5: What existing protocols catch}

\begin{figure}[t]
\centering
\includegraphics[width=\linewidth]{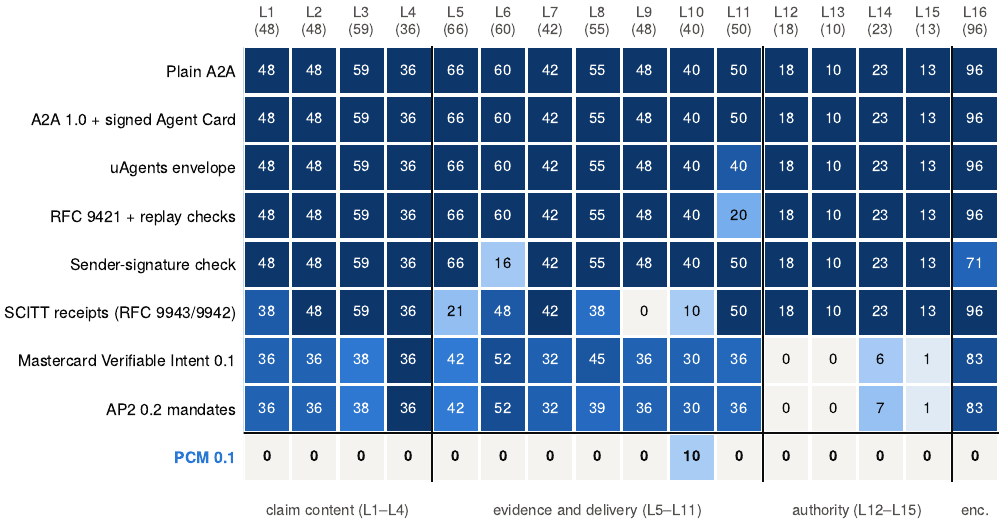}
\caption{\textbf{Lies accepted per class (cell value, out of $n$) for each receiver.} Darker is worse. Authenticating
the sender (Agent Card, uAgents, RFC~9421, sender signature) catches only replays, forged signatures and unparseable wire input. SCITT receipts
catch evidence that is not in the log (L9, part of L5/L8/L10) but cannot tie a claim to its records. AP2 and VI catch
authority lies (L12--L15) but carry no evidence about tests, hosts or files. PCM's only non-zero cell is L10.}
\label{fig:heatmap}
\end{figure}

\begin{table}[t]
\centering\small
\caption{Lies accepted by scope and receiver. Scopes partition the 712 lies. ``Freshness'' is the 10 stale-evidence
lies; other L10 lies (an old checkpoint with later records) fall under ``evidence''.}
\label{tab:scope}
\setlength{\tabcolsep}{4.2pt}
\begin{tabular}{@{}l r rrrr r rr r r@{}}
\toprule
& & \multicolumn{4}{c}{\textit{sender authentication}} & & \multicolumn{2}{c}{\textit{mandates}} & & \\
\cmidrule(lr){3-6}\cmidrule(lr){8-9}
\textbf{Scope} & $n$ & A2A & Card & uAg. & 9421 & SCITT & VI & AP2 & SigOnly & \textbf{PCM}\\
\midrule
claim & 192 & 192 & 192 & 192 & 192 & 182 & 146 & 146 & 192 & \textbf{0}\\
evidence & 265 & 265 & 265 & 265 & 265 & 113 & 208 & 208 & 229 & \textbf{0}\\
authority & 90 & 90 & 90 & 90 & 90 & 90 & 26 & 21 & 82 & \textbf{0}\\
delivery & 50 & 50 & 50 & 40 & 20 & 50 & 36 & 36 & 50 & \textbf{0}\\
encoding & 96 & 96 & 96 & 96 & 96 & 96 & 83 & 83 & 71 & \textbf{0}\\
message & 9 & 9 & 9 & 9 & 9 & 9 & 0 & 0 & 9 & \textbf{0}\\
freshness & 10 & 10 & 10 & 10 & 10 & 10 & 10 & 10 & 10 & \textbf{10}\\
\midrule
\textbf{total} & \textbf{712} & 712 & 712 & 702 & 682 & 550 & 509 & 504 & 643 & \textbf{10}\\
\textit{honest accepted} & \textit{83} & 83 & 83 & 83 & 83 & 83 & 67 & 67 & 83 & 69$^\dagger$\\
\bottomrule
\end{tabular}

\vspace{2pt}
{\footnotesize $^\dagger$ PCM: 69 \vacc{} plus 12 \vunp{} by design; the other receivers have no \vunp{} outcome.}
\end{table}

\Cref{fig:teaser,fig:heatmap} and \cref{tab:scope} compare the receivers. Three patterns stand out.

\emph{Authenticating the sender does not authenticate the content.} The signed Agent Card catches nothing, because it
authenticates a sender who is the liar and leaves messages unsigned. The uAgents envelope catches the 10 replays
addressed to another receiver. It has no nonce cache, and it reads \code{expires} as a reply timeout, so the other 40
replays pass. RFC~9421 with replay checks catches 30 of 50 replays (same nonce, another receiver, outside the window).
It misses the 10 replays after a receiver restart, because its cache lives in memory, and the 10 below the floor,
because it keeps no per-sender floor. It also misses every content lie, which the sender simply re-signs.

\emph{Transparency without semantics catches forged evidence, not false claims.} SCITT receipts catch 162 lies, all at
the evidence layer: proofs for another tree (all 48 of L9), edited records (45 of 66 in L5), forged checkpoint
signatures, unpinned logs, and claims without records. They miss true records cited for false claims (L1--L3), because
SCITT has no rule that ties a claim to a record. They also miss the agent's own reports (L4), because SCITT has no
tiers; records of another run or tenant; every replay; and every grant lie.

\emph{Mandates cover authority, not actions.} AP2 catches 208 lies and VI 203, almost all of them grant and approval
lies: all of L12 and L13, 12 of the 13 in L15, and 132 of the 153 approval lies. None of the 450 lies about tests,
hosts and files carries a mandate, so all of them pass. AP2 and VI also reject 16 honest messages. Eight are delegated
grants, because AP2~0.2 has no agent-to-agent delegation. The other eight are approvals whose message does not disclose
the approved values that the model needs, a limitation of our model. Conversely, 63 approval lies (64 for VI) are
rejected only because the dataset lacks those values. Counting them as accepted gives the most AP2 and VI could accept:
567 (79.6\%) and 573 (80.5\%). Our best estimate of a faithful count is 553 and 559. \Cref{fig:teaser} shows the
measured 504 and 509, the numbers that favour AP2 and VI.

\emph{Stacking does not close the gap.} A receiver that runs the AP2, SCITT and RFC~9421 checks and accepts only what
all three accept still accepts 365 lies (51.3\%, CI 47.6--54.9\%) and rejects 16 honest messages. It accepts all of L4,
most of L1--L3 and L16, and 18 of the 20 replays after a restart or below the floor. The missing ingredient is not
another signature but a \emph{rule that ties each claim to records written by someone other than the sender}.

\subsection{Findings: what the benchmark changed}
\label{sec:eval-findings}

\begin{table}[t]
\centering\small
\caption{Findings of the PCM-Lies benchmark, before (dataset 1, 879 messages) and after (dataset 2, 926 messages, with
the fixes). Three were fixed in the specification and both verifiers, with vectors; two are documented 0.1 limits.}
\label{tab:findings}
\renewcommand{\arraystretch}{1.12}
\begin{tabularx}{\linewidth}{@{}l L{3.6cm} Y Y l@{}}
\toprule
& \textbf{Finding} & \textbf{Before} & \textbf{After} & \textbf{Status}\\
\midrule
F1 & leading-zero IPv4 (\code{0127.0.0.1}) & \vacc: \code{87.0.0.1} proven, though macOS connected to \code{127.0.0.1} (7 lies) &
  \vunp: ``not written as a dotted quad'' & fixed\\
F2 & stale evidence & \vacc{} (10) & \vacc{} (10) & open (limit)\\
F3 & approval of a smaller or other call & \vacc{} (6 lies) & \vunp{} or \vrej{} via \code{covers} & fixed\\
F4 & honest fetch answered 404 & \vrej{} (2 honest) & \vrej{} (2 honest) & open (limit)\\
F5 & \code{1e400} read two ways & verifiers disagree (5) & both refuse as \vbad & fixed\\
\midrule
\multicolumn{2}{@{}l}{H1 lies accepted} & 23/687 (3.3\%, CI 2.2--5.0\%) & 10/712 (1.4\%, CI 0.8--2.6\%) & \\
\multicolumn{2}{@{}l}{\quad of which not freshness} & 13 & \textbf{0} & \\
\multicolumn{2}{@{}l}{H4 verifier agreement} & 874/879 & 926/926 & \\
\bottomrule
\end{tabularx}
\end{table}

A benchmark is only useful if it can find bugs, and PCM-Lies found three (\cref{tab:findings}). \textbf{F1} is a
platform-dependent attack. The executor fetched \code{http://0127.0.0.1/} and macOS connected to \code{127.0.0.1},
but the specification read the host as \code{inet\_aton} does, so a claim of \code{87.0.0.1}, an address never
contacted, was \prov. \textbf{F3} showed that leaving the comparison of \code{approved} with the intent to receivers
fails in practice: a receiver acting on the verdict alone accepted a \$5 approval for a \$50 request. It motivated
\code{covers}. \textbf{F5} was a determinism bug. Each was fixed in the specification and both verifiers and pinned by
vectors. The dataset grew by 47 messages to exercise the fixes, and the benchmark was re-run. Across the 875 messages
the two datasets share, the only verdict changes are \vacc{} $\to$ \vunp{} (26 messages), and the only reason-class
change is F5's.

\subsection{RQ6: Cost}

\begin{figure}[t]
\centering
\includegraphics[width=\linewidth]{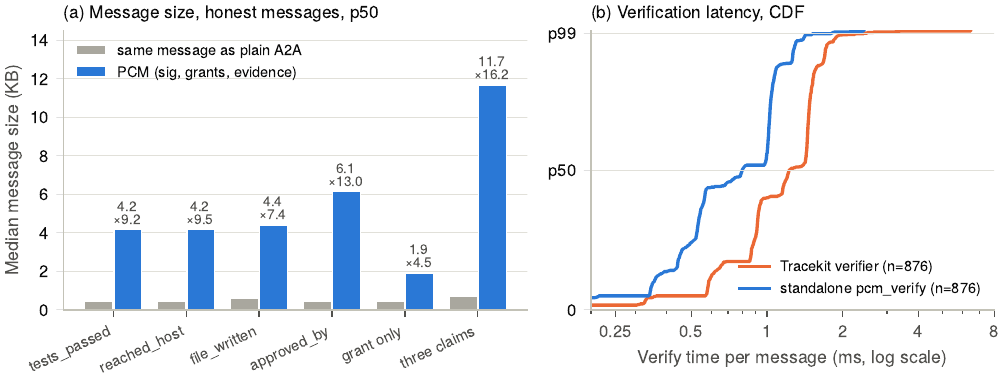}
\caption{\textbf{Cost of PCM.} (a)~Median size of honest messages by claim type, against the same message without
\code{sig}, \code{grants} and evidence (``plain A2A''); labels give the PCM size in KB and the median per-message
ratio. (b)~Distribution of verification time over the 876 messages that parse (the other 50 are raw texts that the
loaders refuse before any check).}
\label{fig:cost}
\end{figure}

\begin{table}[t]
\centering\small
\caption{Message size (honest messages) and verification latency (all parsed messages of each type, lies included), in
ms, single-threaded CPython 3.11 on Apple silicon. Sizes are the UTF-8 bytes of compact, key-sorted JSON (1\,KB $=$
1000\,B).}
\label{tab:latency}
\begin{tabular}{@{}l rr rr rr@{}}
\toprule
& \multicolumn{2}{c}{\textbf{size (KB)}} & \multicolumn{2}{c}{\textbf{\tk{} verifier}} & \multicolumn{2}{c}{\textbf{standalone}}\\
\cmidrule(lr){2-3}\cmidrule(lr){4-5}\cmidrule(l){6-7}
\textbf{Claims} & p50 & vs.\ plain & p50 & p99 & p50 & p99\\
\midrule
\code{tests\_passed} & 4.2 & 9.2$\times$ & 1.42 & 2.25 & 0.99 & 1.16\\
\code{reached\_host} & 4.2 & 9.5$\times$ & 1.45 & 1.67 & 0.99 & 1.23\\
\code{file\_written} & 4.4 & 7.4$\times$ & 1.42 & 1.64 & 0.99 & 1.10\\
\code{approved\_by} & 6.1 & 13.0$\times$ & 1.67 & 2.51 & 1.25 & 1.82\\
grant only & 1.9 & 4.5$\times$ & 0.87 & 1.44 & 0.45 & 0.69\\
three claims & 11.7 & 16.2$\times$ & 0.61 & 3.38 & 0.38 & 2.43\\
\midrule
\textbf{all (876)} & & & \textbf{1.22} & \textbf{2.42} & \textbf{0.80} & \textbf{1.81}\\
\bottomrule
\end{tabular}
\end{table}

\Cref{fig:cost} and \cref{tab:latency} report cost. A single-claim PCM message is 4.2--6.1\,KB at the median, 7--13
times the size of the same request without proof. Most of the overhead is the checkpoint note and the records, each with
a $\lceil\log_2 n\rceil$-hash inclusion proof. \code{approved\_by} is the largest because its evidence carries three
linked records, and four when the approval is cited through the root grant. A message with three claims is 11.7\,KB.
In absolute terms these sizes are small next to the LLM context that produced the message: 11.7\,KB is roughly 3k
tokens, and the evidence never needs to be shown to a model. It is consumed by the verifier. Verification is
dominated by Ed25519 checks and hashing and takes about a millisecond in pure Python: 0.80\,ms at the median and
1.81\,ms at p99 for the standalone verifier. (Latency is measured over every parsed message of a type, lies and ablations included. The three-claim group mixes
four full honest messages, which take about 2.4\,ms each, with their four evidence-stripped twins, which take 0.37\,ms, hence its low median.) Producing evidence costs the
honest sender about one second of waiting for a checkpoint that covers its records (\cref{sec:impl}).

\subsection{Threats to validity}
\label{sec:threats}

\paragraph{Departures from the pre-specified design}
The design called for an improve loop that would repeat ``until H1--H4 pass''. We stopped with H1 and H2 failing
because their causes, F2 and F4, are properties of the 0.1 claim semantics and evidence profile. Fixing them requires
format changes, which we specify for a later draft rather than fold into this evaluation. The dataset grew after the
first run (47 messages for the fixes). Four honest fetches with numeric host spellings were moved to L16, so that H2
measures a sender that follows the specification. L11 gained two replay kinds (outside the window and below the floor).
The sender-signature baseline checks the PCM signer's key, which the sender cannot use, so it catches forged and
stripped signatures and accepts 643 of the lies rather than all of them. H2 was first counted per message and is now
counted per claim, as specified. One message (\code{L15.G.13}) was relabelled before the final run of dataset~1. The
design note has no public timestamp prior to the results.

\paragraph{Internal validity}
Labels come from the generator's record of what happened, never from a verifier. Dataset~1 was generated three times
and gave the same counts in every cell, apart from the relabel. Dataset~2 was generated once. Ids, nonces, timestamps
and signatures change between generations, while the seed (20261028) fixes keys and choices.

\paragraph{External validity}
All runs used one machine (macOS 26.3, arm64, Python 3.11.15). F1 depended on the platform: glibc reads leading-zero
octets as octal, so a Linux-generated dataset labels some L16 messages differently. The signer and executor ran
in-process, the executor was \tk's custody code with an in-process client rather than the HTTP executor service, and
host names were served by the local server without real DNS. The six protocol receivers are models built from
specifications, not runs of their SDKs. They omit transport security (which authenticates the liar too) and VI's
suggested mandate lifetimes. By our reading of their specifications, MCP, IBM ACP, LangChain Agent Protocol, Agora,
Coral and FIPA-ACL reduce to plain A2A on these claims, and Visa TAP and the OpenAI/Stripe Agentic Commerce Protocol to
RFC~9421 with replay checks at best; we did not measure them. Finally, the lie taxonomy is ours. Lies outside it (for
example, an agent that truthfully reports a test suite chosen to be vacuous) are discussed as limits rather than
measured.

\section{Discussion and Limitations}
\label{sec:limits}

\subsection{Evidence is not freshness}

Every PCM~0.1 claim type is existential: ``a call ran and succeeded''. The sender chooses which records to disclose, so
evidence can show that something happened, but never that nothing happened since. This explains all 10 false accepts in
\cref{sec:eval}. Within 0.1, a receiver that acts on current state applies three checks. First, an \textbf{age bound}:
it refuses evidence whose \emph{oldest} record is older than a per-intent bound (seconds or minutes) relative to the
message \code{ts}. It uses the oldest record because a sender can always add newer ones, and it compares only times set
by the log, so the receiver's clock plays no part. Second, an \textbf{after-last-change bound}: when the receiver knows
when the thing it acts on last changed (the commit to merge was pushed, or it sent the request), it refuses older
evidence. Third, for files, a \textbf{digest check} over the bytes it holds, which is the one check that rules out a
later change outright. Both verifiers print the evidence age to support the first check.

For a later draft we have designed two mechanisms (\cref{app:fresh}). \emph{The run since the call} adds a
signer-set \code{head} to the envelope, naming the last record of the sender's run at signing time, and an evidence
form that carries every record of the run from the claimed call up to \code{head}. These records are chained by the
per-run \code{run\_prev\_hash} that the log already maintains, so no record in between can be omitted, and a later
failure of the same suite or a later write of the same path becomes visible. \emph{Bound claims} have the executor
record what a call ran on: the git tree id of a fresh checkout for \code{tests\_passed}, and the path and digest it read
back for \code{file\_written}. A receiver then compares the tree with the commit it will merge. Together they would
close the L10 class at a cost of roughly 1\,KB per later record.

\subsection{A response is not a failure}

An HTTP 404 proves that the host was reached, but the 0.1 evidence profile records only whether a result was
\code{ok}, and the executor records a status of 400 or more as not \code{ok}. Honest claims of this kind are therefore
\fail, which is the cause of both H2 misses. An egress observer's record proves the connection whatever the answer.
PCM~0.2 can record the protocol status, so that \code{reached\_host} rests on any response.

\subsection{Trust in the signer and in executors}

PCM moves trust from the agent to the signer, but does not remove it. A compromised signer, or a caller who can make it
sign as another identity, can mint anything (\cref{sec:keycompromise}). A second, subtler dependency is on executors.
A verifier places a \T2 call by its tool name, so an executor that runs operator-named HTTP actions could record one
under the name \code{Bash}, and an HTTP request that succeeded would then prove \code{tests\_passed}. The specification
forbids logs that offer the evidence profile from recording a tool under a name of another class. A future trust member
would let receivers pin \emph{which} attesting parties may attest each tool class.

\subsection{What a proof does not say}

A \prov{} \code{tests\_passed} proves that a specific command line exited 0 through an executor. It does not prove that
the tests that matter ran: runner flags and configuration files in the working directory can make a suite vacuous.
Receivers therefore compare \code{suite} with the suites they accept. A proven drive-path \code{file\_written} assumes a
Windows executor. A proven \code{approved\_by} without \code{covers} says what alice approved, and that may not be what
is now being requested; that is why requiring schemas accept only covering approvals. PCM~0.1 also cannot prove
absence (``no other call happened''). These limits are stated in the specification so that receivers do not read more
into \prov{} than the record supports.

\subsection{Privacy}
\label{sec:privacy}

A receiver learns the sender's identity, run and tenant, the intent and its args, the payload, and every grant in the
chain, including the approving person's identity and name. From evidence it learns the disclosed records in full
(types, times, tool names, call ids, approvers, policy outcomes, observed hosts), the log's size at the checkpoint, and
the args of opened decisions. It does not learn records that were not sent, or the args of unopened decisions.
Commitments hide those args, but an opening is permanent and, because the message is signed, attributable. A salt also
exposes low-entropy args to brute force, which is why the signer hands salts only to the run's own agent. Senders
should disclose the minimum each claim needs.

\subsection{Implications for LLM agent systems}

PCM is complementary to defences that harden a single agent against prompt injection, such as instruction hierarchies,
dual-LLM and capability-based designs~\cite{debenedetti2025camel,willison2023dualllm}, and benchmarks such as
AgentDojo~\cite{debenedetti2024agentdojo}. Those defences reduce the chance that an agent is subverted. PCM bounds what
a subverted agent can make \emph{others} believe and do. Three integration patterns follow.
\begin{itemize}
\item \textbf{Verifier as a gate.} The receiving framework verifies each PCM part \emph{before} the model sees it and
  drops \vrej{} parts. The model then sees each claim annotated with its result (proven, or ``the sender says''), so
  the planner reasons over structured provenance rather than free text.
\item \textbf{Authority outside the model.} Grants and their caveats are checked by code, so a prompt-injected sender
  cannot talk a receiver into exceeding them: the grant, not the model, sets the limit.
\item \textbf{Auditable pipelines.} When the signer records every message signed and every grant issued, as the specification recommends and \tk{} does, a multi-agent
  run leaves a verifiable, append-only trail of who asserted what, on which evidence, under whose authority.
\end{itemize}
Agents increasingly exchange model state as well as text, for example by fusing one model's KV cache into another's
(cache-to-cache transfer)~\cite{c2c}. A draft extension in the specification records such transfers (sender and
receiver model ids, modality, projector, size and digest) as a signer-decided call, so that a receiver can at least
check that the bytes it received are the bytes the sender declared to its signer before the transfer. Nothing in that
record proves what the state means.

\section{Related Work}
\label{sec:related}

\begin{table}[t]
\centering\footnotesize
\caption{What a receiver can check under each approach. \cmark: yes; \pmark: partly or in a limited setting; \xmark:
no, by our reading of each specification. ``Offline'' means without contacting the sender, the signer or the service.}
\label{tab:related}
\setlength{\tabcolsep}{2.4pt}
\renewcommand{\arraystretch}{1.12}
\begin{tabular}{@{}l ccccccc@{}}
\toprule
& \makecell{Sender\\cannot sign} & \makecell{Claims tied\\to records} & \makecell{Self-report\\$\neq$ proof} &
  \makecell{Person-\\rooted auth.} & \makecell{Attenuating\\delegation} & \makecell{Replay\\across restart} & Offline\\
\midrule
A2A / MCP (+TLS)~\cite{a2a,mcp} & \xmark & \xmark & \xmark & \xmark & \xmark & \xmark & ---\\
Signed Agent Card / uAgents & \xmark & \xmark & \xmark & \xmark & \xmark & \xmark & \cmark\\
RFC 9421 + replay (ANP, TAP) & \xmark & \xmark & \xmark & \xmark & \xmark & \pmark & \cmark\\
SCITT receipts~\cite{rfc9943,rfc9942} & \cmark & \xmark & \xmark & \xmark & \xmark & \xmark & \cmark\\
AP2 / Verifiable Intent~\cite{ap2,mastercardvi} & \pmark & \xmark & \xmark & \cmark & \xmark & \pmark & \cmark\\
Macaroons / Biscuit / UCAN & \xmark & \xmark & \xmark & \xmark & \cmark & \xmark & \cmark\\
OAuth agent delegation~\cite{south2025authenticated} & \pmark & \xmark & \xmark & \cmark & \cmark & \xmark & \pmark\\
Notarized Agents~\cite{figuera2026notarized} & \cmark & \pmark & \cmark & \xmark & \xmark & \xmark & \pmark\\
NabaOS tool receipts~\cite{basu2026toolreceipts} & \pmark & \cmark & \pmark & \xmark & \xmark & \xmark & \xmark\\
\midrule
\textbf{PCM 0.1} & \cmark & \cmark & \cmark & \cmark & \cmark & \cmark & \cmark\\
\bottomrule
\end{tabular}
\end{table}

\paragraph{Proof-carrying code and certified results}
PCM borrows its name and its shape from proof-carrying code~\cite{necula1996safe,necula1997pcc}: the producer ships
evidence, and the consumer runs a small checker instead of trusting the producer. Unlike PCC, PCM's ``proof'' is a
signed record with an inclusion proof rather than a derivation in a logic, so it holds only as far as its signer is
trusted. The two approaches compose. A future claim type could carry a certificate checked by a proof checker,
anchored by a PCM record of the call that produced it.

\paragraph{Transparency logs}
PCM's evidence format is that of Certificate Transparency~\cite{rfc6962,rfc9162}, with C2SP notes, checkpoints and
witness cosignatures~\cite{c2sp-note,c2sp-checkpoint,c2sp-cosig}. CT logs certificates for public audit, whereas PCM
hands a receiver the handful of records of one run that back one claim, from a log that may be private to a
deployment. SCITT~\cite{rfc9943,rfc9942} checks that a signed statement is in a transparency log. Individual drafts
profile it for agent actions, but SCITT itself has no rule that ties a claim to a record, no T1/T2 distinction, no
replay rule and no grants. Our benchmark shows the consequence: true records cited for false claims pass
(\cref{sec:eval}). PCM adds exactly these semantics over the same kind of evidence.

\paragraph{Capability-based authorisation}
Macaroons~\cite{birgisson2014macaroons}, Biscuit~\cite{biscuit} and UCAN~\cite{ucan} support offline attenuation by the
holder. PCM grants also only narrow, but no agent holds a key: the signer signs each step after checking it. A root
exists only after a person's recorded approval of its exact terms, and caveats are typed by the intent's schema and
compared with a fixed table (\cref{tab:implication}) rather than a general logic such as Biscuit's Datalog. The table is
deliberately incomplete (bounds never imply equality), which trades expressiveness for verdicts that are identical
across implementations. South et al.~\cite{south2025authenticated} extend OAuth~2.0 and OpenID Connect with agent
credentials and scoped delegation, which answers on whose behalf an agent acts. PCM is complementary: it adds claims
about what the agent did, and grants that a receiver checks offline.

\paragraph{Agent payment mandates}
AP2~\cite{ap2}, now stewarded by the FIDO Alliance, checks a chain of SD-JWT mandates for a purchase: who authorised it,
within which constraints, until when, bound to the agent's key. Mastercard's Verifiable Intent~\cite{mastercardvi} adds
an issuer-certified user key. Both answer ``was this purchase authorised?'', not ``did the agent do what it says?'', and
AP2~0.2 excludes agent-to-agent delegation. Our models of both catch most authority lies and none of the lies about
tests, hosts or files. Proof-Gated Signing~\cite{ghosh2026pgs} guards a different point of the same pipeline: before a
wallet-controlling agent signs a transaction, an SMT solver checks its value and permission policy over the whole range
of oracle-price uncertainty, and the result is compiled into on-chain post-conditions that the wallet enforces
atomically. It constrains what an agent may \emph{do}; PCM constrains what other agents will \emph{believe} about what
it did.

\paragraph{Agent receipts and provenance}
\tk~\cite{ghosh2026tracekit}, PCM's reference implementation, records each coding-agent session's requests, stated
reasoning and executed actions in a hash-chained, externally anchorable ledger for audit. PCM builds on its attesting
signer and Merkle log, and turns the audit trail into evidence that a \emph{receiving agent} checks per claim.
Notarized Agents~\cite{figuera2026notarized} has the \emph{called service} sign a receipt of each action, encrypt it to
the agent's owner and publish it to a witnessed log, so the owner can audit the agent. PCM needs no cooperation from the
service. The sender's attesting signer and executor record the call, and any receiver checks the records offline.
NabaOS~\cite{basu2026toolreceipts} cross-checks a model's claims against HMAC-signed tool-execution receipts to detect
hallucinated tool results. An HMAC can be checked only with the key that made it, whereas PCM's records are signed with
public keys that any receiver can check, and PCM adds authority and delivery rules. In software supply chains,
in-toto~\cite{torresarias2019intoto} signs statements about steps, and SLSA~\cite{slsa} requires that provenance come
from the build platform rather than the build's own steps. PCM applies the SLSA principle per claim, inside a message,
and adds a third outcome for what cannot be checked.

\paragraph{Security of agent protocols}
A2ABreak~\cite{lotfi2026a2abreak} extracts a 37-state machine from the A2A specification and finds 11 vulnerabilities,
among them context injection through unprotected context identifiers and identity loss across multi-hop delegation.
PCM signs its whole message and makes each delegation step a signed grant that names its issuer and subject, although
parts of an A2A message outside the PCM part remain unsigned. Prompt-injection attacks on
agents~\cite{greshake2023indirect,perez2022ignore,zhan2024injecagent} and benchmarks such as
AgentDojo~\cite{debenedetti2024agentdojo} motivate PCM's threat model. Design-level defences such as
CaMeL~\cite{debenedetti2025camel} constrain what a single agent may do with untrusted data, while PCM constrains what
other agents will believe.

\paragraph{Cross-implementation disagreement}
Parser differentials are a classic source of vulnerabilities: in X.509 and TLS stacks~\cite{kaminsky2010pki}, in URL
parsers~\cite{orange2017url}, in Ed25519 verification~\cite{chalkias2020taming} and in JSON
parsers~\cite{seriot2016json}. PCM treats differential-freeness as a design requirement. Wherever readers differ it
specifies a strict reading, maps divergence to \unprov{}, and pins the outcome with vectors produced by a real signer
(\cref{sec:determinism}).

\section{Conclusion}
\label{sec:conclusion}

Multi-agent systems built on LLMs pass each other claims about what happened and about who allowed what, and today a
receiver can do no better than trust the sender. PCM lets each claim carry evidence written by someone other than the
agent, lets authority travel as person-rooted chains that only narrow, and has every receiver check both offline, in
about a millisecond, with the same answer whatever its implementation. Its three-valued semantics keeps doubt apart from
falsity, so a claim the verifier cannot decide costs the sender the proof rather than the message, and a liar cannot turn its
own report into proof. On 712 lies built from real runs, PCM accepts only the 10 that its specification already says
it cannot catch, while models of today's deployed protocols accept between 504 and 712. Closing that last class,
evidence that a state is still current, is the main work for the next draft, together with a dedicated message key,
key rotation, Unicode hosts, and verified transport bindings for A2A and MCP.

\bigskip
\noindent Code: \url{https://github.com/Cygnux-Labs/PCM-protocol}

\bibliographystyle{plainnat}
\bibliography{refs}

\appendix
\section{A Complete Worked Example}
\label{app:example}

All values below are from the conformance vector \vect{accept-approved-by}, produced by the reference implementation's
signer. \code{...} marks elided members. Agent $A$ is \code{uid:990101}, $B$ is \code{uid:990102}, and alice is
\code{uid:990001}.

\paragraph{Schema document}
Both sides pin this document by its hash.
\begin{lstlisting}[language=json]
{"pcm_schema": 0, "name": "refund.request",
 "fields": {"amount": {"type": "money", "currency": "USD", "max": 1000000},
            "order":  {"type": "int", "min": 1, "max": 999999999}},
 "claims": ["approved_by"], "grant": true}
\end{lstlisting}

\paragraph{Message}
\begin{lstlisting}[language=json]
{"from": {"agent": "uid:990101", "run": "ec130a625bc622738e7f2652097d9535", "tenant": "default"},
 "to": {"agent": "uid:990102"},
 "intent": {"type": "refund.request@sha256:d8636b6fbbd07430d057ea9ead9c27f0df1b918b10d34ab9257a4971d487846c",
            "args": {"amount": 12000, "order": 81}},
 "claims": [ ... the approved_by claim below ... ],
 "grants": [ ... alice's root grant below ... ],
 "pcm": 0, "id": "0ac65318a85ca571b5a29c348280f1d9", "ts": "2026-10-28T02:24:35.496285Z",
 "nonce": "705e9bb546e4876321bcbc748f982bd0",
 "sig": {"key": "example.com/pcm-test-log+0bb91e66",
         "sig": "Xtv6U9fKIx3FG1t+OGuK5ifNbOHf6aOLDSUaoHWa/g8QGzazVirPtLR57furSUouYWn0eSYn037sTIkLEeAQBQ=="}}
\end{lstlisting}

\paragraph{Root grant}
The signer signed this grant after alice approved the agent's \code{pcm.grant} call with exactly these terms.
\begin{lstlisting}[language=json]
{"pcm_grant": 0, "ts": "2026-10-28T02:24:35.477412Z",
 "issuer": {"kind": "person", "id": "uid:990001", "run": "ec130a625bc622738e7f2652097d9535", "tenant": "default",
            "approval_id": "apr-afa8cc12fcd9032a7eb9c9479deedd68", "person": "alice"},
 "subject": "uid:990101",
 "intent": "refund.request@sha256:d8636b6fbbd07430d057ea9ead9c27f0df1b918b10d34ab9257a4971d487846c",
 "caveats": [{"field": "amount", "op": "<=", "value": 20000}, {"field": "order", "op": "==", "value": 81}],
 "expires_at": "2026-10-28T03:24:35.475470Z", "delegable": false, "parent": null,
 "sig": {"key": "example.com/pcm-test-log+0bb91e66",
         "sig": "Jm0YbtyvSJL4GaRkk8BnjWyq1LfF6o6WPlPry8vKCE7Erh3F8nmi1LW94nr2Zu8WWOEN4Y9bqmbSn00cx6L8AA=="}}
\end{lstlisting}

\paragraph{Claim and evidence}
Three records of $A$'s run, each with a four-hash inclusion proof under a checkpoint of tree size 9.
\begin{lstlisting}[language=json]
{"type": "approved_by", "args": {"person": "alice", "call": "grant-1", "covers": true},
 "evidence": {
  "checkpoint": "example.com/pcm-test-log\n9\nQhInJS9+Rjyar8mhP1wGwzx1O2DYcMy8af/XYRCqSb8=\n\n— example.com/pcm-test-log C7keZoQnlibbpH9yOCeEzKJs6Hzc97YYaP1hsfTBxyO0QzEKJRpaXpxBKoLJEIi9d0FMYRNNmWOSl0cOBOf2whMhowQ=\n",
  "records": [
   {"event": {"type": "policy.decision", "seq": 3, "...": "...",
              "data": {"tool_use_id": "grant-1", "tool": "pcm.grant", "decision": "ask",
                       "decision_id": "dec-76e1373f66cddad74ba54eb2ffcf97af", "...": "..."}},
    "hash": "sha256:444ea34333ed7e77be0fe1ca8f57b108b93d79258dd27ec3f2ffe0f5d89409f5",
    "proof": ["GHc2iRz78zZ9UrFNQfRdYDhZyd9ere1mVrgDE2zN4w8=", "a1WP4EFPP5lJvKXGna0/aMjExMY6buoajX2ZZWWR87w=",
              "XV3DvLmIxeoAAS29RasT6tRFgGOFhLjjSzUyQPdNyZs=", "oGluwXsyMyIJMoCTujrGPPsyxEcnsqfn+A06KUVWSuc="]},
   {"event": {"type": "approval.request", "seq": 4, "...": "...",
              "data": {"approval_id": "apr-afa8cc12fcd9032a7eb9c9479deedd68",
                       "decision_id": "dec-76e1373f66cddad74ba54eb2ffcf97af", "...": "..."}},
    "hash": "sha256:39c29ce4973ac0753b77f3ae1a7056a8d06d054fe6b7ea951c423f5b5491288a", "proof": ["..."]},
   {"event": {"schema_version": "tracekit.event.v2", "seq": 5, "type": "approval", "source": "sdk",
              "run_id": "ec130a625bc622738e7f2652097d9535", "tenant": "default", "...": "...",
              "data": {"tool_use_id": "grant-1", "approval_id": "apr-afa8cc12fcd9032a7eb9c9479deedd68",
                       "decision": "approve", "approver": "uid:990001", "self_approved": false,
                       "person": "alice", "...": "..."}},
    "hash": "sha256:3483c8f824e7072e479f93de2a5d9ce2f4e32ed6fc5472e8621a353da9371e1b", "proof": ["..."]}]}}
\end{lstlisting}

\paragraph{Verification result}
For receiver \code{uid:990102}, with a fresh replay state and a 300\,s window:
\begin{lstlisting}[language=json]
{"signature": {"ok": true, "key": "example.com/pcm-test-log+0bb91e66"},
 "replay": {"ok": true, "detail": "fresh"},
 "schema": {"ok": true, "ref": "refund.request@sha256:d8636b6f...", "errors": []},
 "grants": {"result": "ok", "steps": 1, "root": {"kind": "person", "id": "uid:990001", "person": "alice"},
            "problems": []},
 "claims": [{"type": "approved_by", "result": "proven",
             "detail": "approval seq 5 by alice of \"pcm.grant\" call grant-1, which issued this message's root grant",
             "approved": {"tool": "pcm.grant", "args": null}}],
 "verdict": "accept", "reasons": [], "skipped": []}
\end{lstlisting}

\paragraph{Neighbouring vectors}
\Cref{tab:neighbours} shows how small changes to this message move the verdict. Each row is a separate vector.

\begin{table}[h]
\centering\small
\caption{Variations of the worked example and their verdicts (reasons abbreviated).}
\label{tab:neighbours}
\renewcommand{\arraystretch}{1.12}
\begin{tabularx}{\linewidth}{@{}>{\raggedright\arraybackslash\scriptsize}p{6.2cm} Y L{1.9cm}@{}}
\toprule
\textbf{Vector} & \textbf{What differs; reason} & \textbf{Verdict}\\
\midrule
\vect{accept-unproven-t1-result} & adds \code{tests\_passed} backed only by the agent's own report; ``the result is the agent's own report (T1)'' & \vunp\\
\vect{accept-tests-passed-t2} & the executor recorded the result; ``\ldots run by mtls:spiffe://example/executor (T2)'' & \vacc\\
\vect{reject-claim-pytest-or-true} & the executor ran \code{pytest -q \ldots{} || true}; ``not a single test command'' & \vrej\\
\vect{reject-claim-changed-record-hash} & one hex digit of a record hash changed; ``record 1: hash does not match the event'' & \vrej\\
\vect{reject-claim-approved-by-self-approval} & erin approves a call of her own agent's run & \vrej\\
\vect{reject-grant-amount-over-caveat} & amount 50000; ``grant 0: amount 50000 breaks amount <= 20000'' & \vrej\\
\vect{reject-claim-approved-by-covers-other-args} & cites, with \code{covers}, alice's opened approval of a pay of 500 for a pay of 12000 & \vrej\\
\vect{accept-unproven-approved-by-required-without-covers} & the same, without \code{covers}; ``required, not proven to cover this request'' & \vunp\\
\vect{accept-unproven-reached-host-idn} & \code{reached\_host} from a fetch of \code{http://bücher.example/}; ``its URL is not printable ASCII: \ldots'' & \vunp\\
\vect{accept-unproven-reached-host-relative-path} & fetch of \code{/api.example.com/v1/refunds}; ``its URL has no scheme \ldots'' & \vunp\\
\bottomrule
\end{tabularx}
\end{table}

\section{Protocol Data}
\label{app:runners}

\paragraph{Test runners (42 sequences)}
A \code{tests\_passed} suite must begin with one of these word sequences, compared exactly.
\begin{center}\small
\begin{tabular}{@{}llll@{}}
\code{pytest} & \code{py.test} & \code{python -m pytest} & \code{python3 -m pytest}\\
\code{python -m unittest} & \code{python3 -m unittest} & \code{tox} & \code{nox}\\
\code{nose2} & \code{trial} & \code{behave} & \code{jest}\\
\code{vitest} & \code{mocha} & \code{ava} & \code{karma}\\
\code{cucumber} & \code{playwright test} & \code{phpunit} & \code{rspec}\\
\code{ctest} & \code{npm test} & \code{npm run test} & \code{yarn test}\\
\code{yarn run test} & \code{pnpm test} & \code{pnpm run test} & \code{bun test}\\
\code{go test} & \code{cargo test} & \code{cargo nextest run} & \code{dotnet test}\\
\code{deno test} & \code{swift test} & \code{mix test} & \code{bazel test}\\
\code{make test} & \code{make check} & \code{mvn test} & \code{mvn verify}\\
\code{gradle test} & \code{gradle check} & \code{} & \code{}\\
\end{tabular}
\end{center}

\paragraph{Tool placement table}
The table has 54 names; its SHA-256 is
\code{15f8410f513344e4c60327d4fc3dd149}\allowbreak\code{49ebeaf7722f641d985d3ca721c8fe07}.
\begin{center}\small
\renewcommand{\arraystretch}{1.15}
\begin{tabularx}{\linewidth}{@{}l Y@{}}
\toprule
\textbf{Class} & \textbf{Tool names}\\
\midrule
\code{shell} (8) & \code{Bash}, \code{bash}, \code{container.exec}, \code{local\_shell}, \code{run\_shell\_command}, \code{run\_terminal\_cmd}, \code{shell}, \code{terminal}\\
\code{http} (14) & \code{WebFetch}, \code{WebSearch}, \code{fetch}, \code{http\_get}, \code{http\_post}, \code{http\_request}, \code{mcp:fetch/fetch}, \code{requests\_delete}, \code{requests\_get}, \code{requests\_patch}, \code{requests\_post}, \code{requests\_put}, \code{srt:request}, \code{web\_fetch}\\
\code{fs} (7) & \code{Edit}, \code{Glob}, \code{Grep}, \code{MultiEdit}, \code{NotebookEdit}, \code{Read}, \code{Write}\\
\code{payment} (12) & \code{create\_charge}, \code{create\_payment}, \code{create\_payout}, \code{create\_refund}, \code{create\_transfer}, \code{get\_balance}, \code{get\_transaction}, \code{list\_payments}, \code{list\_transactions}, \code{pay\_invoice}, \code{send\_payment}, \code{transfer\_funds}\\
\code{sql} (10) & \code{execute\_query}, \code{execute\_sql}, \code{mcp:mysql/execute\_sql}, \code{mcp:postgres/query}, \code{mcp:sqlite/read\_query}, \code{mcp:sqlite/write\_query}, \code{mcp:supabase/execute\_sql}, \code{query\_database}, \code{run\_sql}, \code{sql\_db\_query}\\
\code{email} (3) & \code{gmail\_send\_message}, \code{send\_email}, \code{send\_gmail\_message}\\
\bottomrule
\end{tabularx}
\end{center}
Only \code{shell} (for \code{tests\_passed}) and \code{http} (for \code{reached\_host}) are read by a 0.1 checker.
\code{file\_written} reads only the tool \code{Write}.

\paragraph{Field patterns}
\begin{center}\small
\begin{tabular}{@{}lll@{}}
\toprule
\textbf{Name} & \textbf{Length} & \textbf{Pattern (whole value, ASCII classes)}\\
\midrule
IDENTITY & $\le 256$ & \code{\^{}[a-z0-9\_]+:[\^{}\textbackslash n]+\$}\\
ID & 1--128 & \code{\^{}[A-Za-z0-9.\_:-]+\$}\\
SCHEMA\_REF & $\le 136$ & \code{\^{}[a-z][a-z0-9\_.-]\{0,63\}@sha256:[0-9a-f]\{64\}\$}\\
TS & $\le 27$ & \code{\^{}[0-9]\{4\}-[0-9]\{2\}-[0-9]\{2\}T[0-9]\{2\}:[0-9]\{2\}:[0-9]\{2\}\textbackslash.[0-9]\{6\}Z\$}\\
DIGEST & 71 & \code{\^{}sha256:[0-9a-f]\{64\}\$}\\
\bottomrule
\end{tabular}
\end{center}

\section{Design for Fresh Evidence}
\label{app:fresh}

This appendix summarises the design that the specification proposes for closing the stale-evidence class (L10) in a
later draft.

\paragraph{The run since the call}
The evidence profile already chains each run: every record carries \code{run\_seq}, its position in its run, and
\code{run\_prev\_hash}, the hash of the run's previous record. The design adds a signer-set envelope member
$\code{head} = \{\code{run\_seq}: n, \code{hash}: \textsc{digest}\}$, the last record of the sender's run when the signer
took the message to sign. It is set like \code{ts}, so the agent cannot choose it. The design also adds an evidence item
marked \code{"since": true}, whose records are the claim's records followed by \emph{every} record of the run after the
first of them, up to \code{head}. The verifier checks that consecutive records have consecutive \code{run\_seq} values
and linked \code{run\_prev\_hash} values, and that the last record matches \code{head}. Because the run chain binds
each record to the next and the message signature binds \code{head}, no record between the call and the message can be
omitted. The checkers then read the later records:
\begin{itemize}
\item \code{tests\_passed}: a later \code{shell} call of the same suite whose result is not \code{ok} fails the claim
  (``the suite failed after it passed''). A later unopened \code{shell} call with no result or a failed one makes the
  claim \unprov{} until the sender opens it.
\item \code{file\_written}: a later write of the claimed path with other content fails the claim. A later write whose
  path is not shown makes it \unprov. This requires the table to separate writing from reading tools within \code{fs}.
\end{itemize}
The cost is about 1\,KB per later record, and a cap bounds how long after the call a sender can claim. The design does
not cover runs of the suite in \emph{other} runs, writes by other processes (the receiver's digest check covers those),
or earlier flaky runs.

\paragraph{Bound claims}
The executor records what a call ran on. For \code{tests\_passed\{suite, tree\}}, the agent names a git commit, and the
executor checks out its tree into a fresh directory the agent cannot write, runs the suite there and records the tree
id. The receiver compares that id with the commit it will merge (\code{git rev-parse C\^{}\{tree\}}). Hashing the
agent's working directory instead would not do, because an ignored \code{conftest.py} changes the result without
changing the id. For \code{file\_written}, the executor reads the file back after writing and records the absolute path
and the digest of the bytes on disk. That record alone then proves the claim without opening the decision, at the cost
of the path appearing in clear in the log.

\paragraph{Responses with an error status}
For \code{reached\_host}, the evidence profile would record whether the call got a response, for example its protocol
status, so that a 404 proves the host was reached.

\section{Models of Real Protocols}
\label{app:protocols}

\Cref{tab:protomodels} summarises what each receiver model checks and the specification sections it rests on. The
code is \code{bench/lies/protocols.py}. Every key is Ed25519, derived from the holder's name and the seed, and a receiver
looks up a public key as its protocol does (a DID document, an Almanac entry, a JWKS).

\begin{table}[h]
\centering\small
\caption{Receiver models used in RQ5.}
\label{tab:protomodels}
\renewcommand{\arraystretch}{1.15}
\begin{tabularx}{\linewidth}{@{}L{2.7cm} Y L{3.4cm}@{}}
\toprule
\textbf{Receiver} & \textbf{What it checks} & \textbf{What it cannot see}\\
\midrule
A2A 1.0 + signed card & Server authenticates by a scheme the card declares; client verifies the card's JWS over its RFC~8785 form (spec \S7.3, \S7.4, \S8.4). Messages, tasks and artefacts are unsigned; \code{messageId} de-duplication is optional (off). & every message-level lie\\
uAgents envelope & Sender signs SHA-256 of sender, target, session, schema digest, payload, expires and nonce; receiver verifies and refuses a target it does not host. & replays to the same target; all content\\
RFC 9421 + replay & Signature over \code{@method}, \code{@target-uri}, \code{content-digest} with created/expires/nonce/keyid; 300\,s window; in-memory (keyid, nonce) cache. & restarts; floors; all content\\
SCITT receipts & Each claim must carry statements, each with a content hash and an inclusion receipt against a tree head a pinned log signed. & claim-to-record semantics, tiers, run binding, replay, grants\\
AP2 0.2 mandates & KB-JWT (\code{cnf}, \code{aud}, nonce, \code{iat}); open mandate signed by a trusted platform, intent, constraints, \code{exp}; checkout mandates of the same args/tool, disclosed. & tests, hosts, files; delegation\\
Verifiable Intent 0.1 & As AP2, with mandates signed by a person's device key that an issuer certifies (L1). & as AP2\\
\bottomrule
\end{tabularx}
\end{table}

\section{Conformance Vector Catalogue}
\label{app:vectors}

The 187 vectors, grouped by area. Colour shows the expected verdict:
\textcolor{pcmaqua!80!black}{$\blacksquare$}~\vacc,
\textcolor{pcmyellow!85!black}{$\blacksquare$}~\vunp,
\textcolor{pcmred}{$\blacksquare$}~\vrej,
\textcolor{pcmviolet}{$\blacksquare$}~\vbad.

\newcommand{\vgroup}[2]{\par\smallskip\noindent{\sffamily\footnotesize\bfseries #1 \textmd{(#2)}}\par\nopagebreak}
\newcommand{\vitem}[2]{{\scriptsize\noindent\hangindent=1em\textcolor{#1}{$\blacksquare$}\,\vect{#2}\par}}
\begin{multicols}{2}
\raggedright
\vgroup{Envelope, parsing and signatures}{10}
\vitem{pcmaqua!80!black}{accept-identity-line-break-chars}
\vitem{pcmaqua!80!black}{accept-jcs-payload}
\vitem{pcmviolet}{bad-input-number-past-double}
\vitem{pcmred}{reject-claim-witness-key-in-logs}
\vitem{pcmred}{reject-envelope-extra-field}
\vitem{pcmred}{reject-envelope-pcm-false}
\vitem{pcmred}{reject-forged-signature}
\vitem{pcmred}{reject-sig-non-canonical-base64}
\vitem{pcmred}{reject-small-order-key}
\vitem{pcmred}{reject-unpinned-key}
\vgroup{Schemas and intents}{7}
\vitem{pcmaqua!80!black}{accept-schema-field-types}
\vitem{pcmred}{reject-schema-bool-as-int}
\vitem{pcmred}{reject-schema-extra-field}
\vitem{pcmred}{reject-schema-field-types}
\vitem{pcmred}{reject-schema-int-as-string}
\vitem{pcmred}{reject-schema-money-over-max}
\vitem{pcmred}{reject-unknown-schema}
\vgroup{Replay and recipient}{10}
\vitem{pcmaqua!80!black}{accept-replay-misaddressed-spends-no-nonce}
\vitem{pcmaqua!80!black}{accept-replay-same-agent-other-log}
\vitem{pcmaqua!80!black}{accept-replay-schema-rejected-spends-no-nonce}
\vitem{pcmaqua!80!black}{accept-skipped-replay-and-recipient}
\vitem{pcmred}{reject-replay-before-started}
\vitem{pcmred}{reject-replay-duplicate-nonce}
\vitem{pcmred}{reject-replay-floor-arrival-order}
\vitem{pcmred}{reject-replay-no-real-time}
\vitem{pcmred}{reject-replay-outside-window}
\vitem{pcmred}{reject-wrong-recipient}
\vgroup{Grants and delegation}{33}
\vitem{pcmaqua!80!black}{accept-delegation-in-set}
\vitem{pcmaqua!80!black}{accept-delegation-less-than}
\vitem{pcmaqua!80!black}{accept-delegation-narrowing}
\vitem{pcmaqua!80!black}{accept-delegation-time-bound}
\vitem{pcmaqua!80!black}{accept-grant-expires-at-ts}
\vitem{pcmred}{reject-delegation-over-child-cap}
\vitem{pcmred}{reject-grant-amount-over-caveat}
\vitem{pcmred}{reject-grant-bad-caveat}
\vitem{pcmred}{reject-grant-below-non-delegable}
\vitem{pcmred}{reject-grant-broken-parent}
\vitem{pcmred}{reject-grant-caveat-bool-for-int}
\vitem{pcmred}{reject-grant-caveat-string-for-int}
\vitem{pcmred}{reject-grant-child-outlives-parent}
\vitem{pcmred}{reject-grant-child-wrong-issuer}
\vitem{pcmred}{reject-grant-cross-tenant}
\vitem{pcmred}{reject-grant-expired}
\vitem{pcmred}{reject-grant-from-another-log}
\vitem{pcmred}{reject-grant-in-set-widened-child}
\vitem{pcmred}{reject-grant-issuer-is-subject}
\vitem{pcmred}{reject-grant-missing-caveat-child}
\vitem{pcmred}{reject-grant-needed}
\vitem{pcmred}{reject-grant-other-intent}
\vitem{pcmred}{reject-grant-root-by-agent}
\vitem{pcmred}{reject-grant-root-with-parent}
\vitem{pcmred}{reject-grant-root-without-person}
\vitem{pcmred}{reject-grant-signature-as-message}
\vitem{pcmred}{reject-grant-subject-not-sender}
\vitem{pcmred}{reject-grant-tampered}
\vitem{pcmred}{reject-grant-time-bound-widened-child}
\vitem{pcmred}{reject-grant-unpinned-key}
\vitem{pcmred}{reject-grant-widen-and-tighten-child}
\vitem{pcmred}{reject-grant-widening-child}
\vitem{pcmred}{reject-replay-grant-failure-spends-nonce}
\vgroup{Evidence (generic)}{15}
\vitem{pcmyellow!85!black}{accept-unproven-claim-without-evidence}
\vitem{pcmyellow!85!black}{accept-unproven-no-evidence}
\vitem{pcmaqua!80!black}{accept-witnessed}
\vitem{pcmred}{reject-claim-changed-record-hash}
\vitem{pcmred}{reject-claim-cut-proof}
\vitem{pcmred}{reject-claim-evidence-from-another-log}
\vitem{pcmred}{reject-claim-evidence-from-another-run}
\vitem{pcmred}{reject-claim-evidence-from-another-tenant}
\vitem{pcmred}{reject-claim-opening-wrong-args}
\vitem{pcmred}{reject-claim-proof-non-canonical-base64}
\vitem{pcmred}{reject-claim-record-listed-twice}
\vitem{pcmred}{reject-claim-record-not-an-event}
\vitem{pcmred}{reject-claim-salt-without-args}
\vitem{pcmred}{reject-claim-unpinned-log-note}
\vitem{pcmred}{reject-claim-witnesses-required-unmet}
\vgroup{\texttt{tests\_passed}}{25}
\vitem{pcmaqua!80!black}{accept-tests-passed-npm-test}
\vitem{pcmaqua!80!black}{accept-tests-passed-python-m-pytest}
\vitem{pcmaqua!80!black}{accept-tests-passed-t2}
\vitem{pcmyellow!85!black}{accept-unproven-t1-result}
\vitem{pcmyellow!85!black}{accept-unproven-unplaced-tool}
\vitem{pcmred}{reject-claim-bash-c}
\vitem{pcmred}{reject-claim-hash-inside-word}
\vitem{pcmred}{reject-claim-pytest-or-true}
\vitem{pcmred}{reject-claim-tests-failed}
\vitem{pcmred}{reject-claim-tests-passed-V}
\vitem{pcmred}{reject-claim-tests-passed-ansi-c-quote}
\vitem{pcmred}{reject-claim-tests-passed-background}
\vitem{pcmred}{reject-claim-tests-passed-co}
\vitem{pcmred}{reject-claim-tests-passed-collect-only}
\vitem{pcmred}{reject-claim-tests-passed-comment}
\vitem{pcmred}{reject-claim-tests-passed-denied-call}
\vitem{pcmred}{reject-claim-tests-passed-dot-slash}
\vitem{pcmred}{reject-claim-tests-passed-h}
\vitem{pcmred}{reject-claim-tests-passed-he}
\vitem{pcmred}{reject-claim-tests-passed-help}
\vitem{pcmred}{reject-claim-tests-passed-malformed-no-evidence}
\vitem{pcmred}{reject-claim-tests-passed-runner-path}
\vitem{pcmred}{reject-claim-tests-passed-unbalanced-quote}
\vitem{pcmred}{reject-claim-tests-passed-version}
\vitem{pcmred}{reject-claim-tests-passed-wrapper}
\vgroup{\texttt{approved\_by}}{26}
\vitem{pcmaqua!80!black}{accept-approved-by}
\vitem{pcmaqua!80!black}{accept-approved-by-covers-direct}
\vitem{pcmaqua!80!black}{accept-approved-by-covers-grant}
\vitem{pcmaqua!80!black}{accept-approved-by-opened-args}
\vitem{pcmyellow!85!black}{accept-unproven-approved-by-covers-not-opened}
\vitem{pcmyellow!85!black}{accept-unproven-approved-by-covers-schema-without-tool}
\vitem{pcmyellow!85!black}{accept-unproven-approved-by-required-without-covers}
\vitem{pcmred}{reject-claim-approval-of-another-call}
\vitem{pcmred}{reject-claim-approved-by-agent-approver}
\vitem{pcmred}{reject-claim-approved-by-args-true-and-1}
\vitem{pcmred}{reject-claim-approved-by-covers-args-true-and-1}
\vitem{pcmred}{reject-claim-approved-by-covers-grant-not-root}
\vitem{pcmred}{reject-claim-approved-by-covers-other-args}
\vitem{pcmred}{reject-claim-approved-by-covers-other-tool}
\vitem{pcmred}{reject-claim-approved-by-decision-of-another-call}
\vitem{pcmred}{reject-claim-approved-by-decision-without-tool}
\vitem{pcmred}{reject-claim-approved-by-other-decision}
\vitem{pcmred}{reject-claim-approved-by-rejected-answer}
\vitem{pcmred}{reject-claim-approved-by-request-names-two-decisions}
\vitem{pcmred}{reject-claim-approved-by-request-of-another-approval}
\vitem{pcmred}{reject-claim-approved-by-self-approval}
\vitem{pcmred}{reject-claim-approved-by-two-decisions}
\vitem{pcmred}{reject-claim-approved-by-unmapped-identity}
\vitem{pcmred}{reject-claim-approved-by-without-decision}
\vitem{pcmred}{reject-claim-decision-without-id}
\vitem{pcmred}{reject-claim-wrong-person}
\vgroup{\texttt{reached\_host}}{48}
\vitem{pcmaqua!80!black}{accept-reached-host-a-label}
\vitem{pcmaqua!80!black}{accept-reached-host-claim-port}
\vitem{pcmaqua!80!black}{accept-reached-host-dotted-quad}
\vitem{pcmaqua!80!black}{accept-reached-host-http}
\vitem{pcmaqua!80!black}{accept-reached-host-ipv4-mapped}
\vitem{pcmaqua!80!black}{accept-reached-host-nat64}
\vitem{pcmaqua!80!black}{accept-reached-host-observed}
\vitem{pcmaqua!80!black}{accept-reached-host-trailing-dot}
\vitem{pcmyellow!85!black}{accept-unproven-reached-host-ambiguous}
\vitem{pcmyellow!85!black}{accept-unproven-reached-host-blob}
\vitem{pcmyellow!85!black}{accept-unproven-reached-host-bracket}
\vitem{pcmyellow!85!black}{accept-unproven-reached-host-bracket-ipv6-userinfo}
\vitem{pcmyellow!85!black}{accept-unproven-reached-host-bracket-no-host}
\vitem{pcmyellow!85!black}{accept-unproven-reached-host-bracket-not-host}
\vitem{pcmyellow!85!black}{accept-unproven-reached-host-bracket-other-host}
\vitem{pcmyellow!85!black}{accept-unproven-reached-host-bracket-trailing}
\vitem{pcmyellow!85!black}{accept-unproven-reached-host-browser-brackets}
\vitem{pcmyellow!85!black}{accept-unproven-reached-host-claim-bad-port}
\vitem{pcmyellow!85!black}{accept-unproven-reached-host-claim-no-host}
\vitem{pcmyellow!85!black}{accept-unproven-reached-host-claim-not-quad}
\vitem{pcmyellow!85!black}{accept-unproven-reached-host-deviation}
\vitem{pcmyellow!85!black}{accept-unproven-reached-host-fullwidth-ipv6}
\vitem{pcmyellow!85!black}{accept-unproven-reached-host-hex-ipv4}
\vitem{pcmyellow!85!black}{accept-unproven-reached-host-idn}
\vitem{pcmyellow!85!black}{accept-unproven-reached-host-ipv4-leading-zero}
\vitem{pcmyellow!85!black}{accept-unproven-reached-host-ipv4-number}
\vitem{pcmyellow!85!black}{accept-unproven-reached-host-ipv6-leading-zero}
\vitem{pcmyellow!85!black}{accept-unproven-reached-host-ipv6-leading-zero-backslash}
\vitem{pcmyellow!85!black}{accept-unproven-reached-host-ipv6-leading-zero-later-octet}
\vitem{pcmyellow!85!black}{accept-unproven-reached-host-label-character}
\vitem{pcmyellow!85!black}{accept-unproven-reached-host-leading-space}
\vitem{pcmyellow!85!black}{accept-unproven-reached-host-mapped-to-nothing}
\vitem{pcmyellow!85!black}{accept-unproven-reached-host-no-connection}
\vitem{pcmyellow!85!black}{accept-unproven-reached-host-non-ascii-path}
\vitem{pcmyellow!85!black}{accept-unproven-reached-host-not-ipv6}
\vitem{pcmyellow!85!black}{accept-unproven-reached-host-nul}
\vitem{pcmyellow!85!black}{accept-unproven-reached-host-percent}
\vitem{pcmyellow!85!black}{accept-unproven-reached-host-relative-path}
\vitem{pcmyellow!85!black}{accept-unproven-reached-host-sensor-connect}
\vitem{pcmyellow!85!black}{accept-unproven-reached-host-short-ipv4}
\vitem{pcmyellow!85!black}{accept-unproven-reached-host-tab-in-escape}
\vitem{pcmyellow!85!black}{accept-unproven-reached-host-two-brackets}
\vitem{pcmyellow!85!black}{accept-unproven-reached-host-two-trailing-dots}
\vitem{pcmyellow!85!black}{accept-unproven-reached-host-unreadable-claim}
\vitem{pcmred}{reject-claim-reached-host-denied-connect}
\vitem{pcmred}{reject-claim-reached-host-file-url}
\vitem{pcmred}{reject-claim-reached-host-other-host}
\vitem{pcmred}{reject-claim-reached-host-too-long}
\vgroup{\texttt{file\_written}}{10}
\vitem{pcmaqua!80!black}{accept-file-written}
\vitem{pcmaqua!80!black}{accept-file-written-drive-path}
\vitem{pcmyellow!85!black}{accept-unproven-file-written-drive-separators}
\vitem{pcmyellow!85!black}{accept-unproven-file-written-relative-call}
\vitem{pcmred}{reject-claim-file-written-backslash}
\vitem{pcmred}{reject-claim-file-written-dotdot}
\vitem{pcmred}{reject-claim-file-written-not-write-tool}
\vitem{pcmred}{reject-claim-file-written-relative}
\vitem{pcmred}{reject-claim-file-written-relative-no-evidence}
\vitem{pcmred}{reject-claim-file-written-wrong-digest}
\vgroup{Claims (generic) and verdicts}{3}
\vitem{pcmyellow!85!black}{accept-unproven-required-claim-missing}
\vitem{pcmred}{reject-claim-malformed-unpinned-note-line}
\vitem{pcmred}{reject-strict-required-unproven}

\end{multicols}

\end{document}